\documentclass[twocolumn]{aastex63}
\usepackage{aas_macros}
\usepackage{color}
\usepackage{soul}
\usepackage{natbib}
\usepackage{hyperref}
\usepackage{amsmath}
\usepackage{xspace}
\usepackage{lineno}
\usepackage{graphicx}
\usepackage{enumitem}
\usepackage{amssymb}
\usepackage{xifthen}
\usepackage{hyperref}
\usepackage[normalem]{ulem}

\hypersetup{    
  colorlinks      = {true},
  linkcolor       = {blue},
  citecolor       = {blue},
  urlcolor        = {blue},
}

\newcommand{\code}[1]{\texttt{#1}\xspace}

\newcommand{\SSSSS}{${S}^5$\xspace}
\newcommand{\gaia}{\textit{Gaia}\xspace}
\newcommand{\Gaia}{\gaia}

\newcommand{\unit}[1]{\ensuremath{\mathrm{\,#1}}\xspace}
\newcommand{\feh}{\unit{[Fe/H]}}

\newcommand{\vhel}         {\mbox{$v_{\mathrm{hel}}$}}
\newcommand{\vgsr}         {\mbox{$v_{\mathrm{gsr}}$}}

\newcommand{\kms}{\unit{km\,s^{-1}}}
\newcommand{\masyr}{\unit{mas\,yr^{-1}}}

\defcitealias{Torrealba16}{T16}
\defcitealias{Torrealba19}{T19}

\shorttitle{Antlia 2 and Crater 2}
\shortauthors{Ji et al.}

\begin{document}

\title{Kinematics of Antlia 2 and Crater 2 from The Southern Stellar Stream Spectroscopic Survey (\SSSSS)}

\author[0000-0002-4863-8842]{Alexander~P.~Ji}
\affiliation{Observatories of the Carnegie Institution for Science, 813 Santa Barbara St., Pasadena, CA 91101, USA}
\affiliation{Department of Astronomy \& Astrophysics, University of Chicago, 5640 S Ellis Avenue, Chicago, IL 60637, USA}
\affiliation{Kavli Institute for Cosmological Physics, University of Chicago, Chicago, IL 60637, USA}

\author[0000-0003-2644-135X]{Sergey~E.~Koposov}
\affiliation{Institute for Astronomy, University of Edinburgh, Royal Observatory, Blackford Hill, Edinburgh EH9 3HJ, UK}
\affiliation{Institute of Astronomy, University of Cambridge, Madingley Road, Cambridge CB3 0HA, UK}
\affiliation{Kavli Institute for Cosmology, University of Cambridge, Madingley Road, Cambridge CB3 0HA, UK}

\author[0000-0002-9110-6163]{Ting~S.~Li}
\affiliation{Observatories of the Carnegie Institution for Science, 813 Santa Barbara St., Pasadena, CA 91101, USA}
\affiliation{Department of Astrophysical Sciences, Princeton University, Princeton, NJ 08544, USA}
\affiliation{Department of Astronomy and Astrophysics, University of Toronto, 50 St. George Street, Toronto ON, M5S 3H4, Canada}
\affiliation{NHFP Einstein Fellow}

\author[0000-0002-8448-5505]{Denis~Erkal}
\affiliation{Department of Physics, University of Surrey, Guildford GU2 7XH, UK}

\author[0000-0002-6021-8760]{Andrew~B.~Pace}
\affiliation{McWilliams Center for Cosmology, Carnegie Mellon University, 5000 Forbes Ave, Pittsburgh, PA 15213, USA}

\author[0000-0002-4733-4994]{Joshua~D.~Simon}
\affiliation{Observatories of the Carnegie Institution for Science, 813 Santa Barbara St., Pasadena, CA 91101, USA}

\author[0000-0002-0038-9584]{Vasily~Belokurov}
\affiliation{Institute of Astronomy, University of Cambridge, Madingley Road, Cambridge CB3 0HA, UK}

\author[0000-0001-8536-0547]{Lara~R.~Cullinane}
\affiliation{Research School of Astronomy and Astrophysics, Australian National University, Canberra, ACT 2611, Australia}
\author[0000-0001-7019-649X]{Gary~S.~Da~Costa}
\affiliation{Research School of Astronomy and Astrophysics, Australian National University, Canberra, ACT 2611, Australia}
\affiliation{Centre of Excellence for All-Sky Astrophysics in Three Dimensions (ASTRO 3D), Australia}
\author[0000-0003-0120-0808]{Kyler~Kuehn}
\affiliation{Lowell Observatory, 1400 W Mars Hill Rd, Flagstaff,  AZ 86001, USA}
\affiliation{Australian Astronomical Optics, Faculty of Science and Engineering, Macquarie University, Macquarie Park, NSW 2113, Australia}
\author[0000-0003-3081-9319]{Geraint~F.~Lewis}
\affiliation{Sydney Institute for Astronomy, School of Physics, A28, The University of Sydney, NSW 2006, Australia}
\author[0000-0002-6529-8093]{Dougal~Mackey}
\affiliation{Research School of Astronomy and Astrophysics, Australian National University, Canberra, ACT 2611, Australia}
\author[0000-0003-2497-091X]{Nora~Shipp}
\affiliation{Department of Astronomy \& Astrophysics, University of Chicago, 5640 S Ellis Avenue, Chicago, IL 60637, USA}
\affiliation{Kavli Institute for Cosmological Physics, University of Chicago, Chicago, IL 60637, USA}
\author[0000-0002-8165-2507]{Jeffrey~D.~Simpson}
\affiliation{School of Physics, UNSW, Sydney, NSW 2052, Australia}
\affiliation{Centre of Excellence for All-Sky Astrophysics in Three Dimensions (ASTRO 3D), Australia}
\author[0000-0003-1124-8477]{Daniel~B.~Zucker}
\affiliation{Department of Physics \& Astronomy, Macquarie University, Sydney, NSW 2109, Australia}
\affiliation{Macquarie University Research Centre for Astronomy, Astrophysics \& Astrophotonics, Sydney, NSW 2109, Australia}
\author[0000-0001-6154-8983]{Terese~T.~Hansen}
\affiliation{George P. and Cynthia Woods Mitchell Institute for Fundamental Physics and Astronom, Texas A\&M University, College Station, TX 77843, USA}
\affiliation{Department of Physics and Astronomy, Texas A\&M University, College Station, TX 77843, USA}
\author[0000-0001-7516-4016]{Joss~Bland-Hawthorn}
\affiliation{Sydney Institute for Astronomy, School of Physics, A28, The University of Sydney, NSW 2006, Australia}
\affiliation{Centre of Excellence for All-Sky Astrophysics in Three Dimensions (ASTRO 3D), Australia}

\collaboration{17}{(\SSSSS Collaboration)}

\correspondingauthor{A.~P.~Ji}
\email{alexji@uchicago.edu}

\begin{abstract}
We present new spectroscopic observations of the diffuse Milky Way satellite galaxies Antlia 2 and Crater 2, taken as part of the Southern Stellar Stream Spectroscopic Survey (\SSSSS).
The new observations approximately double the number of confirmed member stars in each galaxy and more than double the spatial extent of spectroscopic observations in Antlia 2.
A full kinematic analysis, including \Gaia EDR3 proper motions, detects a clear velocity gradient in Antlia 2 and a tentative velocity gradient in Crater 2.
The velocity gradient magnitudes and directions are consistent with particle stream simulations of tidal disruption.
Furthermore, the orbit and kinematics of Antlia 2 require a model that includes the reflex motion of the Milky Way induced by the Large Magellanic Cloud.
We also find that Antlia 2's metallicity was previously overestimated, so it lies on the empirical luminosity-metallicity relation and is likely only now experiencing substantial stellar mass loss.
Current dynamical models of Antlia 2 require it to have lost over 90\% of its stars to tides, in tension with the low stellar mass loss implied by the updated metallicity.
Overall, the new kinematic measurements support a tidal disruption scenario for the origin of these large and extended dwarf spheroidal galaxies.
\end{abstract}


\section{INTRODUCTION}
\label{intro}

Antlia 2 (Ant2) and Crater 2 (Cra2) are dwarf spheroidal (dSph) satellite galaxies of the Milky Way with present-day stellar masses $\sim 10^6 M_\odot$ but unusually large half-light radii, greater than 1 kpc.
Ant2 was discovered by a search through the \gaia DR2 all-sky astrometric survey (\citealt{Torrealba19}, henceforth T19).
Ant2 has an astonishingly large half-light radius of almost 3~kpc, as large as the Large Magellanic Cloud (LMC) but with a tiny fraction ($10^{-4}$) of its stellar mass, implying a surface brightness of ${\sim}32$ mag arcsec$^{-2}$, the lowest average surface brightness for any detected galaxy to date.
Cra2 was discovered using deep photometry in the ATLAS survey (\citealt{Torrealba16}, henceforth T16), with a half-light radius of ${\sim}1$~kpc and surface brightness of ${\sim}30$ mag arcsec$^{-2}$.
\citet{Caldwell17}, surprisingly, found that Cra2 had an exceptionally low velocity dispersion of only 2.7 \kms, suggesting an unusually underdense dark matter halo.
Previously known Milky Way satellites of similar total luminosity (like the Ursa Minor, Draco, and Sextans dSphs), which were discovered decades ago \citep[e.g.,][]{Wilson55,Irwin90}, are typically a few hundred parsecs in size and so of higher surface brightness, making Ant2 and Cra2 two of the lowest surface brightness galaxies in the known universe.
The closest analogues are several satellites of M31, including Andromeda XIX and XXI \citep{Collins20,Collins21}.

The central question is whether these galaxies' extreme properties can be explained by tides.
Galaxies at Ant2 and Cra2's luminosity are usually more compact and have a larger velocity dispersion \citep[e.g.,][]{Caldwell17,Simon19,Torrealba19,Collins20,Collins21}.
Tidal interactions between a dwarf galaxy and the Milky Way can (but do not always) increase the radius and decrease the velocity dispersion of the dwarf galaxy, producing remnants with properties similar to Ant2 and Cra2 \citep[e.g.,][]{Penarrubia08,Errani15,Sanders18,Fattahi18,Fu19}.
Current evidence does suggest both Ant2 and Cra2 have been substantially affected by tides.
\citetalias{Torrealba19} found that Ant2 is off the empirical luminosity-metallicity relation \citep{Kirby13}, suggesting it lost over 90\% of its stellar mass, and also that Ant2 has an orbit bringing it within 30 kpc of the Milky Way center.
\citet{Fu19} found Cra2 also has a pericenter within 30 kpc of the Milky Way center, and its current properties can be explained if Cra2 has lost ${\sim}90\%$ of its stellar mass to tides (also see \citealt{Sanders18}).
The plausibility of this scenario depends on whether these galaxies reside in cuspy or cored dark matter halos \citep{Sanders18,Fu19,Torrealba19,Sameie20}.
Alternatively, others have proposed that the properties of such low surface brightness galaxies are perhaps better explained with Modified Newtonian Dynamics (MOND) than dark matter \citep[e.g.,][]{McGaugh16}. 
However, given the low surface brightness and large on-sky extent of Ant2 and Cra2, additional kinematic measurements are needed.

In this paper, we present new spectroscopic observations of Ant2 and Cra2 from the Southern Stellar Stream Spectroscopic Survey \citep[\SSSSS ;][]{Li19}. Using the wide field of view and high multiplexing of 2dF on AAT, we have roughly doubled the number of known spectroscopic members in Ant2 and Cra2 compared to the existing literature, and we have also doubled the radial extent of observations in Ant2. In addition,
we have included the substantially improved proper motions from \gaia EDR3 \citep{Lindegren20} as part of our analysis.
Our main result is a clear detection of a velocity gradient in Ant2 that strongly suggests it has recently experienced substantial tidal disruption. We also tentatively detect a velocity gradient in Cra2.
Section~\ref{sec:spec} describes our spectroscopic observations, data reduction, and velocity/metallicity measurements.
We update the luminosity and spatial parameters for Ant2 in Section~\ref{sec:spatial} using \gaia EDR3.
Section~\ref{sec:kinematics} gives the results of our 6D kinematic analysis for both galaxies.
Section~\ref{sec:mdf} describes our fit to the metallicity distribution functions.
Section~\ref{sec:dynamics} describes the orbital analysis.
We compare to previous results and discuss implications for galaxy formation, dark matter, and Modified Newtonian Dynamics in Section~\ref{sec:discussion}, then conclude in Section~\ref{sec:summary}. 

\section{Spectroscopic Data}\label{sec:spec}

\subsection{Observations and Target Selection}

Both Ant2 and Cra2 were observed as part of the \emph{Southern Stellar Stream Spectroscopic Survey} (\SSSSS). \SSSSS uses the AAOmega spectrograph on the 3.9~m Anglo-Australian Telescope (AAT), fed by the Two Degree Field (``2dF") fiber positioner facility; see \citet{Li19} for details on the survey strategy, target selection, observation, and reduction of \SSSSS data. 
Although \SSSSS focuses on the physics of stellar streams as the main science goal, we observed Ant2 and Cra2, as both galaxies were postulated to have tidal stripping features in the original discovery papers \citepalias{Torrealba16,Torrealba19}.
Furthermore, the substantial extent of the galaxies on the sky makes them ideal targets for the AAT's large field of view.

Ant2 was observed with five AAT pointings between Feb 27 2020 and Mar 1 2020. Each field was observed with three 40 min exposures. The fields were arranged with a central pointing plus two overlapping pointings along the minor axis and two along the major axis. Although the central pointing was not strictly necessary to achieve complete sky coverage, we included it because the target density in the central field is higher than the available fibers allow for in a single configuration (392 fibers, including sky fibers). 

The targets for Ant2 were selected based on the photometry, parallax and proper motions from \gaia DR2 \citep{gaiadr2,Lindegren18}.
Specifically, we used dereddened photometry calculated assuming the color-dependent extinction corrections from \citet{Babusiaux2018} and $E(B-V)$ values from \citet{Schlegel:1998}.
We first constructed an empirical fiducial isochrone in dereddened $G$ vs $G-RP$ using a list of member stars in the Milky Way's ultra faint dwarf galaxies compiled in \citet{Pace19}.
We shifted the isochrone to the distance modulus of Ant2 ($\mu = 20.6$), then selected stars within 0.15 mag from the isochrone in $G-RP$ and brighter than 19.5 mag in de-reddened $G-$band.
The ultra-faint dwarf galaxy stars are typically more metal-poor than Ant2, but the color selection window is wide enough to not bias the metallicity distribution of Ant2.
In proper motion space, we selected targets with 
$$|\mu_{\alpha*} - \mu_{{\alpha*},0}| < \mathrm{max}(k \sigma_{\mu,\alpha*}, 0.3)$$
$$|\mu_\delta - \mu_{\delta,0}| < \mathrm{max}(k \sigma_{\mu,\delta}, 0.3)$$
where $\mu_{{\alpha*},0}$ and $\mu_{\delta,0}$ are the proper motion of Ant2 from \citetalias{Torrealba19}, and $k$ defines priority classes. Since Ant2 is near the Galactic Plane ($b = 10^\circ$), the foreground contamination is relatively high. In order to maximize the target efficiency, we divided the targets into different priorities, with $k = 0.5,\,1,\,2$ as high, medium, and low priority targets, respectively.
Finally, we restricted the targets to have
$\varpi < 3\sigma_\varpi$,
where $\varpi$ is parallax and $\sigma_\varpi$ is the parallax error.

The selection criteria above provide about 400-500 available targets per AAT field. These targets are then used as input for fiber allocation using the software \code{configure}\footnote{https://www.aao.gov.au/science/software/configure} \citep{Miszalski:2006ef}. Each AAT field contains about 360 fibers assigned to Ant2 targets, 25 sky fibers, and 8 fibers for alignment stars.
 
We observed the two outermost fields along the semi-major axis on Feb 27, 2020, the two outermost fields along the semi-minor axis on Feb 29, and the central field on Mar 1. To maximize the number of Ant2 members, quick reductions were performed after every night and targets with signal-to-noise ratio per pixel (S/N) larger than 6 in the red arm were moved to the lowest priority category for the observations in the next night, when the targets were also in another AAT field. In total, with 5 AAT pointings, we observed 1110 Ant2 targets, of which 950 were marked as stars by our pipeline (the \code{good\_star} flag, \citealt{Li19}), and 508 have S/N $>$ 4 that are used in this work.

\begin{figure*}
    \centering
    \includegraphics[width=\linewidth]{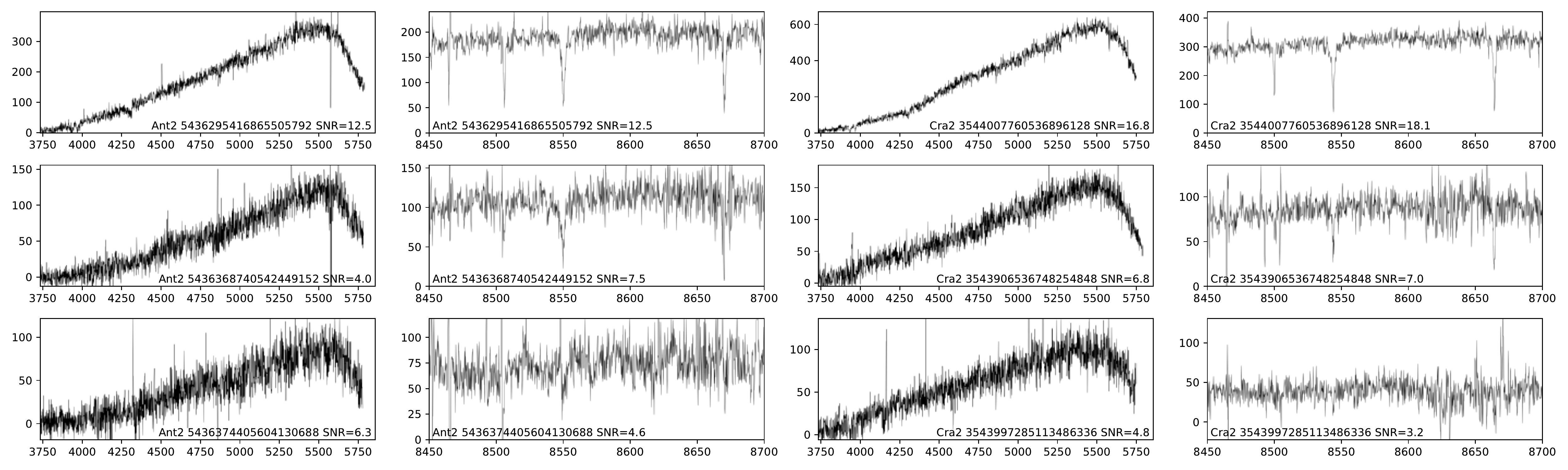}
    \caption{Representative spectra for three random stars in Ant2 (left two columns) and Cra2 (right two columns). The blue and red spectra for a given star are shown separately, with the SNR for each arm indicated on the figure.
    The top row shows high SNR stars with clear absorption lines.
    The middle row shows moderate SNR stars, at the edge of what we consider acceptable for Ca triplet equivalent widths.
    The bottom row shows low SNR stars, enough to obtain a velocity but not a good Ca triplet equivalent width.}
    \label{fig:spectra}
\end{figure*}

For Cra2, since the size of the galaxy on the sky is smaller, only one AAT field centered at Cra2 was observed. We obtained two 2250~s exposures on Feb 29 and two 2700~s exposures on Mar 1 for Cra2, with identical target lists and fiber allocations for both nights. The target selection procedure is very similar to that described above for Ant2, with the exception that only 250 targets are selected with the combined parallax, proper motion (using \citealt{Fritz18} for the mean Cra2 motion), and photometry selection.
Therefore, we added an additional $\sim100$ targets that are outside the isochrone selection window but have proper motions consistent with Cra2. In total we observed 354 targets, of which 301 were marked as stars by our pipeline and 215 have S/N $>$ 3 that are used here.
Note we use a less stringent S/N cut for Cra2 because there is much less contamination in the field, but using an S/N $>4$ cut makes no difference to our results.
Examples of blue and red spectra are shown in Figure~\ref{fig:spectra}.

\subsection{Data Reduction and Analysis}

The data used in this paper are from an internal \SSSSS data release (DR2.2) where the analysis is improved compared to \citet[][previously DR1.4]{Li19} that was publicly released \citep{s5release}.
Previously, we fit the higher-resolution red calcium triplet (CaT) spectrum and the blue low-resolution spectrum separately.
We now use the ability of the \code{rvspecfit} code \citep{rvspecfit} to perform simultaneous modeling of multiple spectra to fit the red and blue spectra as 
well as repeated observations of the same object from different nights, with proper consideration of the heliocentric correction from each observation. 
Since the red spectra have a spectral resolution 8x higher than the blue spectra and therefore give higher precision in RV, we allow a velocity offset
between the blue and red spectra during the fit. However, we emphasize that for many objects stellar parameters like $T_{\rm eff}$, $\log g$ and [Fe/H] gain more 
information from blue arm spectra because of the much larger spectral coverage, i.e., 3700--5700~\AA\ vs. 8400--8800~\AA. We also use a 
photometric prior on $T_{\rm eff}$ similar to the one described in \citet{Li19} that relies on DECam $g-r$ and $r-z$ colors or \gaia $G-RP$ color for sources without DECam photometry. 
Additionally, rather than relying on the original PHOENIX stellar atmosphere grid \citep{Husser13}, that has substantial step sizes 
of $0.5-1$ dex in the [Fe/H] grid and occasional grid gaps, we use a refined grid with smaller step sizes. This grid is obtained by performing 
global Radial Basis Function interpolation on a rectangularly spaced grid without gaps in $T_{\rm eff}$, $\log g$, [Fe/H] and [$\alpha$/Fe].
This grid has a step size of 0.25 for both [Fe/H] and [$\alpha$/Fe]. Because of the grid's regularity, we then use a multi-linear interpolation 
as opposed to linear Delaunay triangulation interpolation. This tends to improve the stellar parameters and metallicities, reducing clustering of 
measured parameters around the grid points. 

The new processing pipeline does not substantially affect the radial velocity measurements other than for objects
with multiple observations, where the accuracy is improved through simultaneous modeling of spectra. The radial velocities and uncertainties are 
re-calibrated the same way as described in \citet{Li19}, including validation of the zero-point against APOGEE DR16 \citep{Jonsson20} and GALAH DR3 \citep{Buder20}.
We adopt the same corrections to the velocities as in \citet{Li19}. Thanks to the use of red and blue spectra, we find the metallicities in DR2.2 are 
more accurate than for DR1.4.
Full catalogs are available in Appendix~\ref{sec:tables} (Tables~\ref{tab:ant2} and \ref{tab:cra2} for Ant2 and Cra2, respectively), including membership probabilities from Section~\ref{sec:kinematics}.

Previous verification tests with high-resolution spectroscopy have found that \code{rvspecfit} is not as accurate or precise as calcium triplet metallicities when distances to stars are known \citep{Li19,Li20,Ji20b,Wan20}.
Thus, we also determine CaT metallicities from equivalent widths and the \citet{Carrera13} calibration, which requires absolute $V$ magnitudes.
Equivalent widths were measured by fitting a Gaussian plus Lorentzian function, with a minimum systematic uncertainty of 0.2{\AA} that typically translates into 0.13 dex \citep{Li17}. Visual inspection of the fits shows that only stars with S/N $>5$ should be considered to have reliable CaT metallicities.
The absolute $V$ magnitudes are determined from \gaia EDR3 $G$, $BP$, and $RP$ photometry, first applying the filter transformations in \citet{Riello20}, then dereddening using \citet{Schlegel:1998,Schlafly11} and adding the distance moduli in Table~\ref{tab:galprops}.
The \gaia bandpasses are large and the filter transformations or dereddening could be uncertain, but we find that using the $V$ magnitude transformations for Pan-STARRS 1 photometry in Cra2 \citep{Chambers16} and the NOIRLab Source Catalog (NSC) DR2 photometry for Ant2 \citep{Nidever20} makes an insignificant $<0.03$ dex difference in the measured metallicities. We thus use the \gaia photometry to be consistent across both galaxies, and because the NSC DR2 photometry is incomplete for Ant2.

In Ant2, both \code{rvspecfit} in \SSSSS and the CaT metallicities are substantially lower than the metallicities inferred in \citetalias{Torrealba19}, by about 0.5 dex.
This is a zero-point offset in the \citetalias{Torrealba19} metallicities, which were fit with an early version of \code{rvspecfit}.
To confirm this, we have re-analyzed the original data used in \citetalias{Torrealba19} using both the new \code{rvspecfit} and the calcium triplet, obtaining a lower metallicity. We have also verified the lower metallicities of a few stars using high-resolution spectroscopy (Ji et al. in prep). Thus, we believe our updated metallicity measurement is more accurate.

For galaxy member stars, the CaT metallicities are preferred over the \code{rvspecfit} spectrum synthesis metallicities, as they have both better accuracy and precision \citep[e.g.,][]{Li20,Wan20}.
However, the \code{rvspecfit} metallicities are applicable to both galaxy and foreground stars, so they are used in mixture models for determining galaxy membership.
Some stars in DR2.2 have erroneous metallicities of $\mbox{[Fe/H]} \sim 0$, which will be remedied in future \SSSSS analyses.
However, the overall performance of \code{rvspecfit} is still quite good, obtaining metallicities only biased high by 0.1 dex compared to the calcium triplet in our two galaxies.
We provide both metallicities in Tables \ref{tab:ant2} and \ref{tab:cra2}.
\begin{figure}
    \centering
    \includegraphics[width=\linewidth]{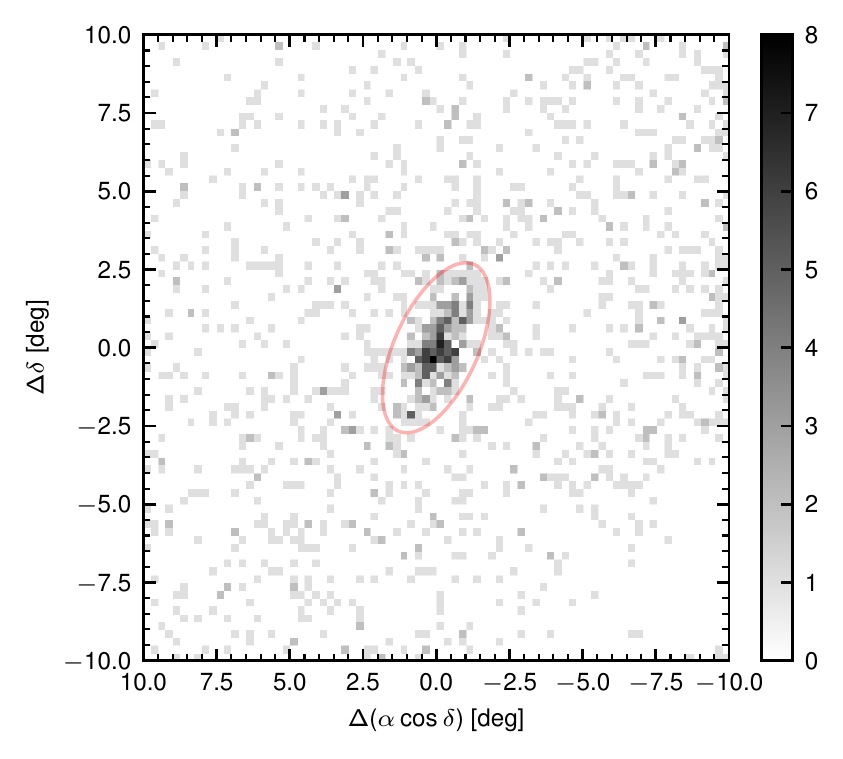}
    \caption{The on-sky density distribution of color-magnitude, proper motion and parallax selected stars for Ant2 from \gaia EDR3. The red ellipse indicates twice the half-light radius. These stars were used for the spatial model described in Section \ref{sec:spatial}.}
    \label{fig:ant2spatial}
\end{figure}

\section{Structural Parameters}\label{sec:spatial}

For Cra2, we adopt the luminosity and spatial parameters from \citetalias{Torrealba16}, a spherical Plummer profile \citep{Dejonghe87} with half light radius of 31.2 arcmin.
\citet{Vivas20} also determined spatial parameters from deeper DECam data but with a smaller field of view, finding evidence for a small ellipticity $e=0.12 \pm 0.02$. We include their measurement in Table~\ref{tab:galprops} but do not use it in membership determination.

Since Ant2 is heavily blended with Milky Way foreground stars, its structural parameters require including \gaia astrometry to remove the contamination.
\citet{Torrealba19} used \gaia DR2 to determine structural parameters.
We now use the better photometry and astrometry in \gaia EDR3 to redetermine the structural parameters of Ant2. 
We use data from a ten degree radius circular area on the sky around Ant2 with $E(B-V)_{SFD}<0.75$ to avoid the possible incompleteness due to high extinction, then select stars with dereddened $G_0 < 20$. We select likely Ant2 members as stars with parallaxes $\omega - 0.01 < 2 \sigma_\omega$,  proper motions in $\alpha$ and $\delta$ consistent with the mean Ant2 proper motion in Table~\ref{tab:galprops} $|\mu - \mu_0|< \text{min}(2\sigma_\mu,0.3)$, and dereddened $G-RP$ colors within 0.075 mag of the red giant branch track. 

The on-sky density of selected stars is shown in Figure~\ref{fig:ant2spatial}.
This spatial density was modeled as an elliptical Plummer profile  
plus a background density with a quadratic spatial gradient.
Stars were sorted into HEALPIX\footnote{\url{http://healpix.sourceforge.net}} equal area bins with \code{nside=512} (${\approx}$6.9 arcmin on a side). The likelihood function is  the standard Poisson likelihood for independent bins, with the Poisson rate given by the equation:
\begin{equation}
\begin{split}
\rho(x,y|x_0,y_0,e,PA,a_{h},I,{\bf b})= \exp(I) \times \nonumber \\
\left(1+\frac{1}{a_{h}^2}
\left[\left(\frac{(x-x_0)\cos(PA)-(y-y_0)\sin(PA)}{1-e}\right)^2+\nonumber \right. \right. \\
\left. \left. \left((y-y_0)\cos(PA)+(x-x_0)\sin(PA)\right)^2\right]\right)^{-2} + \nonumber\\
\exp(b_0+ b_x x+ b_y y + b_{xy} x y + b_{xx} x^2 + b_{yy} y^2)
\end{split}
\end{equation}
where $x,y$  are coordinates of stars in the tangential projection, $x_0,y_0$ is the center of the object, $e$ is the ellipticity, $PA$ the positional angle, $a_{h}$ the size along the major axis, $I$ the logarithm of the central surface brightness, and $b_0,b_x,b_y,b_{xx},b_{xy},b_{yy}$ are background parameters.
The prior is uniform in all parameters aside from $x_0,y_0\sim N(0,1)$. The posterior was sampled using \code{Stan} \citep{stan} and summarized using the 16th/50th/84th percentiles in Table~\ref{tab:galprops}. In the table we also provide the inferred circularized half-light radius $r_h= a_{h}\sqrt{1-e} $.
Overall our measurements are more precise than but consistent with \citetalias{Torrealba19}, except for the ellipticity $e= 0.60 \pm 0.04$, which is about $2\sigma$ larger than in \citetalias{Torrealba19}, who measured $e = 0.38 \pm 0.08$.

We also redetermine the galaxy luminosity broadly following \citet{Munoz18}. We refit the spatial density profile using significantly wider selection criteria in proper motion, parallax and color to avoid incompleteness. Integrating the Plummer profile fit, we find $N=750 \pm 50$ stars brighter than our cutoff of $G_0=20.0$.
Using the luminosity function from a \citet{Dotter08} isochrone (12Gyr, [Fe/H]=$-2.0$, [$\alpha$/Fe]=$+0.4$) and a fixed distance modulus of 20.6 \citepalias{Torrealba19}, this corresponds to $M_V = -9.86 \pm 0.08$.
Using MIST isochrones instead yielded a similar $M_V = -9.83$ \citep{Choi16}.
The surface brightness within the half light radius is then $30.7$ mag arcsec$^{-2}$.
Because we find Ant2's luminosity to be much higher and the radius to be slightly smaller ($r_h=66.3$ arcmin) than \citetalias{Torrealba19} (who had $M_V=-9.03$ and $r_h=76.2$ arcmin), we obtain a surface brightness higher by over 1 mag arcsec$^{-2}$.
Finally, we tested whether two Plummer components were a better fit to the data, but we found no substantial improvement to the log-likelihood with the extra component.

\section{Membership Modeling and Kinematic Properties}\label{sec:kinematics}

\begin{deluxetable*}{lccl}
\tablecolumns{4}
\tablecaption{\label{tab:galprops}Galaxy Properties}
\tablehead{Parameter & Antlia 2 & Crater 2 & Description}
\startdata
  $\alpha$ (deg) & $143.8079 \pm 0.0492$ & 177.310 (2) & Galaxy center RA \\
  $\delta$ (deg) & $-36.6991 \pm 0.0800$ & -18.413 (2) & Galaxy center Dec \\
  $\mu$ (mag) & $20.6 \pm 0.11$ (1) & $20.35 \pm 0.07$ (2,3) & Distance modulus\\
  $d$ (kpc) & 131.8 (1) & 117.5 (2) & Distance \\
  $M_V$ & $-9.86 \pm 0.08$\tablenotemark{a} & $-8.2$ (2) & Luminosity \\
  $r_h$ (arcmin) & $66.3 \pm 4.6$ & $31.2 \pm 2.5$ (2) & Circularized half light radius\\
  $r_h$ (pc) & $2541 \pm 175$ & $1066 \pm 84$ (2) & Circularized half light radius\\
  $a_h$ (arcmin) & $104.6 \pm 8.6$ & \nodata & Half light major axis\\
  $a_h$ (pc) & $4010 \pm 329$ & \nodata & Half light major axis\\
  $e$ ($1-b/a$) & $0.60 \pm 0.04$ & $0.12 \pm 0.02$ (3)\tablenotemark{b} & Ellipticity \\
  $\theta_{\text{PA}}$ (deg) & $154.0 \pm 2.4$ & $135 \pm 4$ (3)\tablenotemark{b} & Position angle (E. of N.)\\
  $\mu_V$ (mag arcsec$^{-2}$) & $30.7 \pm 0.2$ & $30.6 \pm 0.2$ (2) & Surface brightness within half light radius \\
\hline
  $\vhel$ ($\kms$) & $+288.8 ^{+0.4} _{-0.4}$ & $+89.3 ^{+0.3} _{-0.3}$ & Heliocentric radial velocity \\
  $\mu_{\alpha^*}$ ($\masyr$) & $-0.094 ^{+0.007} _{-0.007}$ & $-0.073 ^{+0.021} _{-0.021}$ & Heliocentric proper motion, RA $\cos \delta$\tablenotemark{c}\\
  $\mu_{\delta}$ ($\masyr$) & $+0.103 ^{+0.008} _{-0.008}$ & $-0.123 ^{+0.013} _{-0.013}$ & Heliocentric proper motion, Dec.\tablenotemark{c}\\
\hline
  $\vgsr$ (\kms) & $49.9^{+0.4}_{-0.4}$ & $-81.4 ^{+0.3} _{-0.3}$ & Galactic Standard of Rest, radial velocity \\
  $\mu_{\alpha^*,\text{gsr}}$ (\masyr) & $-0.047^{+0.007}_{-0.007}$ & $+0.133 ^{+0.021} _{-0.021}$ & Galactic Standard of Rest, proper motion, RA $\cos \delta$\tablenotemark{c}\\
  $\mu_{\delta,\text{gsr}}$ (\masyr) & $+0.179 ^{+0.008} _{-0.008}$ & $+0.118 ^{+0.013} _{-0.013}$ & Galactic Standard of Rest, proper motion, Dec\tablenotemark{c}\\
  $\sigma_v$ (\kms) & $5.98^{+0.37}_{-0.36}$ & $2.34 ^{+0.42} _{-0.30}$ & Radial velocity dispersion \\
  $k_v$ (\kms/deg) & $5.72^{+0.60}_{-0.56}$ & $2.19 ^{+1.18} _{-1.20}$ & Linear radial velocity gradient\\
  $k_v$ (\kms/kpc) & $2.49^{+0.26}_{-0.25}$ & $1.07 ^{+0.57}_{-0.58}$ & Linear radial velocity gradient\\
  $\theta_v$ (deg) & $+174 ^{+9} _{-10}$ & $+97 ^{+29} _{-35}$ & RV gradient direction (E. of N.) \\
\hline
  $\sigma_v$, no $k_v$ (\kms) & $7.69^{+0.40}_{-0.37}$ & $2.43 ^{+0.54} _{-0.35}$ & Radial velocity dispersion without linear gradient \\
\hline
  $\left<\feh\right>$ (dex, DR2.2) & $-1.77 ^{+0.08} _{-0.08}$ & $-2.10 ^{+0.08} _{-0.08}$ & Mean metallicity using full spectrum fit \tablenotemark{d} \\
  $\sigma_{\text{Fe}}$ (dex, DR2.2) & $0.66 ^{+0.03} _{-0.03}$ & $0.34 ^{+0.03} _{-0.03}$ & Metallicity dispersion \tablenotemark{d} \\
  $k_{\text{Fe}}$ (dex/deg, DR2.2) & $+0.10 ^{+0.09} _{-0.09}$ & $-0.05 ^{+0.20} _{-0.20}$ & Metallicity radial gradient \\
\hline
  $\left<\feh\right>$ (dex, CaT) & $-1.90 \pm 0.04$ & $-2.16 \pm 0.04$ & Mean metallicity using calcium triplet EW \\
  $\sigma_{\text{Fe}}$ (dex, CaT) & $0.34 \pm 0.03$ & $0.24 \pm 0.05$ & Metallicity dispersion \\
\hline
  $N_{\text{mem}}$ & 283 & 141 & Number of clear member stars in our sample ($P > 0.95$). \\
  $N_{\text{tot}}$ & 288.7 & 141.0 & Sum of membership probabilities \\
\hline
  $M_{\text{dyn}}$ ($M_\odot$) & $10^{7.92 \pm 0.09}$ & $10^{6.74 \pm 0.21}$ & Dynamical mass within half light radius \citep{Wolf10} \\
  $L_V$ ($L_\odot$) & $10^{5.88}$ & $10^{5.21}$ & Luminosity in $L_\odot$ assuming $M_{V,\odot} = +4.83$ \\
  $M_\star$ ($M_\odot$) & $10^{6.22}$ & $10^{5.55}$ & Stellar mass in $M_\odot$ assuming $M_\star/L_V = 2.2$ \\
  $M/L$ ($M_\odot/L_\odot$) & 100 & 31 & Mass to light ratio within half light radius \\
\enddata
\tablerefs{(1) \citealt{Torrealba19}, (2) \citealt{Torrealba16}, (3) \citealt{Vivas20}}
\tablenotetext{a}{This is the statistical uncertainty in Poisson star counts. The actual uncertainty may be higher, see text.}
\tablenotetext{b}{Not used in the Cra2 spatial likelihood, see text.}
\tablenotetext{c}{The systematic error in each proper motion component is not included but is an additional 0.023 \masyr \citep{Lindegren20}.}
\tablenotetext{d}{Calcium triplet metallicities are preferred over the DR2.2 values. The metallicity dispersion from the full spectrum fit is particularly influenced by a few outliers with inaccurate metallicities.}
\tablecomments{Values without references are determined in this work.}
\end{deluxetable*}

\begin{figure*}
    \centering
    \includegraphics[width=\linewidth]{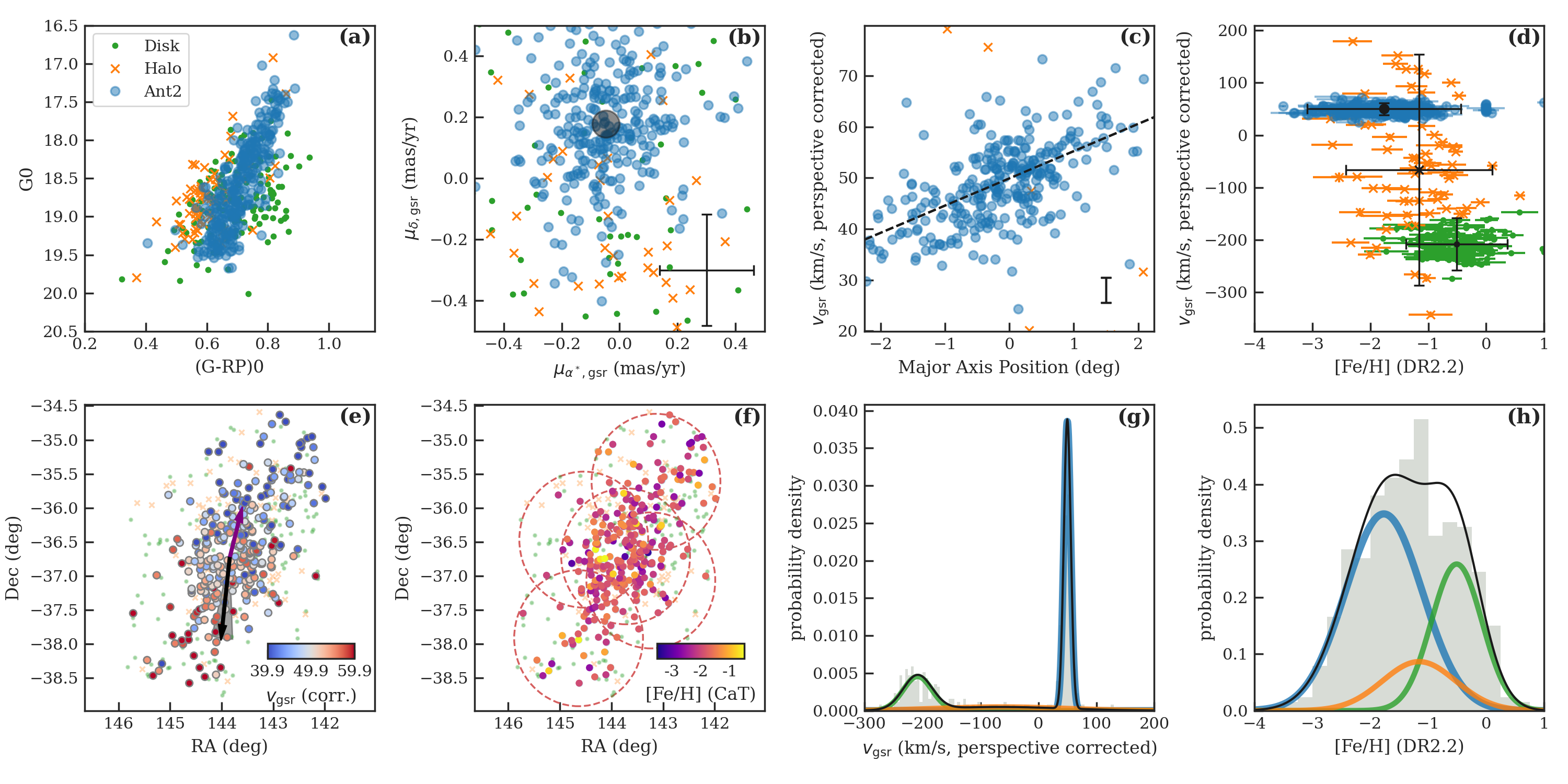}
    \caption{Antlia 2 mixture model fit.
    \emph{(a)} color-magnitude diagram of dereddened \gaia EDR3 photometry \citep[using dereddening prescription from][]{Babusiaux2018}, color-coded by membership (blue circles = Ant2, orange crosses = halo, green dots = disk). The CMD is not used in determining members.
    \emph{(b)} \gaia EDR3 proper motions, corrected to Galactic Standard of Rest. Large black dot indicates the mean value for Ant2. Black error bars indicate the median proper motion uncertainty in the whole sample.
    \emph{(c)} $\vgsr$ corrected for perspective motion plotted against major axis ($\theta_{PA}=154^\circ$). The best-fit linear gradient is shown as a dashed black line (projected to the major axis from the actual velocity gradient direction). Black error bar indicates the median velocity uncertainty in the whole sample.
    \emph{(d)} $\vgsr$ and \SSSSS DR2.2 [Fe/H] (from full spectrum fit), color-coded by membership. Black symbols indicate the component mean, while error bars indicate 2${\times}$ the intrinsic standard deviation of that component.
    \emph{(e)} position of spectroscopically observed stars. Member stars are color-coded by $\vgsr$, corrected for the perspective motion effect. The purple arrow indicates the GSR proper motion. The black arrow indicates the direction of the velocity gradient from low to high velocity, with the underlying grey shaded wedge indicating the $1\sigma$ uncertainty.
    \emph{(f)} position of spectroscopically observed stars, with member stars color-coded by [Fe/H] from the \emph{calcium triplet}. Large dotted circles indicate the AAT pointings.
    \emph{(g)} histogram of $\vgsr$, showing the best-fit individual components in blue, orange, and green; and the sum in black.
    \emph{(h)} histogram of \SSSSS DR2.2 [Fe/H] from full spectrum fit, showing the best-fit individual components in blue, orange, and green; and the sum in black.
    }
    \label{fig:ant2summary}
\end{figure*}

\begin{figure*}
    \centering
    \includegraphics[width=\linewidth]{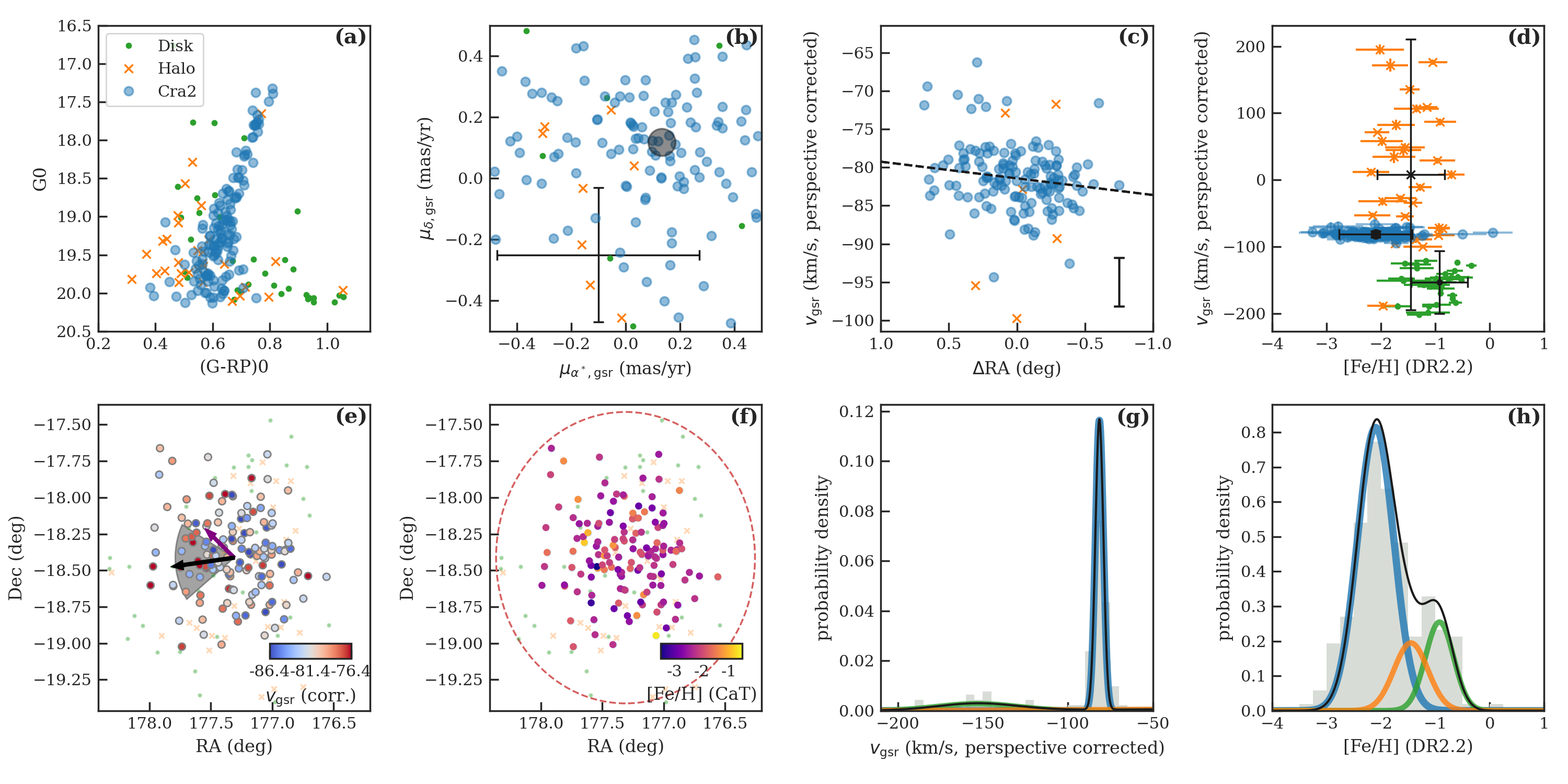}
    \caption{Crater 2 mixture model fit. The panels are the same as Figure~\ref{fig:ant2summary}, except panel (c) has $\Delta$RA on the x-axis.
    }
    \label{fig:cra2summary}
\end{figure*}

Stellar membership and kinematic properties are measured using a combination of sky position, radial velocity and metallicity from \code{rvspecfit}, and \gaia EDR3 proper motions.
We impose quality cuts of S/N $> 3$ for Cra2, and S/N $>4$ for Ant2 because there is more contamination for the latter.
If multiple exposures were taken for one star, we use the S/N of the best individual exposure for the quality cut.
We also require a velocity error $< 10\kms$, as visual inspection shows stars with larger velocity errors are often substantially affected by sky subtraction residuals.
No CMD information was included, but it was already used during spectroscopic target selection.
We have verified that ignoring any one of the spatial positions, metallicities, or proper motions does not significantly change the membership or fitted parameters of our mixture model.

Before fitting, we identified possible binary stars using previous observation epochs.
Stars were considered binary candidates if their radial velocities differed by more than $3\sigma$ in different epochs.
For Ant2, we used our re-analyzed observations of the data taken in \citetalias{Torrealba19}, finding 7 likely binary stars among the Ant2 members.
For Cra2, we used literature velocities from \citet{Caldwell17} and \citet{Fu19}. There is a 1 \kms systematic offset in our velocities of the matched stars, but after removing this offset 3 likely binary stars were identified.
These stars were given velocity errors of 999 \kms to effectively remove their velocities from the fit, and they are marked as binary candidates in Tables~\ref{tab:ant2} and \ref{tab:cra2}.
We expect the true binary fraction to be closer to 50\% \citep[e.g.,][]{Spencer18}, which could affect Cra2 and its very low velocity dispersion, but it is unlikely to affect Ant2 with its higher velocity dispersion.
Similarly, stars with metallicity uncertainties larger than 0.5 dex were given errors of 99 dex, because these metallicities are usually due to fitting sky residuals.

These data were fit with a three-component mixture model: one galaxy component and two foreground components (for the halo and disk foregrounds).
The foreground components are assumed to be uniform both spatially and in proper motion space.
The foreground radial velocities and metallicities are each modeled as 1-D Gaussians.
For the galaxies, we model their spatial, 3D velocity, and metallicities including a linear velocity gradient and radial metallicity gradient (the detailed likelihoods are described elsewhere, e.g., \citealt{Walker16,Caldwell17,Pace20}).
The spatial components are modeled as Plummer profiles using the spatial parameters from Table~\ref{tab:galprops} (Section~\ref{sec:spatial} for Ant2, \citetalias{Torrealba16} for Cra2). We hold the spatial parameters fixed to the photometric values to avoid biases from the spectroscopic selection.
The galaxy radial velocity is modeled as having a linear velocity gradient with Gaussian intrinsic velocity dispersion around the gradient.
The galaxy [Fe/H] is modeled as a Gaussian with a radial gradient for the mean.
The galaxy proper motion is modeled to be a single value for the whole galaxy with no dispersion (current proper motion uncertainties cannot resolve a tangential velocity dispersion).
Note that our spectroscopic selection function for proper motions is rather complicated and depends on the proper motion uncertainties (Section~\ref{sec:spec}), but we have verified that implementing a proper motion background model that more accurately reflects the exact selection makes no difference to the final fitted parameters or member stars.
The posterior is sampled using \code{emcee} \citep{emcee} with 64 walkers and 10,000 steps per chain.

Our coverage for both Ant2 and Cra2 spans at least a 1 degree radius on the sky. Thus, the corrections for both the differential Solar reflex motion and the perspective motion (requiring a known proper motion) are crucial to obtain accurate kinematics.
For the Solar reflex motion, we perform all our fits with velocities that have been corrected to the Galactic Standard of Rest (GSR, $\vgsr$), assuming the local standard of rest velocity of $232.8 \kms$ \citep{McMillan17}, a relative Solar velocity of $(11.1,\, 12.24,\, 7.25) \kms$ \citep{Schonrich10}, and other parameters set to the default values in \code{astropy} version 4.0 (in particular a distance of 8.122 kpc to the Galactic center, \citealt{GRAVITY18}).
The radial velocity differences due to a differential reflex correction across Ant2 and Cra2 are ${\pm}1.0$ and ${\pm}1.7\kms$, for fields of view of ${\pm}2.2$ deg and ${\pm}0.7$ deg, respectively.
The proper motions over the same spatial extent are differentially affected by ${\pm}0.010$\masyr and ${\pm}0.014$\masyr for Ant2, and less than ${\pm}0.005$\masyr for Cra2.

Additionally, we include the perspective motion effect on the measured radial velocities and proper motions \citep[e.g.,][]{Kaplinghat08,Walker08}. The perspective motion effect is important for radial velocities (up to ${\pm}3.0$ and ${\pm}1.5 \kms$ for Ant2 and Cra2, respectively, across their fields of view), but not for proper motions (less than 0.005 $\masyr$).
Practically, we parameterize each galaxy's velocity in GSR coordinates using the radial velocity $\vgsr$ and the proper motions $\mu_{\alpha^*,\text{gsr}}$ and $\mu_{\delta,\text{gsr}}$ at the center of the galaxy to define a Cartesian vector.
To compare to data, this Cartesian velocity is converted to the observed values $\vhel$, $\mu_{\alpha^*}$, and $\mu_\delta$ for each star at its $\alpha,\delta$ using \code{astropy} reference frames.
We have also accounted for correlations between proper motions and parallax in \gaia, using the known distance to the galaxies (see Appendix~\ref{sec:pmcorrection}). This does not substantially affect Ant2, but decreases $\mu_\delta$ for Cra2 by 0.01 $\masyr$.

The results of the mixture model fits are provided in Table~\ref{tab:galprops}, with membership probabilities included in Tables~\ref{tab:ant2} and \ref{tab:cra2}.
Figures~\ref{fig:ant2summary}~and~\ref{fig:cra2summary} visually show the model parameters compared to several data dimensions. In these figures, all observed radial velocities and proper motions have been corrected to the Galactic Standard of Rest (GSR), with the effect of perspective motion and Solar reflex motion removed.
In panels (a), (b), (c), and (d), the observed stars have been color-coded by the component for which they have over 50\% membership probability. Members in Ant2 and Cra2 are shown as large blue circles, while the foreground disk and halo are shown as small green points and orange crosses.
Panels (d) and (g) show that the radial velocity alone very cleanly separates both galaxies from the foreground populations. The metallicity, proper motion, and spatial information play a relatively minor role (though the latter two are important parts of our spectroscopic selection).
Note that our CMD color selection is wide enough to pick up stars substantially redder and bluer than the member stars (panels (a)), so we do not expect a metallicity bias from our selection.

\begin{figure}
    \centering
    \includegraphics[width=\linewidth]{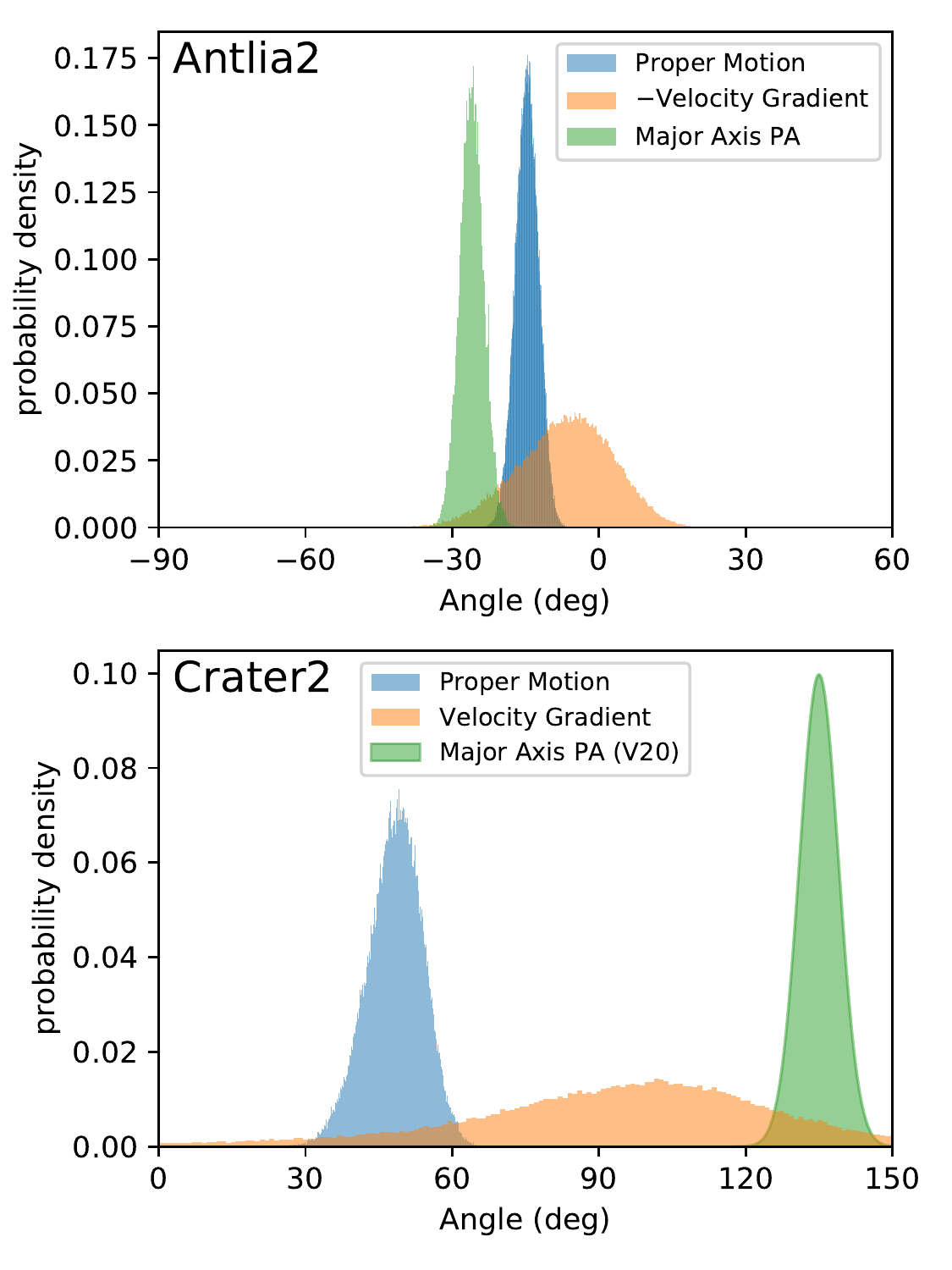}
    \caption{Posterior distributions for important angles in Ant2 (top panel) and Cra2 (bottom panel).
    All angles are East of North.
    The proper motion direction has been reflex corrected.
    In the top panel, the velocity gradient has been flipped by 180$^\circ$.
    In the bottom panel, the major axis PA is taken from \citet{Vivas20} and assumed to be Gaussian.
    All angles are roughly aligned in Ant2, but the small difference between the proper motion and spatial orientation is significant. This is an effect of the Large Magellanic Cloud (Section~\ref{sec:dynamics}).
    }
    \label{fig:angles}
\end{figure}

Panels (c) and (e) show the radial velocity as a function of position.
We first consider Ant2, which displays a clear linear velocity gradient of $k_v = 5.72 ^{+0.60}_{-0.56}$ \kms/deg ($2.49^{+0.26}_{-0.25}$ \kms/kpc), a $>5\sigma$ detection that is visually apparent in Figure~\ref{fig:ant2summary}c,e.
The best-fit direction of the gradient (pointing from low to high $\vgsr$) is indicated by the black arrow in panels (e), and in both galaxies the gradient direction is roughly on the same axis as the proper motion vector (magenta arrows in panels (e)).
This is shown more quantitatively in Figure~\ref{fig:angles}, where the posterior distributions of the best-fit position angles of the (reflex-corrected) proper motion, velocity gradient, and elliptical position angle are shown.
In Ant2 the best-fit angles are all fairly similar, but there is a significant difference between the major axis orientation and proper motion direction (see Section~\ref{sec:dynamics}).

\begin{figure}
    \centering
    \includegraphics[width=\linewidth]{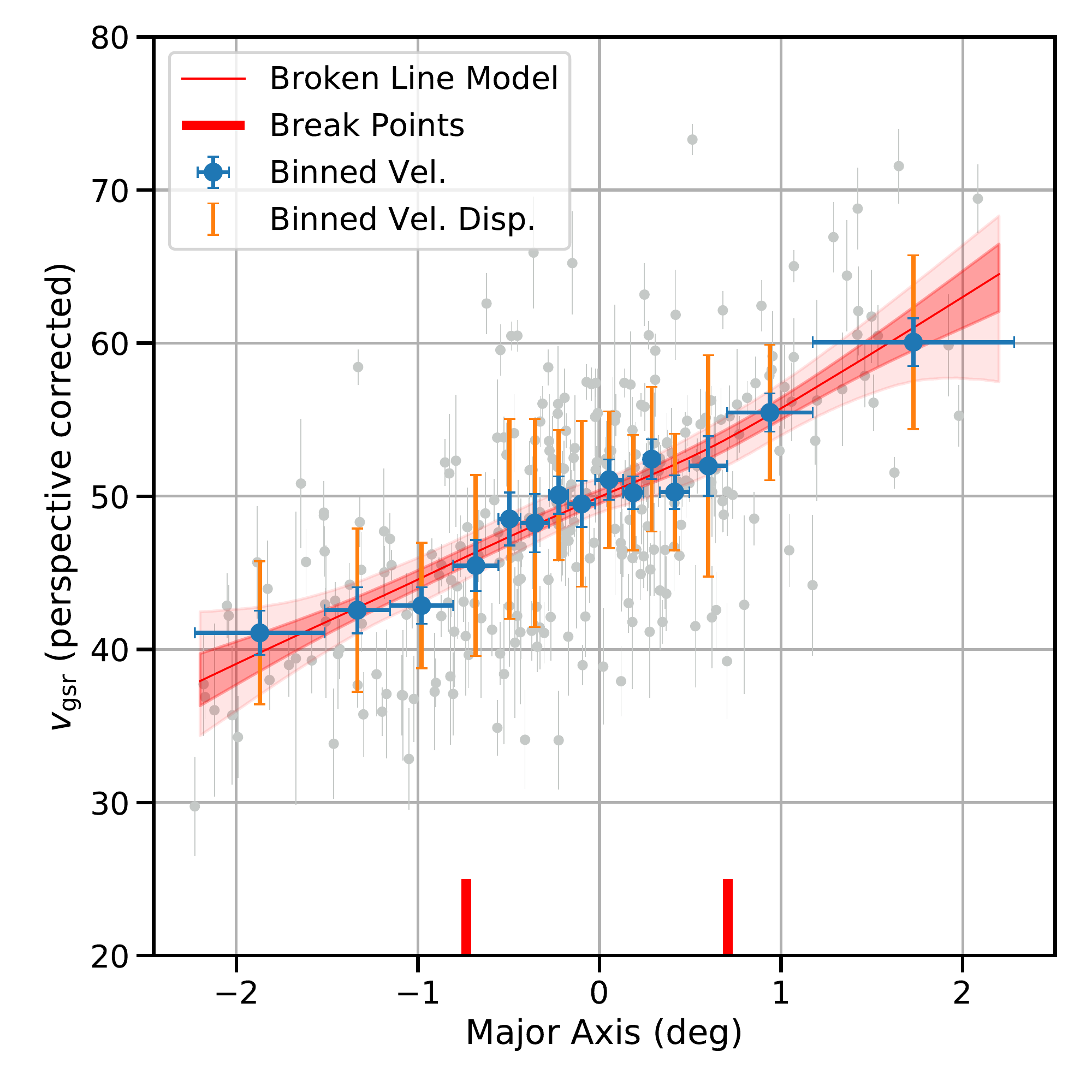}
    \caption{Testing for a nonlinear velocity trend in Ant2.
    The data for each member star is shown in light grey.
    We consider two models.
    First, we fit fifteen separate bins of velocities and velocity dispersions: the blue circles and error bars are the mean and error on the mean in each bin, the orange error bars are the velocity dispersion.
    Second, we fit three continuous line segments with a constant velocity dispersion. The red line is the mean model, the shaded region indicates 1$\sigma$ and 2$\sigma$ confidence intervals for the mean, and the break points are indicated by short red lines at the bottom of the figure. 
    While the inner 1 degree of the galaxy may have a slight flattening, a single velocity gradient is still a good description of the current data.
    }
    \label{fig:ant2vel}
\end{figure}

Simulations of tidally disrupting systems often show an ``S''-shaped velocity profile centered around the disrupting progenitor \citep[e.g.,][also see Section~\ref{sec:particlesim}]{Erkal17}.
We briefly investigate the presence of such structure in Figure~\ref{fig:ant2vel}.
First, we separate stars into 15 bins of roughly equal size along the major axis, fitting a velocity mean (blue points) and dispersion (orange bars) within each bin.
Second, we model the velocities as a set of three line segments with two variable breakpoints (red lines at the bottom of the figure) with a constant velocity dispersion, so as to be able to detect any sharp changes in slopes.
The binned velocities do suggest there may be a small flattening at the center of Ant2. The broken line model suggests the data at positive major axis may have higher slope than the rest of Ant2. But overall, neither of these models appears substantially more compelling than a single linear slope given current data, and we leave more detailed modeling to the community.

Next, we consider the possible velocity gradient in Cra2.
Figure~\ref{fig:cra2summary}c,e shows that there may be a slight gradient aligned roughly in the RA direction, but it is detected at only 1.8$\sigma$ ($k_v = 2.19^{+1.18}_{-1.20}$ \kms/deg) and its direction is very poorly constrained (Figure~\ref{fig:angles}).
We consider this a tentative detection of a velocity gradient, sufficient for us to later broadly discuss but not make claims about the implications of a velocity gradient in Cra2.
Examining just the proper motion and spatial orientation, Figure~\ref{fig:angles} does show a clear difference between the direction of the precise \Gaia EDR3 proper motion and the orientation of the elliptical fit to the galaxy's shape by \citet{Vivas20}.
As noted before, \citet{Vivas20} detected a small but significant ellipticity of $0.12 \pm 0.02$ in Cra2 using deep DECam observations of RGB and RRL stars, while \citetalias{Torrealba16} found no ellipticity with shallower but more spatially extended photometry. Additional deep observations with wider spatial coverage are likely needed to confirm the shape and orientation of Cra2, so we do not discuss it further here.

\section{Metallicity Distribution and Luminosity-Metallicity Relation}\label{sec:mdf}

After galaxy members are identified, we fit the CaT metallicities with metallicity distribution function (MDF) models to provide a first look at the formation history of these galaxies.
Only stars with member probability $>0.95$ and S/N $>5$ are included here, due to the increased S/N required to fit equivalent widths of individual lines.
The sample sizes for the MDF are thus reduced from 283 and 141 stars to 161 and 65 stars for Ant2 and Cra2, respectively.
Note our membership determination does include [Fe/H] information from \code{rvspecfit} assuming a Gaussian MDF, and in principle this could affect the MDF fits derived here. However, removing [Fe/H] from membership determination affects the membership of $\leq 3$ stars, which we have checked does not change the results here. This is in part because $\sigma_{\text{Fe}}$ is overestimated in DR2.2 due to a few bad measurements from \code{rvspecfit} (see panels d and h of Figures~\ref{fig:ant2summary} and \ref{fig:cra2summary}).

To interpret the MDFs, we fit the leaky box, pre-enriched, and extra gas models as described by \citet{Kirby11,Kirby13}.
The leaky box is the classic analytic model characterized by the effective yield $p_{\rm eff}$.
The pre-enriched box model adds a minimum metallicity floor $\feh_0$, while
the extra gas model \citep{LyndenBell75} adds pristine gas to a leaky box parameterized by $M$, where $M=1$ reproduces the leaky box and $M > 1$ adds extra pristine gas to the leaky box, creating a more peaked MDF with a lighter metal-poor tail.
The likelihood includes the metallicity uncertainties by convolving the model MDF with the uncertainty for each star \citep{Kirby11}.
The posterior is sampled using \code{dynesty} \citep{dynesty}\footnote{Code available at \url{https://github.com/alexji/mdfmodels}.}.
The priors are log uniform for $p$ from $10^{-3}$ to $10^{-1}$ for all three models; uniform in $\feh_0$ from $-5$ to $-2$ for the pre-enriched model; and uniform in $M$ from 1 to 30 for the extra gas model.
Additionally, we fit a Gaussian MDF with a mean $\mu$ and intrinsic spread $\sigma$, with a uniform prior for $\mu$ from $-4$ to $-1$ and $\log\sigma$ from $-2$ to $+1$.
The Gaussian fit results are adopted as the CaT metallicity in Table~\ref{tab:galprops}.

\begin{figure}
    \centering
    \includegraphics[width=\linewidth]{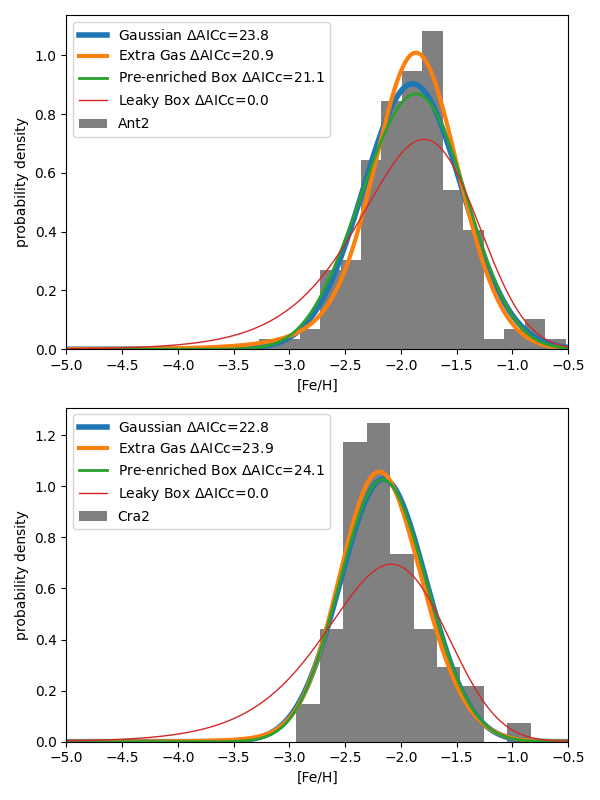}
    \caption{Calcium triplet metallicity distributions for Ant2 (top panel) and Cra2 (bottom panel), along with best-fit models for four MDF forms, convolved by the median uncertainty. The AICc is shown for each model in the legend, where larger ${\Delta}$AICc values indicate a better fit.
    Neither galaxy is well-described by a leaky box model (thin red line).
    However, the other three models are all reasonable fits.
    }
    \label{fig:mdffit}
\end{figure}

The best-fit values are tabulated in Table~\ref{tab:mdfmodels} and shown visually in Fig~\ref{fig:mdffit}.
The models in Fig~\ref{fig:mdffit} are convolved by the median [Fe/H] uncertainty.
The posteriors are all well-behaved (i.e., with a single well defined posterior peak), except for the extra gas model in Cra2 where there is only a lower limit on the $M$ parameter. This is because larger $M$ values make the MDF very sharply peaked, but the typical [Fe/H] uncertainty of ${\approx}0.25$ dex in Cra2 is unable to resolve such a narrow peak.

We compare the model fits using the corrected Akaike Information Criterion (AICc), which is essentially a likelihood ratio (see details in \citealt{Kirby20,Jenkins21}).
Neither Ant2 or Cra2 is well-described by a leaky box model, but this is the only model that can be ruled out.
The observed MDFs do not clearly distinguish between the extra gas, pre-enriched, or Gaussian models, primarily because the metallicity uncertainties are quite large.
For reference, Table~\ref{tab:mdfmodels} also includes the results of similar model fits to other Milky Way satellites from \citet{Kirby13,Jenkins21}. Ant2 has an MDF that is overall fairly similar to Sextans, while Cra2 is overall substantially more metal-poor and more similar to the dwarf galaxy Ursa Minor.
Note that \code{dynesty} does compute the Bayesian evidence, which can also be used for model comparison.
This gives the same qualitative conclusions as the AICc, but we use the AICc here to enable direct comparison in Table~\ref{tab:mdfmodels} to previous results \citep{Kirby13,Kirby20}.

\begin{deluxetable*}{l|c|ccc|ccc|ccc|c}
\tablecolumns{12}
\tabletypesize{\footnotesize}
\tablecaption{\label{tab:mdfmodels} Chemical Properties}
\tablehead{& Leaky Box & \multicolumn{3}{c}{Pre-enriched} & \multicolumn{3}{c}{Extra Gas} & \multicolumn{3}{c}{Gaussian} & \\
dSph & $p_{\rm{eff}}\ (Z_\odot)$ & $p_{\rm{eff}}\ (Z_\odot)$ & [Fe/H]$_0$ & $\Delta$AICc & $p_{\rm{eff}}\ (Z_\odot)$ & $M$ & $\Delta$AICc & $\mu$ & $\sigma$ & $\Delta$AICc & Ref}
\startdata
Antlia2 & $0.019^{+0.002}_{-0.002}$ & $0.014^{+0.002}_{-0.002}$ & $-2.85^{+0.13}_{-0.16}$ & $21.1$ & $0.016^{+0.001}_{-0.001}$ & $5.3^{+2.3}_{-1.6}$ & $20.9$ & $-1.90^{+0.03}_{-0.04}$ & $0.34^{+0.03}_{-0.03}$ & $23.8$ & This Work \\
Crater2 & $0.010^{+0.002}_{-0.001}$ & $0.004^{+0.001}_{-0.001}$ & $-2.67^{+0.10}_{-0.13}$ & $24.1$ & $0.008^{+0.001}_{-0.001}$ & $21.4^{+6.0}_{-8.0}$ & $23.9$ & $-2.16^{+0.04}_{-0.04}$ & $0.24^{+0.05}_{-0.04}$ & $22.8$ & This Work \\
\hline
Fornax & $0.106^{+0.005}_{-0.005}$ & $0.082^{+0.005}_{-0.004}$ & $-2.05^{+0.06}_{-0.06}$ & $124.0$ & $0.111^{+0.003}_{-0.003}$ & $9.3^{+1.5}_{-1.3}$ & $306.9$ & $-1.04^{+0.01}_{-0.01}$ & $0.33$ & \nodata & K13 \\
Leo I & $0.041^{+0.002}_{-0.002}$ & $0.030^{+0.002}_{-0.002}$ & $-2.33^{+0.05}_{-0.06}$ & $178.4$ & $0.043^{+0.001}_{-0.001}$ & $7.9^{+1.2}_{-1.0}$ & $353.3$ & $-1.45^{+0.01}_{-0.01}$ & $0.32$ & \nodata & K13 \\
Sculptor & $0.029^{+0.002}_{-0.002}$ & $0.027^{+0.002}_{-0.002}$ & $-3.39^{+0.18}_{-0.26}$ & $10.7$ & $0.029^{+0.002}_{-0.002}$ & $1.4^{+0.2}_{-0.2}$ & $5.3$ & $-1.68^{+0.01}_{-0.01}$ & $0.46$ & \nodata & K13 \\
Leo II & $0.028^{+0.002}_{-0.002}$ & $0.024^{+0.002}_{-0.002}$ & $-2.92^{+0.11}_{-0.13}$ & $25.5$ & $0.028^{+0.002}_{-0.002}$ & $3.3^{+0.7}_{-0.5}$ & $45.2$ & $-1.63^{+0.01}_{-0.01}$ & $0.40$ & \nodata & K13 \\
Sextans & $0.016^{+0.002}_{-0.002}$ & $0.013^{+0.002}_{-0.002}$ & $-3.17^{+0.16}_{-0.23}$ & $12.0$ & $0.014^{+0.001}_{-0.001}$ & $3.3^{+1.8}_{-1.0}$ & $10.4$ & $-1.94^{+0.01}_{-0.01}$ & $0.47$ & \nodata & K13 \\
Ursa Minor & $0.011^{+0.001}_{-0.001}$ & $0.007^{+0.001}_{-0.001}$ & $-2.92^{+0.09}_{-0.10}$ & $41.9$ & $0.009^{+0.001}_{-0.001}$ & $11.0^{+5.6}_{-4.5}$ & $44.3$ & $-2.13^{+0.01}_{-0.01}$ & $0.43$ & \nodata & K13 \\
Draco & $0.014^{+0.001}_{-0.001}$ & $0.011^{+0.001}_{-0.001}$ & $-3.06^{+0.09}_{-0.10}$ & $37.7$ & $0.013^{+0.001}_{-0.001}$ & $4.2^{+1.3}_{-0.9}$ & $44.7$ & $-1.98^{+0.01}_{-0.01}$ & $0.42$ & \nodata & K13 \\
CVn I & $0.019^{+0.002}_{-0.002}$ & $0.016^{+0.002}_{-0.002}$ & $-3.10^{+0.15}_{-0.20}$ & $13.4$ & $0.017^{+0.002}_{-0.001}$ & $2.6^{+1.0}_{-0.7}$ & $9.6$ & $-1.91^{+0.01}_{-0.01}$ & $0.44$ & \nodata & K13 \\
Bootes I & $0.005^{+0.001}_{-0.001}$ & $0.005^{+0.001}_{-0.001}$ & $-3.74^{+0.18}_{-0.18}$ & $2.9$ & $0.005^{+0.001}_{-0.001}$ & $4.5^{+3.2}_{-1.8}$ & $6.4$ & $-2.33^{+0.05}_{-0.05}$ & $0.27$ & \nodata & J21 \\
\enddata
\tablerefs{(K13) \citealt{Kirby13}, (J21) \citealt{Jenkins21}}
\tablecomments{The $\Delta$AICc values are compared to the Leaky Box model (larger positive values mean more favored).}
\end{deluxetable*}

No spatial metallicity gradients were found for either Ant2 or Cra2 in either the CaT or DR2.2 metallicities.
While multiple chemodynamic populations are often found in dSphs, they typically require datasets of $\mathcal{O}(1000)$ stars \citep[e.g.,][]{Kordopatis16,Pace20} or adding CMD or high-resolution abundance information \citep[e.g.,][]{Lemasle12}.
Our data of $140-290$ metallicities alone are thus probably insufficient to rule out the presence of such populations, though \citet{Walker19} note there are two distinct main sequence turnoffs in Cra2, and more extreme gradients can be detected with fewer stars \citep{Chiti21}.

\begin{figure}
    \centering
    \includegraphics[width=\linewidth]{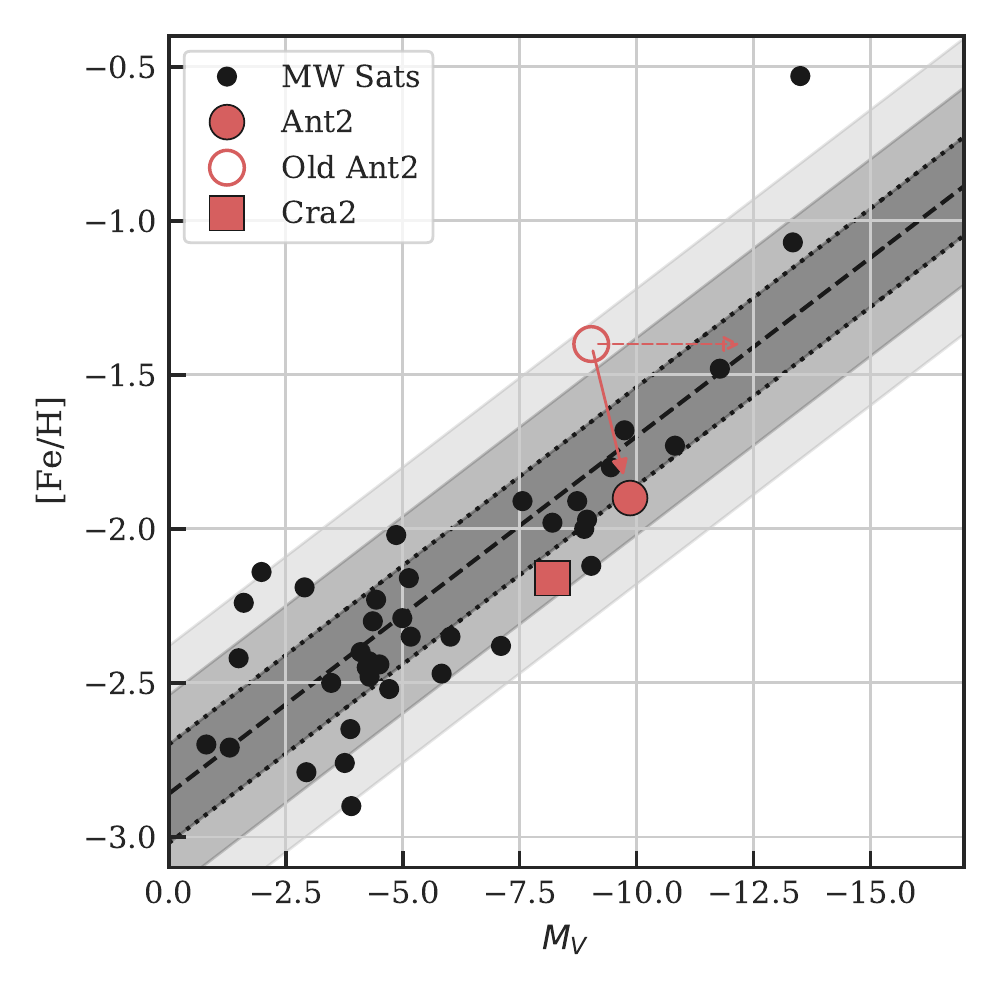}
    \caption{Luminosity-metallicity relationship.
    Milky Way satellite data (black points) are taken from \citet{Simon19}, and black dashed/dotted lines are the luminosity metallicity relation from \citet{Kirby13}.
    The shaded regions indicate 1, 2, and 3$\sigma$ deviations from the LZR.
    Each tick mark on the $x$-axis is 2.5 mag, or $10{\times}$ difference in luminosity.
    The previous Ant2 metallicity (open red circle, from \citetalias{Torrealba19}) was 0.5 dex too high, and comparison to the mass-metallicity relation suggested it had lost ${\gtrsim}90\%$ of its stellar mass from tidal interactions (dashed arrow).
    Our revised luminosity and metallicity (solid arrow and circle) clearly shows Ant2 lies on the mass metallicity relation, and thus has probably not yet lost most of its stellar mass.}
    \label{fig:LZR}
\end{figure}

Figure~\ref{fig:LZR} shows the updated mean CaT metallicities of these galaxies compared to the luminosity-metallicity relation \citep[LZR,][]{Kirby13}. Both Ant2 and Cra2 lie within the scatter of this relation.
For Ant2, this is a stark contrast from the measurement by \citetalias{Torrealba19} (open red circle).
The original measurement placed Ant2 far to the left of the LZR, such that it was likely originally much more massive and probably lost at least 90\% of its stellar mass to tidal interactions (dashed arrow).
The origin of the difference is a substantial zero-point offset in the \citetalias{Torrealba19} metallicities which is not present in recent versions of \code{rvspecfit}. 
The CaT metallicities are about 0.1 dex lower than the \code{rvspecfit} metallicities, but they
should be on the same scale as most MW satellite galaxies in \citet{Kirby13} and the \citet{Simon19} literature compilation, and so they are more suitable for comparisons to the LZR.

Ant2 and Cra2 are both currently about $1.0-1.5\sigma$ below the \citet{Kirby13} LZR.
For their progenitors to remain within 2$\sigma$ of the LZR, Ant2 and Cra2 could have lost at most 66\% and 42\% of their initial stars, respectively.
For these galaxies to have lost 90\% and 99\% of their initial stellar mass corresponds to ${\sim}3\sigma$ and ${\sim}5\sigma$ deviations from the LZR, respectively.
Thus, we conclude that neither Ant2 nor Cra2 has lost ${\gtrsim}90$\% of their stars to tidal disruption yet, though it would not be too surprising for them to have lost about half their stars.

\section{Orbits and Dynamical Modeling}\label{sec:dynamics}

\subsection{Orbit} \label{sec:orbit}

Using the kinematics in Table~\ref{tab:galprops}, we integrate the orbits of Ant2 and Cra2 backwards both in a static Milky Way potential and a potential that includes the interaction of the Milky Way and the LMC. Including the LMC is particularly important since previous studies have shown it has a significant effect on the orbits of these two dwarfs \citep{Erkal20}.

For the static case, we use the Milky Way potential from \cite{McMillan17} which consists of four disks, a bulge, and an NFW dark matter halo. We use \code{galpot} \citep{Dehnen98} to compute the forces from this potential and advance the orbits with a leapfrog integrator. We account for uncertainties in the Milky Way potential, as well as the Solar position and velocity, by sampling 10,000 realizations of the \citet{McMillan17} posterior chains. We also sample the present-day position and velocity of Ant2 and Cra2 within their uncertainties.
We then integrate backwards for 5 Gyr. The orbital parameters for each dwarf are shown in Table~\ref{tab:orbits}.

Next, we consider the Ant2 and Cra2 orbits in the presence of the LMC. We model the Milky Way potential as described above and model the LMC as a Hernquist profile \citep{Hernquist90} with a mass of $1.5\times10^{11} M_\odot$ and a scale radius of 17.14 kpc. This profile is motivated by \citet{Erkal19}, who measured the mass of the LMC from its effect on the Orphan Stream. To account for the motion of the Milky Way in response to the LMC \citep[e.g.][]{Gomez15,Erkal20b,Petersen21}, we model the Milky Way and LMC as particles sourcing their respective potentials \citep[as in][]{Erkal19,Erkal20,Vasiliev21}. The dynamical friction on the LMC is modeled using the approximations in \cite{Jethwa16}. In addition to uncertainties in the Milky Way potential, Solar kinematics, and observed properties of Ant2/Cra2, we also sample from the uncertainties in the LMC's present-day position and velocity using its observed proper motion \citep{Kallivayalil13}, distance \citep{Pietrzyski13}, and radial velocity \citep{vanderMarel02}. For each of these 10,000 realizations, we integrate Ant2 and Cra2 backwards in the combined presence of the LMC and Milky Way for up to 5 Gyr. Since the LMC is believed to be on its first approach to the Milky Way \citep[e.g.][]{Besla07,Kallivayalil13}, we end the simulation if the LMC reaches its apocenter with respect to the Milky Way. We do this to avoid including realizations where the LMC re-approaches the Milky Way, which could alter the orbits of Ant2 and Cra2.  

The results are summarized in Table~\ref{tab:orbits}.
The overall orbital geometries for both Ant2 and Cra2 are qualitatively similar to previous work \citep{Sanders18,Caldwell17,Fu19,Torrealba19}, with Ant2 approaching apocenter and Cra2 just past apocenter.
We thus focus our attention on the differences in pericenters, the most relevant quantity for tidal disruption. 
For Ant2, the static Milky Way model has a pericenter of $52.4 ^{+10.5}_{-7.9}$ kpc, while the LMC model has a pericenter of $38.6 ^{+7.2}_{-5.8}$ kpc.
The new pericenters are larger than previously inferred (${\sim}15$ kpc from \citealt{Chakrabarti19}, $37^{+20}_{-15}$ kpc from \citealt{Torrealba19}) due to an updated \Gaia EDR3 proper motion (Section~\ref{sec:pmcomp}).
For Cra2, the static Milky Way model has a pericenter of $33.2 ^{+5.9}_{-5.1}$ kpc, while the LMC model has a pericenter of $21.7^{+5.1}_{-4.0}$ kpc.
The static model pericenter is similar to those previously inferred ($37.7^{+18.0}_{-13.3}$ kpc from \citealt{Fu19}).
In general, the pericenters of both dwarfs are closer when including the LMC, in agreement with previous studies \citep[e.g.][]{Erkal20}.
Since the LMC has been shown to have a significant effect on structures in the Milky Way \citep[e.g.][]{Erkal19,Erkal20b,Petersen21}, we consider these to be our fiducial result going forward.

To explore how these dwarfs have been affected by the Milky Way, we estimate the tidal radius using \citet{King1962}:
\begin{equation}
r_t = \bigg( \frac{GM_{\rm sat}}{\Omega^2 - \frac{d^2 \phi}{dr^2}} \bigg)^{\frac{1}{3} },
\end{equation}
where $M_{\rm sat}$ is the mass of the satellite, $\Omega$ is the angular frequency of the satellite, and $\frac{d^2\phi}{dr^2}$ is the second derivative of the Milky Way potential with respect to radius. For the satellite mass, we use the dynamical mass within the half-light radius ($M_{\rm dyn}$) based on the \citet{Wolf10} estimator (see Table~\ref{tab:galprops}). This is lower than the total mass so may underestimate the tidal radius, but since $r_t \propto M^{1/3}$ the difference will not be large. We compute this tidal radius at pericenter for each of the 10,000 orbit realizations, reported in Table~\ref{tab:orbits}. To account for the uncertainty in the dynamical mass, we sample the dynamical mass for each orbit realization. As expected, the tidal radius for each dwarf is substantially smaller for the orbits in the presence of the LMC, since these have smaller pericenters. Furthermore, the tidal radii of Ant2 and Cra2 at pericenter are significantly smaller than their current half-light radii, 59\% and 36\% respectively. This suggests both galaxies may have experienced substantial tidal disruption, which we explore next.

\begin{deluxetable}{lcc}
\tablecolumns{3}
\tablecaption{\label{tab:orbits}Orbital Properties}
\tablehead{Parameter & Antlia 2 & Crater 2}
\startdata
Current half-light radius (pc) & $2541 \pm 175$ & $1066 \pm 84$ \\
Current tidal radius (pc) & $4774^{+400}_{-366}$ & $1791^{+330}_{-278}$ \\
\cutinhead{With LMC}
Pericenter (kpc) & $38.6^{+7.2}_{-5.8}$ & $21.7^{+5.4}_{-3.9}$ \\
Apocenter (kpc) & $136.2^{+7.8}_{-7.0}$ & $136.7^{+8.0}_{-6.4}$ \\
Eccentricity & $0.56^{+0.04}_{-0.05}$ & $0.73^{+0.04}_{-0.05}$ \\
Orbital period (Myr) & $2566^{+424}_{-323}$ & $2316^{+239}_{-222}$\\
Time since last pericenter (Myr) & $826^{+99}_{-81}$ & $1490^{+325}_{-258}$ \\
Tidal radius at pericenter (pc) & $1500^{+282}_{-223}$ & $386^{+120}_{-81}$ \\
\cutinhead{No LMC}
Pericenter (kpc) & $52.4^{+10.5}_{-7.9}$ & $33.2^{+5.9}_{-5.1}$ \\
Apocenter (kpc) & $144.1^{+9.8}_{-8.5}$ & $135.9^{+7.0}_{-6.0}$ \\
Eccentricity & $0.47^{+0.04}_{-0.05}$ & $0.61^{+0.05}_{-0.04}$ \\
Orbital period & $2505^{+489}_{-331}$ & $2100^{+320}_{-221}$ \\
Time since last pericenter (Myr) & $870^{+95}_{-82}$ & $1501^{+311}_{-200}$ \\
Tidal radius at pericenter (pc) & $1932^{+387}_{-292}$ & $539^{+134}_{-104}$ \\
\enddata
\end{deluxetable}

\subsection{Tidal Disruption Simulations} \label{sec:particlesim}

Since both Ant2 and Cra2 may be tidally disrupting, we now investigate the expected tidal debris using the modified Lagrange Cloud stripping technique of \cite{Gibbons14} as implemented in \cite{Erkal19}\footnote{We acknowledge \code{gala} \citep{gala} whose implementation of the \citet{Fardal15} mock stream generator was used in initial explorations.}. This technique works by rewinding the orbits of Ant2 and Cra2 for 5 Gyr, and then generating a stream during the forward integration.
We initialize the progenitors of Ant2 and Cra2 to be Plummer spheres with masses and scale radii of $(10^{7.92} M_\odot, 1\rm{kpc})$ and $(10^{6.74} M_\odot, 500\rm{pc})$, respectively. These masses are equal to the inferred dynamical mass within the half light radius from Table~\ref{tab:galprops}.

The simulation setup is similar to Section~\ref{sec:orbit} except we keep the potential and other parameters fixed. For the Milky Way we select the same posterior chain of \cite{McMillan17} that was used in \cite{Li20}. This realization was chosen since \cite{Li20} found that it gave a good fit to the AAU stream. In this potential, the Milky Way halo is lighter than the best-fit model in \citep{McMillan17} ($M_{200} = 8.27\times10^{11} M_\odot$) so the LMC has a realistic past orbit and has recently completed its first approach to the Milky Way. For the solar distance and motion we use 8.122 kpc and (11.1,245.04,7.25) km~s$^{-1}$, respectively, as in Section~\ref{sec:kinematics}. For the LMC, we use the same $1.5\times10^{11} M_\odot$ Hernquist profile as in Section~\ref{sec:orbit}. We perform simulations both with and without the LMC.

\begin{figure*}
    \centering
    \includegraphics[width=0.95\linewidth]{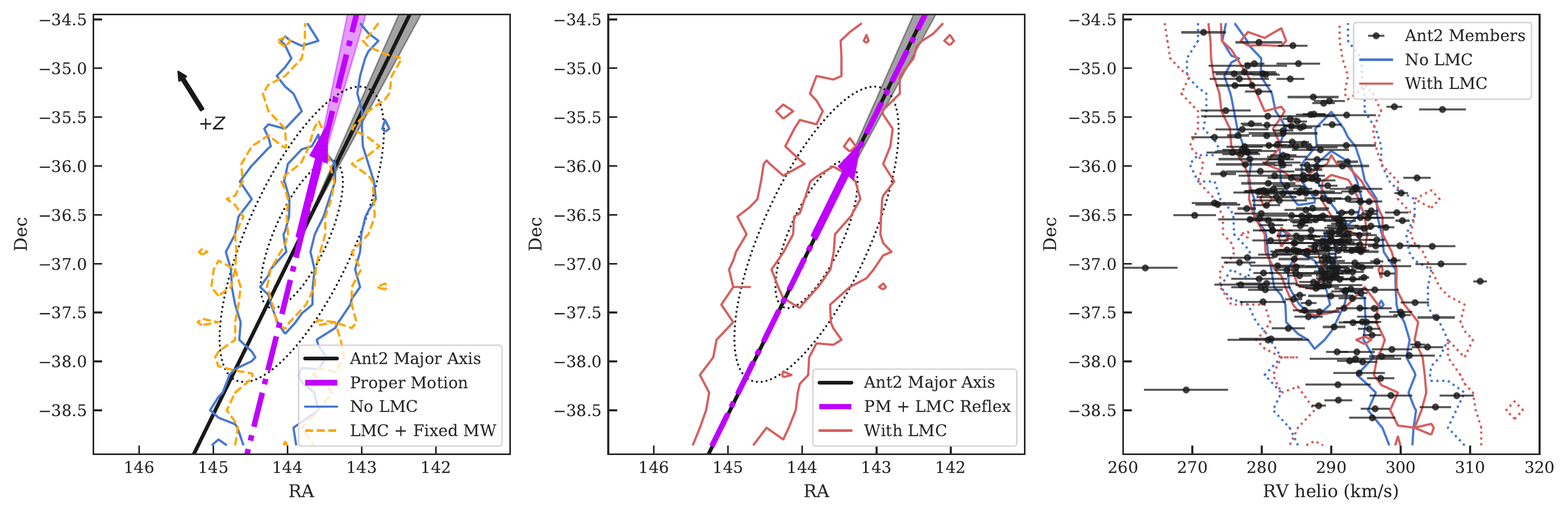}
    \caption{
    Tidal features of Ant2 with and without LMC. 
    \textit{Left:} Spatial orientation of Ant2 ($1\sigma$ and $2\sigma$ dotted black ellipses, black line indicates major axis with $1\sigma$ shaded uncertainty) compared to a stream model without the LMC (blue contours, 68\% and 95\%).
    The stream model is aligned with the observed GSR proper motion (purple arrow and shaded $1\sigma$ uncertainty), which is misaligned with the major axis.
    A model with the LMC but a fixed MW barycenter (orange dashed contours) is essentially identical to no LMC.
    For reference, the black arrow indicates the $+Z$ direction projected onto the sky, which is the approximate direction of influence from the LMC.
    We do not plot individual Ant2 member stars for clarity.
    \textit{Center:} same as left panel, but with a stream model including a $1.5 \times 10^{11} M_\odot$ LMC and the MW's reflex motion (red contours). There is a very good match between the observed spatial orientation and the tidal material in this model.
    The purple arrow includes an additional 28 km~s$^{-1}$ reflex correction in the $-Z$ direction from the LMC to the proper motion, which aligns the PM with the observed spatial orientation.
    \textit{Right:} Dec vs heliocentric radial velocity for data (black points) and models (contours). The velocity gradient in the model agrees well with the observed data when including the LMC. The solid model contours are 68\% and 95\%, while the dotted contours are 5\% to show the full direction of the gradient.
    }
    \label{fig:streammodelant2}
\end{figure*}

\begin{figure*}
    \centering
    \includegraphics[width=0.95\linewidth]{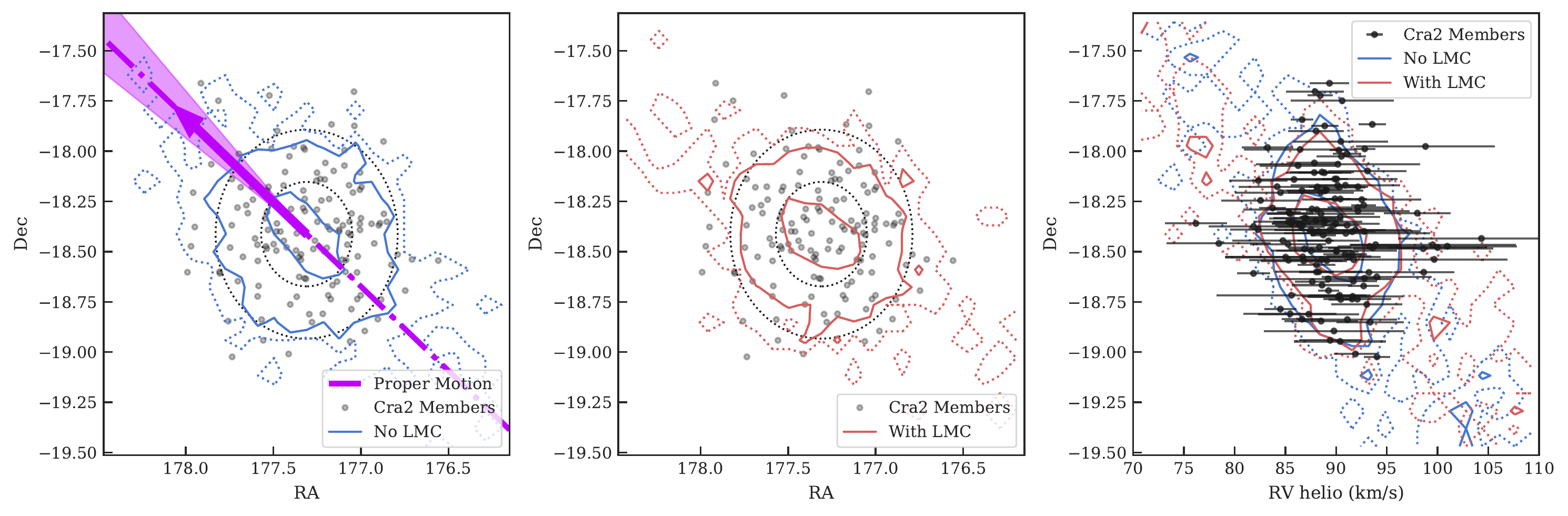}
    \caption{
    Tidal features of Cra2 with and without LMC. 
    \textit{Left:} Spatial extent of Cra2 ($1\sigma$ and $2\sigma$ dotted black ellipses) compared to a stream model without the LMC (blue contours, 5\%, 68\%, and 95\%).
    The GSR proper motion is shown by a purple arrow and shaded $1\sigma$ uncertainty. Note that most of the simulated particles shown are close to the main body of Cra2 since the majority of the particles stripped after the previous pericenter have moved along the stream. 
    \textit{Center:} same as left panel, but with a stream model including the LMC and the MW's reflex motion (red contours).
    \textit{Right:} Dec vs heliocentric radial velocity for Cra2 data (black points) and models (contours).
    }
    \label{fig:streammodelcra2}
\end{figure*}

The Ant2 and Cra2 results are shown in Figures~\ref{fig:streammodelant2} and \ref{fig:streammodelcra2}.
For Cra2, the current observations do not extend  sufficiently far to probe potential tidal features, so we do not comment further. 
However, our ${\sim}5$ degree coverage of Ant2 clearly extends into the region where a tidal stream would be expected.
Indeed, there is striking qualitative agreement between both the orientation and radial velocity gradient of Ant2 and the mock tidal stream. 
We did not do any fine tuning of parameters, which strongly suggests that tidal effects are responsible for the spatial extent and kinematic properties of Ant2.

The left panel of Figure~\ref{fig:streammodelant2} shows that in a static MW potential, there is a misalignment between the simulated orientation of Ant2 tidal features (blue contours) and the actual observed orientation of Ant2.
This corresponds to an offset of ${\sim}12^\circ$ between the reflex-corrected proper motion and the spatial orientation of Ant2, which is statistically significant (Fig~\ref{fig:angles})\footnote{Some of the most-recently stripped material from Ant2 may not lie along the stream orbit, which could introduce minor projection effects as the Sun's position is not exactly in the Ant2 orbital plane. The observed effect would be small, since Ant2 is currently very distant so our current position is near the orbital plane (a maximum ${\sim}3.5^\circ$ effect).}.
Such misalignments are not expected if Ant2 is orbiting in a static potential, but can easily arise in dynamic potentials \citep[e.g.,][]{Erkal19,Shipp19,Vasiliev21}.
The middle panel shows a simulation in the dynamic potential including a $1.5 \times 10^{11} M_\odot$ LMC.
The stream orientation (red contours) is much closer to the observed orientation of Ant2.
Furthermore, the radial velocity gradients in the model with and without the LMC are 6.1 and 4.1 \kms/deg, respectively;  we measure a gradient of $5.7 \pm 0.6 $\kms/deg that matches the LMC model.

At first, this may seem curious, because Ant2's orbit has never taken it close to the LMC \citep{Torrealba19}. Indeed, our simulations show that a plane fit to Ant2's stream has almost the same orientation today as during Ant2's previous pericenter ($2.7^\circ$ difference), when the LMC was still far from the Milky Way. Instead, we find the misalignment is due to the \emph{reflex motion of the Milky Way induced by the LMC} \citep[e.g.,][]{Gomez15,GaravitoCamargo19,Erkal20b}. In order to explore this effect, we looked at the velocity change the LMC has imparted on Ant2 and the Milky Way since Ant2's previous pericenter, when the material in the stream was stripped. For the potential realization used to generate the stream, this pericenter occurred ${\sim}930$ Myr ago. Since then, the LMC has accelerated the Milky Way by $(5.10, 9.93, -43.68)$ km/s and Ant2 by $(-0.32, 16.94, -7.47)$ km/s in the $(X,Y,Z)$ direction; and the Milky Way has moved by $(-3.09, 0.13, 31.21)$ kpc. Note that the LMC has had a much larger effect on the Milky Way, since the LMC has passed much closer to the Milky Way than to Ant2. Thus, we see the effect of the LMC is to (1) accelerate the Milky Way approximately downwards relative to Ant2, effectively adding an upwards (${+}Z$) component to Ant2's velocity from our perspective; and (2) to move the Milky Way upwards, effectively changing our viewpoint of the Ant2 stream.

To explore this in the data, the purple arrow in the middle panel of Figure~\ref{fig:streammodelant2} shows the result of adding 28 \kms in the $-Z$ Galactocentric direction to the 3D velocity of Ant2, then reprojecting this to the expected on-sky orientation. The 28 \kms offset roughly mimics the reflex velocity of the Milky Way to the LMC today \citep{Erkal20b,Petersen21,Vasiliev21}, and this simple correction matches the stream model and the observed orientation of Ant2. This velocity shift is likely also encapsulating the effect of the Milky Way's movement since Ant2's pericenter, which would change our perspective of the Ant2 stream.

As a final check, we ran a stream simulation including the LMC but artificially fixing the position of the Milky Way. No offset in the Ant2 stream was produced (orange contours in left panel of Figure~\ref{fig:streammodelant2}). We thus conclude that Ant2's kinematic properties are well-explained by tidal disruption after accounting for the motion of the Milky Way induced by the LMC's passage. 

Finally, we note that while we have mostly focused our analysis on the effect of the LMC on Ant2, the same general picture also holds for Cra2. Since its most recent pericenter with the Milky Way, Cra2 has been accelerated by the LMC by $(0.75,12.84,-8.24)$ km/s while the Milky Way has been accelerated by $(5.97,19.47,-45.04)$ km/s. Thus, the main effect of the LMC on Cra2's stream will similarly be due to the motion of the Milky Way.

\section{Discussion}\label{sec:discussion}

\subsection{Comparison to previous results}

\subsubsection{Gaia DR2 vs EDR3 proper motions}\label{sec:pmcomp}

\begin{figure*}
    \centering
    \includegraphics[width=0.8\linewidth]{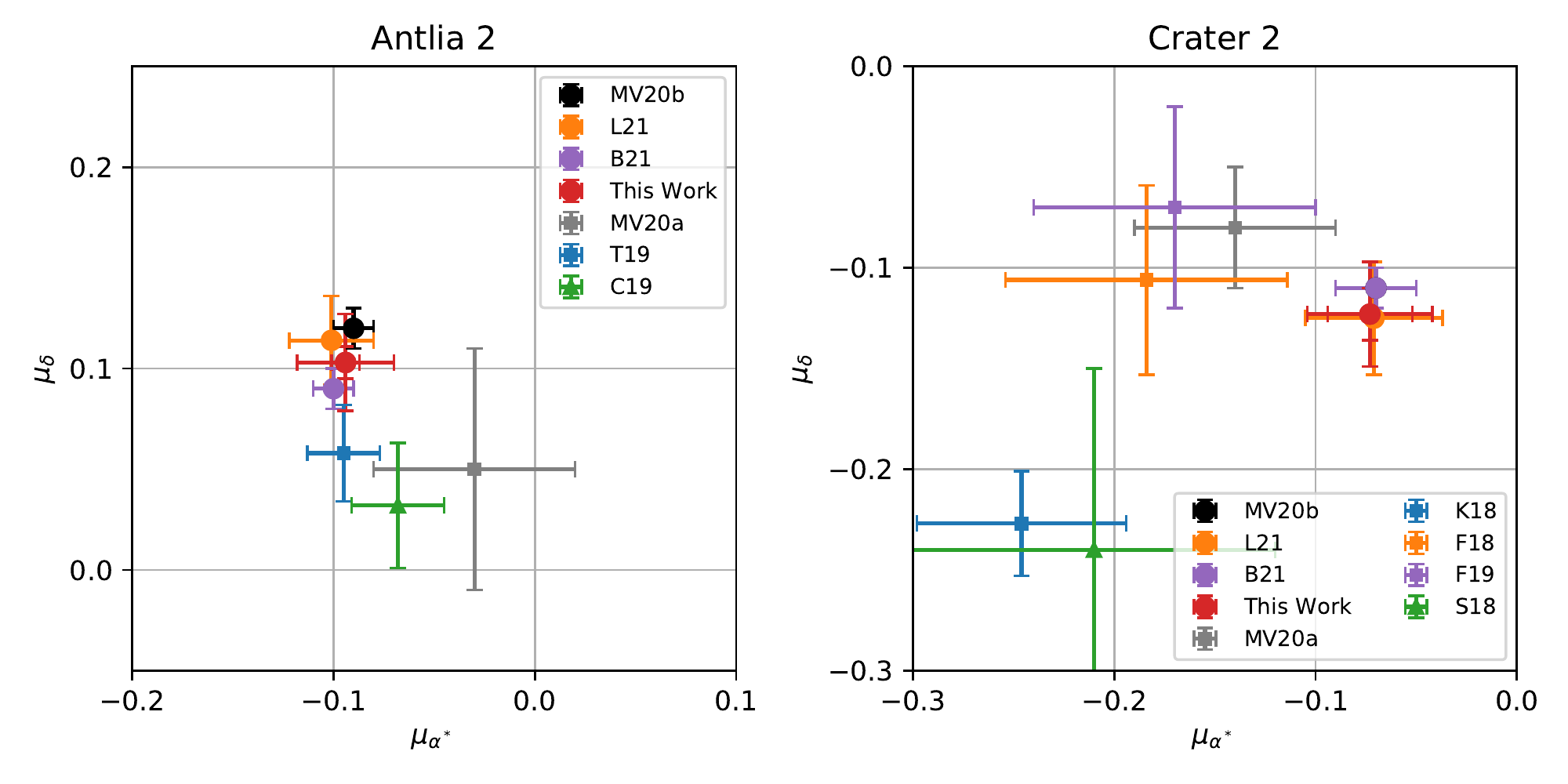}
    \caption{Proper motions of Ant2 (left) and Cra2 (right) from this work compared to literature.
    The large circles are \gaia EDR3 measurements and all find similar results, including the red circle from this work, the black circle from \citet{McConnachie20b} (denoted MV20b), the orange circle from \citet{Li2021pm}, and the purple circle from \citet{Battaglia21}.
    The two sets of red error bars indicate with and without the 0.023 \masyr systematic uncertainty per component for \gaia EDR3 \citep{Lindegren20}.
    The points in squares are proper motion measurements using \gaia DR2 
    (Ant2: grey square MV20a = \citealt{McConnachie20}, blue square = \citetalias{Torrealba19}; Cra2: grey square MV20a = \citealt{McConnachie20}, blue square K18 = \citealt{Kallivayalil18}, orange square F18 = \citealt{Fritz18}, purple square F19 = \citealt{Fu19}).
    In Ant2, the green triangle C19 is a \gaia DR2 measurement by \citet{Chakrabarti19} that has a substantial chance of a very low pericenter of ${\lesssim}15$ kpc. This scenario is disfavored by the new proper motions.
    In Cra2, the green triangle S18 is a prediction from \citet{Sanders18} about the range of proper motions for Cra2 to have experienced very substantial tidal disruption in an NFW halo. This is now barely consistent with our new measurement.
    }
    \label{fig:pmcomp}
\end{figure*}

\gaia EDR3 has more precise proper motions and better control of systematic effects than \gaia DR2 \citep{Lindegren20}.
Figure~\ref{fig:pmcomp} shows the new (heliocentric) proper motions for EDR3 compared to previous literature measurements and/or predictions.
Our results agree well with \citet{McConnachie20b}, \citet{Li2021pm}, and \citet{Battaglia21} who also use \gaia EDR3, while the results using \gaia DR2 are generally consistent with each other but offset from the EDR3 results.

The green triangles in Figure~\ref{fig:pmcomp} indicate proper motions associated with specific scenarios proposed by previous studies.
For Ant2, \citet{Chakrabarti19} measured a proper motion consistent with other DR2 measurements but with a substantial tail that included pericenters as low as 15 kpc. Such low pericenters could have excited perturbations in the outer gas disk of the Milky Way.
Ant2's new proper motions in EDR3 now suggest a much larger pericenter of 38.6 kpc (52.4 kpc when not including the LMC), making it less likely to be the source of those perturbations.
For Cra2, \citet{Sanders18} predicted a range of proper motions for Cra2 to experience substantial tidal disruption if it resides in an NFW halo. This value is now disfavored, though the predicted range is very large and the new proper motion measurement is at the edge of the allowed range.

\subsubsection{Previous Antlia 2 Studies}
\citetalias{Torrealba19} have presented the only other spectroscopic study of Ant2, also using AAT/2dF.
The current paper includes all of the data in \citetalias{Torrealba19}, as well as five additional new fields. All data are reduced and analyzed consistently here. Our new data double the number of member stars and extend out to ${\gtrsim}2$ half light radii.
\citetalias{Torrealba19} tentatively detected an increase in the velocity dispersion with galaxy radius, with a dispersion of ${\sim}5\kms$ within 0.5 deg increasing to ${\sim}7\kms$ at larger radii. They argued at the time that this could be due to a velocity gradient, though the data did not yet support that. Our new data extending out 2x further now clearly show there is a linear velocity gradient across the galaxy in excess of the perspective rotation (panels c and e of Figure~\ref{fig:ant2summary}) with an intrinsic scatter of $6.0 \pm 0.4 \kms$.

\citetalias{Torrealba19} also found that Ant2 had a mean metallicity of $-1.4$, but here we find a substantially lower mean metallicity of $-1.9$ (Figure~\ref{fig:LZR}). The corrected mean metallicity now clearly places Ant2 on the luminosity-metallicity relation \citep{Kirby13}.
This makes the previous conclusion that Ant2 has lost over $90$\% of its stellar mass now disfavored at about $3\sigma$.

\subsubsection{Previous Crater 2 Studies}

\citet{Caldwell17} and \citet{Fu19} are the two previous spectroscopic studies of Cra2.
These two studies had much smaller sample sizes, but spanned a similar spatial extent and had precise velocity uncertainties of ${\sim}1-2\kms$.
Because we allowed a fairly low S/N cut for our members, our median velocity uncertainty for Cra2 members is ${\sim}3\kms$, of which 41 have velocity precision $<2\kms$ similar to previous studies.
We thus have better statistics for large scale trends (e.g. velocity gradients), but only the brighter stars with better S/N and ${\sim}1\kms$ uncertainties contribute to resolving the very low velocity dispersion in Cra2.

Our kinematic model finds tentative evidence for a velocity gradient in Cra2 of $2.2 \pm 1.2 \kms~\text{deg}^{-1}$, 
slightly more significant than the gradient found by \citet{Caldwell17} ($k_v = 1.8^{+1.8}_{-1.2}$\kms~deg$^{-1}$ or a 95\% upper limit of 3.6 \kms~deg$^{-1}$) but still detected at less than $2\sigma$ significance.
The reality of this gradient needs to be elucidated with future data extending to larger radii.
Including the velocity gradient, we find the velocity dispersion of Cra2 is $2.34^{+0.42}_{-0.30} \kms$. If we ignore the velocity gradient, we obtain a dispersion of $2.43^{+0.54}_{-0.35}\kms$. Both values are somewhat smaller than (but consistent with) the value of $2.7 \kms$ found by \citet{Caldwell17} and \citet{Fu19}.

We note there is a ${\approx}1\kms$ zero-point offset in our Cra2 velocity compared to \citet{Caldwell17} and \citet{Fu19}. A similar offset has been found comparing \SSSSS velocities to high-resolution velocities \citep{Ji20b}. The native AAT velocities in fact would match those previous studies, but the zero-point of \SSSSS velocities has been shifted by 1.1 \kms to match that of several large surveys (APOGEE, \gaia, GALAH; \citealt{Li19}).
Regardless, the results presented in this paper are not affected by a 1 \kms global velocity offset.

We find a lower mean metallicity of $\mbox{[Fe/H]}=-2.16 \pm 0.03$ compared to \citet{Caldwell17} ($-1.98 \pm 0.1$) and \citet{Fu19} ($-1.95 \pm 0.06$).
The \citet{Caldwell17} metallicities are from analyzing a small spectral range that tends to overestimate metallicities \citep[e.g.,][]{Ji16d}, but the \citet{Fu19} results are also determined from the calcium triplet with the \citet{Carrera13} relation so they should be identical to ours.
Upon investigation, we found an error in the \citet{Fu19} conversion between CaT equivalent widths and [Fe/H]. Metallicities calculated using their equivalent widths and the \citet{Carrera13} calibration agree with our current AAT measurements. We thus trust that our lower [Fe/H] for Cra2 is accurate.

\subsection{Tidal Disruption of Antlia 2}

The evidence in this paper strongly suggests that Ant2 is currently undergoing tidal disruption, and its extended low surface brightness properties are indicative of the early stages of creating a stellar stream.
Ant2 clearly displays a velocity gradient aligned with its orbit and position angle (Figures~\ref{fig:ant2summary}, \ref{fig:angles}).
The magnitude of the observed velocity gradient matches that predicted in tidal disruption simulations (Figure~\ref{fig:streammodelant2}).
We also see suggestions of a central overdensity with a different orientation and relatively flatter velocity dispersion that might indicate an embedded progenitor system, though these are not statistically significant (Figures~\ref{fig:ant2spatial}, \ref{fig:ant2vel}).
However, Ant2 currently lies within the mass-metallicity relationship, and it is unlikely to have lost more than half its stars so far (Figure~\ref{fig:LZR}, Section~\ref{sec:mdf}).
We appear to have caught Ant2 at a special time, right as it is being tidally disrupted but before it has lost most of its stars.

A key question is whether to interpret the observed velocity gradient as tidal effects or as rotation, since both tidal effects and solid-body rotation have a similar functional form for modeling the line-of-sight velocities.
In the case of Ant2, the gradient is very likely due to tides, because the reflex-corrected proper motion is well-aligned with the velocity gradient and spatial extent of the galaxy, as expected if the gradient is due to tides. Furthermore, our particle stream simulations provide a good match to the observations with no tuning applied (Figures~\ref{fig:streammodelant2}).
Additionally, statistics of dSph galaxies suggest they tend not to have much rotation, with $v_{\text{rot}}/\sigma_v < 0.5$ for 80\% of satellite dSphs \citep{Wheeler17} while Ant2 would have $v_{\text{rot}}/\sigma_v \sim 1$ at its half light radius  (although it has been proposed that galaxies like Cra2 and Ant2 form in high spin halos, e.g., \citealt{Dalcanton97b,Amorisco16}).
Finally, Ant2's orbit takes it close enough to the Milky Way that its tidal radius at pericenter is substantially smaller than its half light radius today (Table~\ref{tab:orbits}).
We cannot rule out the small chance that the rotation axis and velocity just happen to match that expected from Milky Way tides, but such a scenario is quite fine-tuned compared to the tidal scenario.
In the future, these can be observationally distinguished with more precise proper motions that reveal internal motion \citep[e.g.,][]{Zivick20}.
Finding a substantial extended stream along Ant2's orbit would also further support the tidal disruption scenario.

An offset between a stream's spatial orientation and its reflex-corrected proper motion direction is evidence of a time-dependent gravitational potential.
The biggest time-dependent perturbation in the MW is from the LMC, which directly impacts many stellar streams \citep{Shipp19,Erkal19} and also can indirectly affect systems by moving the Milky Way \citep{Vasiliev21}.
Ant2 is now the second example, after the Sagittarius stream \citep{Vasiliev21}, of the indirect effect of the LMC.
The middle panel of Figure~\ref{fig:streammodelant2} shows that for Ant2, this effect is well-approximated by just a reflex correction. This suggests that streams can be used to measure both the direct influence of the LMC and the induced reflex motion of the Milky Way.

Finally, the discussion of Ant2's velocity gradient may also apply to Cra2, although more of the spatial extent of Cra2 needs to be observationally probed to confirm the velocity gradient and its direction.
The tentative velocity gradient in Cra2 points towards the same direction as the proper motion (as opposed to Ant2 where it points opposite), as expected in the tidal disruption scenario, because Cra2 is past apocenter while Ant2 is just nearing apocenter.

\subsection{Tidal Evolution and Dark Matter Halo Profiles}

In standard $\Lambda$CDM galaxy formation theory, galaxies typically follow a positive correlation between galaxy size and velocity dispersion \citep[e.g.,][]{Fattahi18}.
The unusually low velocity dispersions and large sizes of Ant2 and Cra2 pose a potential challenge to this picture \citep[e.g.,][]{McGaugh16,Caldwell17}, but a natural explanation within $\Lambda$CDM is strong tidal stripping \citep[e.g.,][]{Frings17,Sanders18,Fattahi18,Fu19,Torrealba19,Applebaum21}.
Tidally removing substantial amounts of mass can simultaneously increase the size and lower the velocity dispersion of a galaxy, especially if the dark matter halo has been cored through stellar feedback \citep[e.g.,][]{Errani15} or the halo has unusually low concentration \citep[e.g.,][]{Rey19,Sameie20}.
A full investigation of the Ant2 and Cra2 progenitors is beyond the scope of this paper, but we briefly summarize past work on this topic in the context of our new measurements.

\subsubsection{Crater 2}
\citet{Sanders18} performed a suite of controlled tidal disruption simulations of stars in an NFW halo to find the conditions necessary for reproducing the velocity dispersion and physical size of Cra2 in $\Lambda$CDM. They argued that for Cra2 to be consistent with $\Lambda$CDM, its observed GSR proper motion should have a total magnitude less than 0.2 \masyr.
We find a GSR proper motion magnitude of $0.18 \pm 0.02 \masyr$, which is right at the threshold value.
The proposed scenario also requires Cra2 to have lost ${\sim}$99\% of its total mass and ${\sim}70\%$ of its stellar mass to tides, so its original progenitor would be ${\sim}1.3$ mag more luminous and a $2.5{\sigma}$ outlier from the luminosity-metallicity relation.

Changing from a cuspy NFW to a cored density profile for the dark matter halo makes the tidal stripping scenario more likely.
\citet{Fu19} use tidal evolution tracks from \citet{Errani15} to show that tidally stripping 70-90\% of the total mass from a galaxy like today's Sculptor or Ursa Minor dSphs would result in a system with the size and velocity dispersion of Cra2.
Similar results are also broadly found in the APOSTLE simulations \citep{Fattahi18} and controlled simulations \citep{Sanders18}.
Presumably the lower total mass loss in a tidally disrupting cored halo would correspond to a lower stellar mass loss, which would then reduce the tension between Cra2's progenitor and the luminosity-metallicity relation.

\subsubsection{Antlia 2}
\citetalias{Torrealba19} argued Ant2 likely resides in a cored dark matter halo, as it was difficult to tidally produce a galaxy of this size if embedded in a dark matter cusp.
Our new radius measurement of 2.5 kpc is smaller than the previous 3 kpc, but it is still large enough to not substantially change their conclusions.
\citet{Sameie20} argued with a different suite of idealized simulations that a CDM cusp is plausible, though it requires a dark matter halo of unusually low concentration (also see \citealt{Amorisco19}).
However, they agree with \citetalias{Torrealba19} that a cored halo is preferred.
Like other dwarf galaxies, the origin of a dark matter core in Ant2 is not clear.
Baryonic feedback could certainly produce a core in a galaxy of this mass \citepalias{Torrealba19}.
Different dark matter models are another possibility, for instance by adding a self-interaction cross section \citep{Sameie20} or a soliton core from fuzzy dark matter \citep{Broadhurst20}.

Crucially, the models by \citetalias{Torrealba19} and \citet{Sameie20} have Ant2 losing ${\sim}99$\% and $90$\% of its stellar mass, respectively.
Given our update to the Ant2 mean metallicity, these large stellar mass losses now imply Ant2's progenitor was $3-5\sigma$ below the luminosity-metallicity relationship, a clear discrepancy.
Detailed dynamical modeling is now needed to see if Ant2's tidal disruption but low stellar mass loss can be accommodated within $\Lambda$CDM, or if alternate theories are now preferred.

\subsection{External Field Effect in MOND}

The external field effect (EFE) is a prediction of Modified Newtonian Dynamics (MOND) that originates from the nonlinear combination of accelerations in MOND and should apply to satellites of the Milky Way.
Based on its position ${\approx}120$ kpc away from the Galactic center and its very large half light radius of ${\approx}1$ kpc, \citet{McGaugh16} predicted the EFE in Cra2 should cause it to have a very low velocity dispersion of $2.1^{+0.9}_{-0.6}\kms$, where the uncertainty is due to differences in the assumed mass-to-light ratio.
This was confirmed by Cra2's very low velocity dispersion of $2.7 \kms$ from \citet{Caldwell17} and \citet{Fu19}. Including the possible velocity gradient in Cra2, our results ($2.35^{+0.4}_{-0.3}\kms$) are even closer to this prediction.

Ant2 is another good candidate to test the EFE. Its Galactocentric distance is ${\sim}130$kpc, with a radius ${\approx}2.8{\times}$ and a stellar mass ${\approx}4.7{\times}$ that of Cra2.
Since both the internal and external accelerations for Ant2 are similar to Cra2, the EFE prediction for Ant2's velocity dispersion should be similar to that for Cra2, increasing by ${\approx}35\%$ due to the different physical parameters and distance to $2.8^{+1.3}_{-0.8}\kms$ (eqn~2 in \citealt{McGaugh16}).
Ant2's velocity dispersion of $6.0^{+0.4}_{-0.4}\kms$ is thus in conflict with the EFE prediction.

The similar EFE predictions but very different velocity dispersions for Ant2 and Cra2 could pose a challenge for MOND. However, given the clear presence of tidal effects in Ant2 and possible velocity gradient in Cra2, it is important to explore the influence of tides before drawing further conclusions.

\subsection{Comparison to Extragalactic Low Surface Brightness Galaxies}

Low Surface Brightness galaxies (LSBs), with central surface brightnesses ${\gtrsim}23$ mag arcsec$^{-2}$ -- comparable to the sky background -- have long been of interest because they are a substantial part of the galaxy population but are hard to observationally detect and characterize \citep[e.g.,][]{Dalcanton97}.
Ant2 and Cra2 are the nearest LSBs with unusually large radii, though most Milky Way satellite galaxies are technically LSBs due to their low stellar masses.
There are a few LSBs of similar stellar masses as Ant2 and Cra2 in the Local Group and Local Volume also detected in resolved stars (such as And XIX, \citealt{Martin16,Collins20}, And XXI, \citealt{Collins21}, and Coma P, \citealt{Ball18,Brunker19}), and an increasingly large number of relatively luminous, distant, and unresolved LSBs detected in deep photometric surveys \citep[e.g.,][]{vanDokkum15,Koda15,Greco18,Danieli20,Lim20,Tanoglidis21}.
The most extreme LSBs are now often called ultra-diffuse galaxies (UDGs, with $r_e > 1.5$ kpc, central surface brightness $>24$ mag arcsec$^{-2}$, and dwarf spheroidal like morphologies; \citealt{vanDokkum15}).
LSBs have been of particular interest because their population properties are only now becoming well-characterized \citep[e.g.,][]{Greco18,Danieli19,Danieli20,Tanoglidis21,KadoFong21},
their formation mechanisms are still hotly debated \citep[e.g.,][]{Amorisco16,Carleton19,Jiang19,Tremmel20,Wright21,Jackson21,Applebaum21},
and they are good systems to test dark matter theories \citep[e.g.,][]{McGaugh98a,Danieli19b,Emsellem19,Muller20}.

As one of the closest LSBs/UDGs, our observations of Ant2 offer some possible insights for interpreting other, more distant LSBs.
First, the original observations of Ant2 were limited in radial extent. \citetalias{Torrealba19} suggested their observations were likely due to an intrinsic velocity gradient, though they could not rule out a varying velocity dispersion.
The larger radial extent of our Ant2 observations now clearly detect the gradient. 
This may be relevant for recent UDG kinematic measurements using IFU observations of UDG stellar bodies \citep[e.g.,][]{Danieli19b,Emsellem19,Forbes21}, which do not necessarily probe a very large distance away from the centers.
A second concern is interpreting velocity gradients. In Ant2 we detect a clear velocity gradient, but this is consistent with both a linearly rising rotation curve along the major axis and tidal disturbances. It is in large part because we have the proper motion of Ant2 that we can argue that the tidal disturbances are more likely. This was also emphasized for And XIX by \citet{Collins20}, who tentatively detected a velocity gradient but could not distinguish between rotation and tides.

\section{Summary}
\label{sec:summary}

We present new AAT/2dF spectroscopy in the Milky Way satellite dwarf galaxies Antlia 2 and Crater 2, roughly doubling the number of radial velocities and metallicities compared to the literature.
We perform a detailed kinematic analysis including astrometry from \gaia EDR3. We also update the spatial parameters for Ant2 with \gaia EDR3 data, which results in a much higher luminosity.
The new galactic properties are given in Table~\ref{tab:galprops} and visualized in Figures~\ref{fig:ant2summary} and \ref{fig:cra2summary}.
We fit simple chemical evolution models to the metallicity distribution functions, though our sample size and metallicity precision are not sufficient to distinguish between different formation channels (Figure~\ref{fig:mdffit}).

Ant2 displays a clear velocity gradient roughly aligned with its major axis and reflex-corrected proper motion (Figure~\ref{fig:angles}). Cra2 has a low significance detection of a velocity gradient as well, also roughly aligned with its proper motion.
These gradients suggest that tidal interactions with the Milky Way affect these galaxies' kinematics.

Our observations of Crater 2 largely corroborate conclusions from previous studies \citep{Caldwell17,Fu19}, although we find it is 0.2 dex more metal-poor than those studies.
However, our observations of Ant2 suggest qualitatively new interpretations compared to previous studies \citep{Torrealba19}.
First, the observed velocity gradient and spatial orientation are a remarkable match to tidal disruption simulations, strongly suggesting that Ant2 is impacted by tides and possibly embedded in its own stellar stream.
Second, the spatial orientation of Ant2 and its stream is only accurately reproduced if we include the effect of the LMC, not because of a direct impact but because the LMC moves the Milky Way's barycenter.
The clear kinematic signatures of tides contrast with our updated metallicity measurement, which places Ant2 on the luminosity-metallicity relation (Figure~\ref{fig:LZR}) and suggests Ant2 has not lost most of its stellar mass yet.
It remains to be seen whether these facts can be reconciled in standard $\Lambda$CDM galaxy formation, as all models of Ant2 to date lose $>90$\% of their stars to tidal disruption.
But taken all together, the detailed chemodynamics provided by \gaia and our AAT spectroscopy continue to support a tidal origin for the large sizes and low densities of Ant2 and Cra2.

\acknowledgments

We thank Sukanya Chakrabarti, Shany Danieli, Kristen McQuinn, Adrian Price-Whelan, and Martin Rey for helpful discussions; and our referee Mario Mateo for helpful comments.
APJ acknowledges support from a Carnegie Fellowship and the Thacher Research Award in Astronomy.
SK is partially supported by NSF grants AST-1813881, AST-1909584 and Heising-Simons foundation grant 2018-1030. 
TSL is supported by NASA through Hubble Fellowship grant HST-HF2-51439.001 awarded by the Space Telescope Science Institute, which is operated by the Association of Universities for Research in Astronomy, Inc., for NASA, under contract NAS5-26555.
ABP is supported by NSF grant AST-1813881.  DBZ acknowledges the support of the Australian Research Council through Discovery Project grant DP180101791, and this research has also been supported in part by the Australian Research Council Centre of Excellence for All Sky Astrophysics in 3 Dimensions (ASTRO 3D), through project number CE170100013.

This paper includes data obtained with the Anglo-Australian Telescope in Australia. We acknowledge the traditional owners of the land on which the AAT stands, the Gamilaraay people, and pay our respects to elders past and present.

This research has made use of the SIMBAD database, operated at CDS, Strasbourg, France \citep{Simbad}.
This research has made use of NASA’s Astrophysics Data System Bibliographic Services.

This paper made use of the Whole Sky Database (wsdb) created by Sergey Koposov and maintained at the Institute of Astronomy, Cambridge by Sergey Koposov, Vasily Belokurov and Wyn Evans with financial support from the Science \& Technology Facilities Council (STFC) and the European Research Council (ERC).

This work has made use of data from the European Space Agency (ESA) mission
{\it Gaia} (\url{https://www.cosmos.esa.int/gaia}), processed by the {\it Gaia}
Data Processing and Analysis Consortium (DPAC,
\url{https://www.cosmos.esa.int/web/gaia/dpac/consortium}). Funding for the DPAC
has been provided by national institutions, in particular the institutions
participating in the {\it Gaia} Multilateral Agreement.


{\it Facilities:} 
{Anglo-Australian Telescope (AAOmega+2dF)}

{\it Software:} 
{\code{numpy} \citep{numpy}, 
\code{scipy} \citep{scipy},
\code{matplotlib} \citep{matplotlib}, 
\code{seaborn} \citep{seaborn},
\code{astropy} \citep{astropy,astropy:2018},
\code{RVSpecFit} \citep{rvspecfit}
\code{q3c} \citep{Koposov2006}, 
\code{emcee} \citep{emcee},
\code{Stan} \citep{stan},
\code{dynesty} \citep{dynesty},
\code{gala} \citep{gala,gala:zenodo},
\code{healpy} \citep{Gorski05,Zonca2019},
\code{galpot} \citep{Dehnen98}
}

\bibliographystyle{aasjournal}

\newpage

\begin{thebibliography}{}
\expandafter\ifx\csname natexlab\endcsname\relax\def\natexlab#1{#1}\fi
\providecommand{\url}[1]{\href{#1}{#1}}

\bibitem[{{Amorisco}(2019)}]{Amorisco19}
{Amorisco}, N.~C. 2019, \mnras, 489, L22

\bibitem[{{Amorisco} \& {Loeb}(2016)}]{Amorisco16}
{Amorisco}, N.~C., \& {Loeb}, A. 2016, \mnras, 459, L51

\bibitem[{{Applebaum} {et~al.}(2021){Applebaum}, {Brooks}, {Christensen},
  {Munshi}, {Quinn}, {Shen}, \& {Tremmel}}]{Applebaum21}
{Applebaum}, E., {Brooks}, A.~M., {Christensen}, C.~R., {et~al.} 2021, \apj,
  906, 96

\bibitem[{{Astropy Collaboration} {et~al.}(2013){Astropy Collaboration},
  {Robitaille}, {Tollerud}, {Greenfield}, {Droettboom}, {Bray}, {Aldcroft},
  {Davis}, {Ginsburg}, {Price-Whelan}, {Kerzendorf}, {Conley}, {Crighton},
  {Barbary}, {Muna}, {Ferguson}, {Grollier}, {Parikh}, {Nair}, {Unther},
  {Deil}, {Woillez}, {Conseil}, {Kramer}, {Turner}, {Singer}, {Fox}, {Weaver},
  {Zabalza}, {Edwards}, {Azalee Bostroem}, {Burke}, {Casey}, {Crawford},
  {Dencheva}, {Ely}, {Jenness}, {Labrie}, {Lim}, {Pierfederici}, {Pontzen},
  {Ptak}, {Refsdal}, {Servillat}, \& {Streicher}}]{astropy}
{Astropy Collaboration}, {Robitaille}, T.~P., {Tollerud}, E.~J., {et~al.} 2013,
  \aap, 558, A33

\bibitem[{{Ball} {et~al.}(2018){Ball}, {Cannon}, {Leisman}, {Adams}, {Haynes},
  {J{\'o}zsa}, {McQuinn}, {Salzer}, {Brunker}, {Giovanelli}, {Hallenbeck},
  {Janesh}, {Janowiecki}, {Jones}, \& {Rhode}}]{Ball18}
{Ball}, C., {Cannon}, J.~M., {Leisman}, L., {et~al.} 2018, \aj, 155, 65

\bibitem[{{Battaglia} {et~al.}(2021){Battaglia}, {Taibi}, {Thomas}, \&
  {Fritz}}]{Battaglia21}
{Battaglia}, G., {Taibi}, S., {Thomas}, G.~F., \& {Fritz}, T.~K. 2021, arXiv
  e-prints, arXiv:2106.08819

\bibitem[{{Besla} {et~al.}(2007){Besla}, {Kallivayalil}, {Hernquist},
  {Robertson}, {Cox}, {van der Marel}, \& {Alcock}}]{Besla07}
{Besla}, G., {Kallivayalil}, N., {Hernquist}, L., {et~al.} 2007, \apj, 668, 949

\bibitem[{{Broadhurst} {et~al.}(2020){Broadhurst}, {De Martino}, {Luu},
  {Smoot}, \& {Tye}}]{Broadhurst20}
{Broadhurst}, T., {De Martino}, I., {Luu}, H.~N., {Smoot}, G.~F., \& {Tye}, S.
  H.~H. 2020, \prd, 101, 083012

\bibitem[{{Brunker} {et~al.}(2019){Brunker}, {McQuinn}, {Salzer}, {Cannon},
  {Janowiecki}, {Leisman}, {Rhode}, {Adams}, {Ball}, {Dolphin}, {Giovanelli},
  \& {Haynes}}]{Brunker19}
{Brunker}, S.~W., {McQuinn}, K. B.~W., {Salzer}, J.~J., {et~al.} 2019, \aj,
  157, 76

\bibitem[{{Buder} {et~al.}(2020){Buder}, {Sharma}, {Kos}, {Amarsi},
  {Nordlander}, {Lind}, {Martell}, {Asplund}, {Bland-Hawthorn}, {Casey}, {De
  Silva}, {D'Orazi}, {Freeman}, {Hayden}, {Lewis}, {Lin}, {Schlesinger},
  {Simpson}, {Stello}, {Zucker}, {Zwitter}, {Beeson}, {Buck}, {Casagrande},
  {Clark}, {Cotar}, {Da Costa}, {de Grijs}, {Feuillet}, {Horner}, {Khanna},
  {Kafle}, {Liu}, {Montet}, {Nandakumar}, {Nataf}, {Ness}, {Spina}, {Traven},
  {Tepper-Garcia}, {Ting}, {Vogrincic}, {Wittenmyer}, {Zerjal}, \& {the GALAH
  collaboration}}]{Buder20}
{Buder}, S., {Sharma}, S., {Kos}, J., {et~al.} 2020, arXiv e-prints,
  arXiv:2011.02505

\bibitem[{{Caldwell} {et~al.}(2017){Caldwell}, {Walker}, {Mateo}, {Olszewski},
  {Koposov}, {Belokurov}, {Torrealba}, {Geringer-Sameth}, \&
  {Johnson}}]{Caldwell17}
{Caldwell}, N., {Walker}, M.~G., {Mateo}, M., {et~al.} 2017, \apj, 839, 20

\bibitem[{{Carleton} {et~al.}(2019){Carleton}, {Errani}, {Cooper},
  {Kaplinghat}, {Pe{\~n}arrubia}, \& {Guo}}]{Carleton19}
{Carleton}, T., {Errani}, R., {Cooper}, M., {et~al.} 2019, \mnras, 485, 382

\bibitem[{Carpenter {et~al.}(2017)Carpenter, Gelman, Hoffman, Lee, Goodrich,
  Betancourt, Brubaker, Guo, Li, \& Riddell}]{stan}
Carpenter, B., Gelman, A., Hoffman, M., {et~al.} 2017, Journal of Statistical
  Software, Articles, 76, 1.
\newblock \url{https://www.jstatsoft.org/v076/i01}

\bibitem[{{Carrera} {et~al.}(2013){Carrera}, {Pancino}, {Gallart}, \& {del
  Pino}}]{Carrera13}
{Carrera}, R., {Pancino}, E., {Gallart}, C., \& {del Pino}, A. 2013, \mnras,
  434, 1681

\bibitem[{{Chakrabarti} {et~al.}(2019){Chakrabarti}, {Chang}, {Price-Whelan},
  {Read}, {Blitz}, \& {Hernquist}}]{Chakrabarti19}
{Chakrabarti}, S., {Chang}, P., {Price-Whelan}, A.~M., {et~al.} 2019, \apj,
  886, 67

\bibitem[{{Chambers} {et~al.}(2016){Chambers}, {Magnier}, {Metcalfe},
  {Flewelling}, {Huber}, {Waters}, {Denneau}, {Draper}, {Farrow}, {Finkbeiner},
  {Holmberg}, {Koppenhoefer}, {Price}, {Rest}, {Saglia}, {Schlafly}, {Smartt},
  {Sweeney}, {Wainscoat}, {Burgett}, {Chastel}, {Grav}, {Heasley}, {Hodapp},
  {Jedicke}, {Kaiser}, {Kudritzki}, {Luppino}, {Lupton}, {Monet}, {Morgan},
  {Onaka}, {Shiao}, {Stubbs}, {Tonry}, {White}, {Ba{\~n}ados}, {Bell},
  {Bender}, {Bernard}, {Boegner}, {Boffi}, {Botticella}, {Calamida},
  {Casertano}, {Chen}, {Chen}, {Cole}, {Deacon}, {Frenk}, {Fitzsimmons},
  {Gezari}, {Gibbs}, {Goessl}, {Goggia}, {Gourgue}, {Goldman}, {Grant},
  {Grebel}, {Hambly}, {Hasinger}, {Heavens}, {Heckman}, {Henderson}, {Henning},
  {Holman}, {Hopp}, {Ip}, {Isani}, {Jackson}, {Keyes}, {Koekemoer}, {Kotak},
  {Le}, {Liska}, {Long}, {Lucey}, {Liu}, {Martin}, {Masci}, {McLean}, {Mindel},
  {Misra}, {Morganson}, {Murphy}, {Obaika}, {Narayan}, {Nieto-Santisteban},
  {Norberg}, {Peacock}, {Pier}, {Postman}, {Primak}, {Rae}, {Rai}, {Riess},
  {Riffeser}, {Rix}, {R{\"o}ser}, {Russel}, {Rutz}, {Schilbach}, {Schultz},
  {Scolnic}, {Strolger}, {Szalay}, {Seitz}, {Small}, {Smith}, {Soderblom},
  {Taylor}, {Thomson}, {Taylor}, {Thakar}, {Thiel}, {Thilker}, {Unger},
  {Urata}, {Valenti}, {Wagner}, {Walder}, {Walter}, {Watters}, {Werner},
  {Wood-Vasey}, \& {Wyse}}]{Chambers16}
{Chambers}, K.~C., {Magnier}, E.~A., {Metcalfe}, N., {et~al.} 2016, arXiv
  e-prints, arXiv:1612.05560

\bibitem[{{Chiti} {et~al.}(2021){Chiti}, {Frebel}, {Simon}, {Erkal}, {Chang},
  {Necib}, {Ji}, {Jerjen}, {Kim}, \& {Norris}}]{Chiti21}
{Chiti}, A., {Frebel}, A., {Simon}, J.~D., {et~al.} 2021, Nature Astronomy,
  arXiv:2012.02309

\bibitem[{{Choi} {et~al.}(2016){Choi}, {Dotter}, {Conroy}, {Cantiello},
  {Paxton}, \& {Johnson}}]{Choi16}
{Choi}, J., {Dotter}, A., {Conroy}, C., {et~al.} 2016, \apj, 823, 102

\bibitem[{{Collins} {et~al.}(2020){Collins}, {Tollerud}, {Rich}, {Ibata},
  {Martin}, {Chapman}, {Gilbert}, \& {Preston}}]{Collins20}
{Collins}, M. L.~M., {Tollerud}, E.~J., {Rich}, R.~M., {et~al.} 2020, \mnras,
  491, 3496

\bibitem[{{Collins} {et~al.}(2021){Collins}, {Read}, {Ibata}, {Rich}, {Martin},
  {Pe{\~n}arrubia}, {Chapman}, {Tollerud}, \& {Weisz}}]{Collins21}
{Collins}, M. L.~M., {Read}, J.~I., {Ibata}, R.~A., {et~al.} 2021, arXiv
  e-prints, arXiv:2102.11890

\bibitem[{{Dalcanton} {et~al.}(1997{\natexlab{a}}){Dalcanton}, {Spergel},
  {Gunn}, {Schmidt}, \& {Schneider}}]{Dalcanton97}
{Dalcanton}, J.~J., {Spergel}, D.~N., {Gunn}, J.~E., {Schmidt}, M., \&
  {Schneider}, D.~P. 1997{\natexlab{a}}, \aj, 114, 635

\bibitem[{{Dalcanton} {et~al.}(1997{\natexlab{b}}){Dalcanton}, {Spergel}, \&
  {Summers}}]{Dalcanton97b}
{Dalcanton}, J.~J., {Spergel}, D.~N., \& {Summers}, F.~J. 1997{\natexlab{b}},
  \apj, 482, 659

\bibitem[{{Danieli} \& {van Dokkum}(2019)}]{Danieli19}
{Danieli}, S., \& {van Dokkum}, P. 2019, \apj, 875, 155

\bibitem[{{Danieli} {et~al.}(2019){Danieli}, {van Dokkum}, {Conroy}, {Abraham},
  \& {Romanowsky}}]{Danieli19b}
{Danieli}, S., {van Dokkum}, P., {Conroy}, C., {Abraham}, R., \& {Romanowsky},
  A.~J. 2019, \apjl, 874, L12

\bibitem[{{Danieli} {et~al.}(2020){Danieli}, {Lokhorst}, {Zhang}, {Merritt},
  {van Dokkum}, {Abraham}, {Conroy}, {Gilhuly}, {Greco}, {Janssens}, {Li},
  {Liu}, {Miller}, \& {Mowla}}]{Danieli20}
{Danieli}, S., {Lokhorst}, D., {Zhang}, J., {et~al.} 2020, \apj, 894, 119

\bibitem[{{Dehnen} \& {Binney}(1998)}]{Dehnen98}
{Dehnen}, W., \& {Binney}, J. 1998, \mnras, 294, 429

\bibitem[{{Dejonghe}(1987)}]{Dejonghe87}
{Dejonghe}, H. 1987, \mnras, 224, 13

\bibitem[{{Dotter} {et~al.}(2008){Dotter}, {Chaboyer}, {Jevremovi{\'c}},
  {Kostov}, {Baron}, \& {Ferguson}}]{Dotter08}
{Dotter}, A., {Chaboyer}, B., {Jevremovi{\'c}}, D., {et~al.} 2008, \apjs, 178,
  89

\bibitem[{{Emsellem} {et~al.}(2019){Emsellem}, {van der Burg}, {Fensch},
  {Je{\v{r}}{\'a}bkov{\'a}}, {Zanella}, {Agnello}, {Hilker}, {M{\"u}ller},
  {Rejkuba}, {Duc}, {Durrell}, {Habas}, {Lelli}, {Lim}, {Marleau}, {Peng}, \&
  {S{\'a}nchez-Janssen}}]{Emsellem19}
{Emsellem}, E., {van der Burg}, R. F.~J., {Fensch}, J., {et~al.} 2019, \aap,
  625, A76

\bibitem[{{Erkal} \& {Belokurov}(2020)}]{Erkal20}
{Erkal}, D., \& {Belokurov}, V.~A. 2020, \mnras, 495, 2554

\bibitem[{{Erkal} {et~al.}(2017){Erkal}, {Koposov}, \& {Belokurov}}]{Erkal17}
{Erkal}, D., {Koposov}, S.~E., \& {Belokurov}, V. 2017, \mnras, 470, 60

\bibitem[{{Erkal} {et~al.}(2019){Erkal}, {Belokurov}, {Laporte}, {Koposov},
  {Li}, {Grillmair}, {Kallivayalil}, {Price-Whelan}, {Evans}, {Hawkins},
  {Hendel}, {Mateu}, {Navarro}, {del Pino}, {Slater}, {Sohn}, \& {Orphan Aspen
  Treasury Collaboration}}]{Erkal19}
{Erkal}, D., {Belokurov}, V., {Laporte}, C.~F.~P., {et~al.} 2019, \mnras, 487,
  2685

\bibitem[{{Erkal} {et~al.}(2020){Erkal}, {Deason}, {Belokurov}, {Xue},
  {Koposov}, {Bird}, {Liu}, {Simion}, {Yang}, {Zhang}, \& {Zhao}}]{Erkal20b}
{Erkal}, D., {Deason}, A.~J., {Belokurov}, V., {et~al.} 2020, arXiv e-prints,
  arXiv:2010.13789

\bibitem[{{Errani} {et~al.}(2015){Errani}, {Penarrubia}, \&
  {Tormen}}]{Errani15}
{Errani}, R., {Penarrubia}, J., \& {Tormen}, G. 2015, \mnras, 449, L46

\bibitem[{{Fardal} {et~al.}(2015){Fardal}, {Huang}, \& {Weinberg}}]{Fardal15}
{Fardal}, M.~A., {Huang}, S., \& {Weinberg}, M.~D. 2015, \mnras, 452, 301

\bibitem[{{Fattahi} {et~al.}(2018){Fattahi}, {Navarro}, {Frenk}, {Oman},
  {Sawala}, \& {Schaller}}]{Fattahi18}
{Fattahi}, A., {Navarro}, J.~F., {Frenk}, C.~S., {et~al.} 2018, \mnras, 476,
  3816

\bibitem[{{Forbes} {et~al.}(2021){Forbes}, {Gannon}, {Romanowsky}, {Alabi},
  {Brodie}, {Couch}, \& {Ferr{\'e}-Mateu}}]{Forbes21}
{Forbes}, D.~A., {Gannon}, J.~S., {Romanowsky}, A.~J., {et~al.} 2021, \mnras,
  500, 1279

\bibitem[{{Foreman-Mackey} {et~al.}(2013){Foreman-Mackey}, {Hogg}, {Lang}, \&
  {Goodman}}]{emcee}
{Foreman-Mackey}, D., {Hogg}, D.~W., {Lang}, D., \& {Goodman}, J. 2013, \pasp,
  125, 306

\bibitem[{{Frings} {et~al.}(2017){Frings}, {Macci{\`o}}, {Buck}, {Penzo},
  {Dutton}, {Blank}, \& {Obreja}}]{Frings17}
{Frings}, J., {Macci{\`o}}, A., {Buck}, T., {et~al.} 2017, \mnras, 472, 3378

\bibitem[{{Fritz} {et~al.}(2018){Fritz}, {Battaglia}, {Pawlowski},
  {Kallivayalil}, {van der Marel}, {Sohn}, {Brook}, \& {Besla}}]{Fritz18}
{Fritz}, T.~K., {Battaglia}, G., {Pawlowski}, M.~S., {et~al.} 2018, \aap, 619,
  A103

\bibitem[{{Fu} {et~al.}(2019){Fu}, {Simon}, \& {Alarc{\'o}n Jara}}]{Fu19}
{Fu}, S.~W., {Simon}, J.~D., \& {Alarc{\'o}n Jara}, A.~G. 2019, \apj, 883, 11

\bibitem[{{Gaia Collaboration} {et~al.}(2018{\natexlab{a}}){Gaia
  Collaboration}, {Brown}, {Vallenari}, {Prusti}, {de Bruijne}, {Babusiaux},
  {Bailer-Jones}, {Biermann}, {Evans}, {Eyer}, {Jansen}, {Jordi}, {Klioner},
  {Lammers}, {Lindegren}, {Luri}, {Mignard}, {Panem}, {Pourbaix}, {Randich},
  {Sartoretti}, {Siddiqui}, {Soubiran}, {van Leeuwen}, {Walton}, {Arenou},
  {Bastian}, {Cropper}, {Drimmel}, {Katz}, {Lattanzi}, {Bakker}, {Cacciari},
  {Casta{\~n}eda}, {Chaoul}, {Cheek}, {De Angeli}, {Fabricius}, {Guerra},
  {Holl}, {Masana}, {Messineo}, {Mowlavi}, {Nienartowicz}, {Panuzzo},
  {Portell}, {Riello}, {Seabroke}, {Tanga}, {Th{\'e}venin}, {Gracia-Abril},
  {Comoretto}, {Garcia-Reinaldos}, {Teyssier}, {Altmann}, {Andrae}, {Audard},
  {Bellas-Velidis}, {Benson}, {Berthier}, {Blomme}, {Burgess}, {Busso},
  {Carry}, {Cellino}, {Clementini}, {Clotet}, {Creevey}, {Davidson}, {De
  Ridder}, {Delchambre}, {Dell'Oro}, {Ducourant},
  {Fern{\'a}ndez-Hern{\'a}ndez}, {Fouesneau}, {Fr{\'e}mat}, {Galluccio},
  {Garc{\'\i}a-Torres}, {Gonz{\'a}lez-N{\'u}{\~n}ez}, {Gonz{\'a}lez-Vidal},
  {Gosset}, {Guy}, {Halbwachs}, {Hambly}, {Harrison}, {Hern{\'a}ndez},
  {Hestroffer}, {Hodgkin}, {Hutton}, {Jasniewicz}, {Jean-Antoine-Piccolo},
  {Jordan}, {Korn}, {Krone-Martins}, {Lanzafame}, {Lebzelter}, {L{\"o}ffler},
  {Manteiga}, {Marrese}, {Mart{\'\i}n-Fleitas}, {Moitinho}, {Mora}, {Muinonen},
  {Osinde}, {Pancino}, {Pauwels}, {Petit}, {Recio-Blanco}, {Richards},
  {Rimoldini}, {Robin}, {Sarro}, {Siopis}, {Smith}, {Sozzetti}, {S{\"u}veges},
  {Torra}, {van Reeven}, {Abbas}, {Abreu Aramburu}, {Accart}, {Aerts},
  {Altavilla}, {{\'A}lvarez}, {Alvarez}, {Alves}, {Anderson}, {Andrei},
  {Anglada Varela}, {Antiche}, {Antoja}, {Arcay}, {Astraatmadja}, {Bach},
  {Baker}, {Balaguer-N{\'u}{\~n}ez}, {Balm}, {Barache}, {Barata}, {Barbato},
  {Barblan}, {Barklem}, {Barrado}, {Barros}, {Barstow}, {Bartholom{\'e}
  Mu{\~n}oz}, {Bassilana}, {Becciani}, {Bellazzini}, {Berihuete}, {Bertone},
  {Bianchi}, {Bienaym{\'e}}, {Blanco-Cuaresma}, {Boch}, {Boeche}, {Bombrun},
  {Borrachero}, {Bossini}, {Bouquillon}, {Bourda}, {Bragaglia}, {Bramante},
  {Breddels}, {Bressan}, {Brouillet}, {Br{\"u}semeister}, {Brugaletta},
  {Bucciarelli}, {Burlacu}, {Busonero}, {Butkevich}, {Buzzi}, {Caffau},
  {Cancelliere}, {Cannizzaro}, {Cantat-Gaudin}, {Carballo}, {Carlucci},
  {Carrasco}, {Casamiquela}, {Castellani}, {Castro-Ginard}, {Charlot},
  {Chemin}, {Chiavassa}, {Cocozza}, {Costigan}, {Cowell}, {Crifo}, {Crosta},
  {Crowley}, {Cuypers}, {Dafonte}, {Damerdji}, {Dapergolas}, {David}, {David},
  {de Laverny}, {De Luise}, {De March}, {de Martino}, {de Souza}, {de Torres},
  {Debosscher}, {del Pozo}, {Delbo}, {Delgado}, {Delgado}, {Di Matteo},
  {Diakite}, {Diener}, {Distefano}, {Dolding}, {Drazinos}, {Dur{\'a}n},
  {Edvardsson}, {Enke}, {Eriksson}, {Esquej}, {Eynard Bontemps}, {Fabre},
  {Fabrizio}, {Faigler}, {Falc{\~a}o}, {Farr{\`a}s Casas}, {Federici},
  {Fedorets}, {Fernique}, {Figueras}, {Filippi}, {Findeisen}, {Fonti},
  {Fraile}, {Fraser}, {Fr{\'e}zouls}, {Gai}, {Galleti}, {Garabato},
  {Garc{\'\i}a-Sedano}, {Garofalo}, {Garralda}, {Gavel}, {Gavras}, {Gerssen},
  {Geyer}, {Giacobbe}, {Gilmore}, {Girona}, {Giuffrida}, {Glass}, {Gomes},
  {Granvik}, {Gueguen}, {Guerrier}, {Guiraud}, {Guti{\'e}rrez-S{\'a}nchez},
  {Haigron}, {Hatzidimitriou}, {Hauser}, {Haywood}, {Heiter}, {Helmi}, {Heu},
  {Hilger}, {Hobbs}, {Hofmann}, {Holland}, {Huckle}, {Hypki}, {Icardi},
  {Jan{\ss}en}, {Jevardat de Fombelle}, {Jonker}, {Juh{\'a}sz}, {Julbe},
  {Karampelas}, {Kewley}, {Klar}, {Kochoska}, {Kohley}, {Kolenberg},
  {Kontizas}, {Kontizas}, {Koposov}, {Kordopatis}, {Kostrzewa-Rutkowska},
  {Koubsky}, {Lambert}, {Lanza}, {Lasne}, {Lavigne}, {Le Fustec}, {Le
  Poncin-Lafitte}, {Lebreton}, {Leccia}, {Leclerc}, {Lecoeur-Taibi},
  {Lenhardt}, {Leroux}, {Liao}, {Licata}, {Lindstr{\o}m}, {Lister}, {Livanou},
  {Lobel}, {L{\'o}pez}, {Managau}, {Mann}, {Mantelet}, {Marchal}, {Marchant},
  {Marconi}, {Marinoni}, {Marschalk{\'o}}, {Marshall}, {Martino}, {Marton},
  {Mary}, {Massari}, {Matijevi{\v{c}}}, {Mazeh}, {McMillan}, {Messina},
  {Michalik}, {Millar}, {Molina}, {Molinaro}, {Moln{\'a}r}, {Montegriffo},
  {Mor}, {Morbidelli}, {Morel}, {Morris}, {Mulone}, {Muraveva}, {Musella},
  {Nelemans}, {Nicastro}, {Noval}, {O'Mullane}, {Ord{\'e}novic},
  {Ord{\'o}{\~n}ez-Blanco}, {Osborne}, {Pagani}, {Pagano}, {Pailler},
  {Palacin}, {Palaversa}, {Panahi}, {Pawlak}, {Piersimoni}, {Pineau}, {Plachy},
  {Plum}, {Poggio}, {Poujoulet}, {Pr{\v{s}}a}, {Pulone}, {Racero}, {Ragaini},
  {Rambaux}, {Ramos-Lerate}, {Regibo}, {Reyl{\'e}}, {Riclet}, {Ripepi}, {Riva},
  {Rivard}, {Rixon}, {Roegiers}, {Roelens}, {Romero-G{\'o}mez}, {Rowell},
  {Royer}, {Ruiz-Dern}, {Sadowski}, {Sagrist{\`a} Sell{\'e}s}, {Sahlmann},
  {Salgado}, {Salguero}, {Sanna}, {Santana-Ros}, {Sarasso}, {Savietto},
  {Schultheis}, {Sciacca}, {Segol}, {Segovia}, {S{\'e}gransan}, {Shih},
  {Siltala}, {Silva}, {Smart}, {Smith}, {Solano}, {Solitro}, {Sordo}, {Soria
  Nieto}, {Souchay}, {Spagna}, {Spoto}, {Stampa}, {Steele},
  {Steidelm{\"u}ller}, {Stephenson}, {Stoev}, {Suess}, {Surdej}, {Szabados},
  {Szegedi-Elek}, {Tapiador}, {Taris}, {Tauran}, {Taylor}, {Teixeira},
  {Terrett}, {Teyssandier}, {Thuillot}, {Titarenko}, {Torra Clotet}, {Turon},
  {Ulla}, {Utrilla}, {Uzzi}, {Vaillant}, {Valentini}, {Valette}, {van Elteren},
  {Van Hemelryck}, {van Leeuwen}, {Vaschetto}, {Vecchiato}, {Veljanoski},
  {Viala}, {Vicente}, {Vogt}, {von Essen}, {Voss}, {Votruba}, {Voutsinas},
  {Walmsley}, {Weiler}, {Wertz}, {Wevers}, {Wyrzykowski}, {Yoldas},
  {{\v{Z}}erjal}, {Ziaeepour}, {Zorec}, {Zschocke}, {Zucker}, {Zurbach}, \&
  {Zwitter}}]{gaiadr2}
{Gaia Collaboration}, {Brown}, A.~G.~A., {Vallenari}, A., {et~al.}
  2018{\natexlab{a}}, \aap, 616, A1

\bibitem[{{Gaia Collaboration} {et~al.}(2018{\natexlab{b}}){Gaia
  Collaboration}, {Babusiaux}, {van Leeuwen}, {Barstow}, {Jordi}, {Vallenari},
  {Bossini}, {Bressan}, {Cantat-Gaudin}, {van Leeuwen}, \&
  et~al.}]{Babusiaux2018}
{Gaia Collaboration}, {Babusiaux}, C., {van Leeuwen}, F., {et~al.}
  2018{\natexlab{b}}, \aap, 616, A10

\bibitem[{{Garavito-Camargo} {et~al.}(2019){Garavito-Camargo}, {Besla},
  {Laporte}, {Johnston}, {G{\'o}mez}, \& {Watkins}}]{GaravitoCamargo19}
{Garavito-Camargo}, N., {Besla}, G., {Laporte}, C. F.~P., {et~al.} 2019, \apj,
  884, 51

\bibitem[{{Gibbons} {et~al.}(2014){Gibbons}, {Belokurov}, \&
  {Evans}}]{Gibbons14}
{Gibbons}, S.~L.~J., {Belokurov}, V., \& {Evans}, N.~W. 2014, \mnras, 445, 3788

\bibitem[{{G{\'o}mez} {et~al.}(2015){G{\'o}mez}, {Besla}, {Carpintero},
  {Villalobos}, {O'Shea}, \& {Bell}}]{Gomez15}
{G{\'o}mez}, F.~A., {Besla}, G., {Carpintero}, D.~D., {et~al.} 2015, \apj, 802,
  128

\bibitem[{{G{\'o}rski} {et~al.}(2005){G{\'o}rski}, {Hivon}, {Banday},
  {Wandelt}, {Hansen}, {Reinecke}, \& {Bartelmann}}]{Gorski05}
{G{\'o}rski}, K.~M., {Hivon}, E., {Banday}, A.~J., {et~al.} 2005, \apj, 622,
  759

\bibitem[{{Gravity Collaboration} {et~al.}(2018){Gravity Collaboration},
  {Abuter}, {Amorim}, {Anugu}, {Baub{\"o}ck}, {Benisty}, {Berger}, {Blind},
  {Bonnet}, {Brandner}, {Buron}, {Collin}, {Chapron}, {Cl{\'e}net}, {Coud{\'e}
  Du Foresto}, {de Zeeuw}, {Deen}, {Delplancke-Str{\"o}bele}, {Dembet},
  {Dexter}, {Duvert}, {Eckart}, {Eisenhauer}, {Finger}, {F{\"o}rster
  Schreiber}, {F{\'e}dou}, {Garcia}, {Garcia Lopez}, {Gao}, {Gendron},
  {Genzel}, {Gillessen}, {Gordo}, {Habibi}, {Haubois}, {Haug}, {Hau{\ss}mann},
  {Henning}, {Hippler}, {Horrobin}, {Hubert}, {Hubin}, {Jimenez Rosales},
  {Jochum}, {Jocou}, {Kaufer}, {Kellner}, {Kendrew}, {Kervella}, {Kok},
  {Kulas}, {Lacour}, {Lapeyr{\`e}re}, {Lazareff}, {Le Bouquin}, {L{\'e}na},
  {Lippa}, {Lenzen}, {M{\'e}rand}, {M{\"u}ler}, {Neumann}, {Ott}, {Palanca},
  {Paumard}, {Pasquini}, {Perraut}, {Perrin}, {Pfuhl}, {Plewa}, {Rabien},
  {Ram{\'\i}rez}, {Ramos}, {Rau}, {Rodr{\'\i}guez-Coira}, {Rohloff}, {Rousset},
  {Sanchez-Bermudez}, {Scheithauer}, {Sch{\"o}ller}, {Schuler}, {Spyromilio},
  {Straub}, {Straubmeier}, {Sturm}, {Tacconi}, {Tristram}, {Vincent}, {von
  Fellenberg}, {Wank}, {Waisberg}, {Widmann}, {Wieprecht}, {Wiest},
  {Wiezorrek}, {Woillez}, {Yazici}, {Ziegler}, \& {Zins}}]{GRAVITY18}
{Gravity Collaboration}, {Abuter}, R., {Amorim}, A., {et~al.} 2018, \aap, 615,
  L15

\bibitem[{{Greco} {et~al.}(2018){Greco}, {Greene}, {Strauss}, {Macarthur},
  {Flowers}, {Goulding}, {Huang}, {Kim}, {Komiyama}, {Leauthaud}, {Leisman},
  {Lupton}, {Sif{\'o}n}, \& {Wang}}]{Greco18}
{Greco}, J.~P., {Greene}, J.~E., {Strauss}, M.~A., {et~al.} 2018, \apj, 857,
  104

\bibitem[{{Hernquist}(1990)}]{Hernquist90}
{Hernquist}, L. 1990, \apj, 356, 359

\bibitem[{Hunter(2007)}]{matplotlib}
Hunter, J.~D. 2007, Computing in Science \& Engineering, 9, 90.
\newblock
  \url{http://scitation.aip.org/content/aip/journal/cise/9/3/10.1109/MCSE.2007.55}

\bibitem[{{Husser} {et~al.}(2013){Husser}, {Wende-von Berg}, {Dreizler},
  {Homeier}, {Reiners}, {Barman}, \& {Hauschildt}}]{Husser13}
{Husser}, T.~O., {Wende-von Berg}, S., {Dreizler}, S., {et~al.} 2013, \aap,
  553, A6

\bibitem[{{Irwin} {et~al.}(1990){Irwin}, {Bunclark}, {Bridgeland}, \&
  {McMahon}}]{Irwin90}
{Irwin}, M.~J., {Bunclark}, P.~S., {Bridgeland}, M.~T., \& {McMahon}, R.~G.
  1990, \mnras, 244, 16P

\bibitem[{{Jackson} {et~al.}(2021){Jackson}, {Kaviraj}, {Martin}, {Devriendt},
  {Slyz}, {Silk}, {Dubois}, {Yi}, {Pichon}, {Volonteri}, {Choi}, {Kimm},
  {Kraljic}, \& {Peirani}}]{Jackson21}
{Jackson}, R.~A., {Kaviraj}, S., {Martin}, G., {et~al.} 2021, \mnras,
  arXiv:2010.02219

\bibitem[{{Jenkins} {et~al.}(2020){Jenkins}, {Li}, {Pace}, {Ji}, {Koposov}, \&
  {Mutlu-Pakdil}}]{Jenkins21}
{Jenkins}, S., {Li}, T.~S., {Pace}, A.~B., {et~al.} 2020, arXiv e-prints,
  arXiv:2101.00013

\bibitem[{{Jethwa} {et~al.}(2016){Jethwa}, {Erkal}, \& {Belokurov}}]{Jethwa16}
{Jethwa}, P., {Erkal}, D., \& {Belokurov}, V. 2016, \mnras, 461, 2212

\bibitem[{{Ji} {et~al.}(2016){Ji}, {Frebel}, {Ezzeddine}, \& {Casey}}]{Ji16d}
{Ji}, A.~P., {Frebel}, A., {Ezzeddine}, R., \& {Casey}, A.~R. 2016, \apjl, 832,
  L3

\bibitem[{{Ji} {et~al.}(2020){Ji}, {Li}, {Hansen}, {Casey}, {Koposov}, {Pace},
  {Mackey}, {Lewis}, {Simpson}, {Bland-Hawthorn}, {Cullinane}, {Da Costa},
  {Hattori}, {Martell}, {Kuehn}, {Erkal}, {Shipp}, {Wan}, \& {Zucker}}]{Ji20b}
{Ji}, A.~P., {Li}, T.~S., {Hansen}, T.~T., {et~al.} 2020, \aj, 160, 181

\bibitem[{{Jiang} {et~al.}(2019){Jiang}, {Dekel}, {Freundlich}, {Romanowsky},
  {Dutton}, {Macci{\`o}}, \& {Di Cintio}}]{Jiang19}
{Jiang}, F., {Dekel}, A., {Freundlich}, J., {et~al.} 2019, \mnras, 487, 5272

\bibitem[{Jones {et~al.}(2001)Jones, Oliphant, Peterson, {et~al.}}]{scipy}
Jones, E., Oliphant, T., Peterson, P., {et~al.} 2001, {SciPy}: Open source
  scientific tools for {Python}, , .
\newblock \url{http://www.scipy.org/}

\bibitem[{{J{\"o}nsson} {et~al.}(2020){J{\"o}nsson}, {Holtzman}, {Allende
  Prieto}, {Cunha}, {Garc{\'\i}a-Hern{\'a}ndez}, {Hasselquist}, {Masseron},
  {Osorio}, {Shetrone}, {Smith}, {Stringfellow}, {Bizyaev}, {Edvardsson},
  {Majewski}, {M{\'e}sz{\'a}ros}, {Souto}, {Zamora}, {Beaton}, {Bovy}, {Donor},
  {Pinsonneault}, {Poovelil}, \& {Sobeck}}]{Jonsson20}
{J{\"o}nsson}, H., {Holtzman}, J.~A., {Allende Prieto}, C., {et~al.} 2020, \aj,
  160, 120

\bibitem[{{Kado-Fong} {et~al.}(2021){Kado-Fong}, {Petrescu}, {Mohammad},
  {Greco}, {Greene}, {Adams}, {Huang}, {Leisman}, {Munshi}, {Tanoglidis}, \&
  {Van Nest}}]{KadoFong21}
{Kado-Fong}, E., {Petrescu}, M., {Mohammad}, M., {et~al.} 2021, arXiv e-prints,
  arXiv:2106.05288

\bibitem[{{Kallivayalil} {et~al.}(2013){Kallivayalil}, {van der Marel},
  {Besla}, {Anderson}, \& {Alcock}}]{Kallivayalil13}
{Kallivayalil}, N., {van der Marel}, R.~P., {Besla}, G., {Anderson}, J., \&
  {Alcock}, C. 2013, \apj, 764, 161

\bibitem[{{Kallivayalil} {et~al.}(2018){Kallivayalil}, {Sales}, {Zivick},
  {Fritz}, {Del Pino}, {Sohn}, {Besla}, {van der Marel}, {Navarro}, \&
  {Sacchi}}]{Kallivayalil18}
{Kallivayalil}, N., {Sales}, L.~V., {Zivick}, P., {et~al.} 2018, \apj, 867, 19

\bibitem[{{Kaplinghat} \& {Strigari}(2008)}]{Kaplinghat08}
{Kaplinghat}, M., \& {Strigari}, L.~E. 2008, \apjl, 682, L93

\bibitem[{{King}(1962)}]{King1962}
{King}, I. 1962, \aj, 67, 471

\bibitem[{{Kirby} {et~al.}(2013){Kirby}, {Cohen}, {Guhathakurta}, {Cheng},
  {Bullock}, \& {Gallazzi}}]{Kirby13}
{Kirby}, E.~N., {Cohen}, J.~G., {Guhathakurta}, P., {et~al.} 2013, \apj, 779,
  102

\bibitem[{{Kirby} {et~al.}(2020){Kirby}, {Gilbert}, {Escala}, {Wojno},
  {Guhathakurta}, {Majewski}, \& {Beaton}}]{Kirby20}
{Kirby}, E.~N., {Gilbert}, K.~M., {Escala}, I., {et~al.} 2020, \aj, 159, 46

\bibitem[{{Kirby} {et~al.}(2011){Kirby}, {Lanfranchi}, {Simon}, {Cohen}, \&
  {Guhathakurta}}]{Kirby11}
{Kirby}, E.~N., {Lanfranchi}, G.~A., {Simon}, J.~D., {Cohen}, J.~G., \&
  {Guhathakurta}, P. 2011, \apj, 727, 78

\bibitem[{{Koda} {et~al.}(2015){Koda}, {Yagi}, {Yamanoi}, \&
  {Komiyama}}]{Koda15}
{Koda}, J., {Yagi}, M., {Yamanoi}, H., \& {Komiyama}, Y. 2015, \apjl, 807, L2

\bibitem[{{Koposov} \& {Bartunov}(2006)}]{Koposov2006}
{Koposov}, S., \& {Bartunov}, O. 2006, in Astronomical Society of the Pacific
  Conference Series, Vol. 351, Astronomical Data Analysis Software and Systems
  XV, ed. C.~{Gabriel}, C.~{Arviset}, D.~{Ponz}, \& S.~{Enrique}, 735

\bibitem[{{Koposov}(2019)}]{rvspecfit}
{Koposov}, S.~E. 2019, {RVSpecFit: Radial velocity and stellar atmospheric
  parameter fitting}, , , ascl:1907.013

\bibitem[{{Kordopatis} {et~al.}(2016){Kordopatis}, {Amorisco}, {Evans},
  {Gilmore}, \& {Koposov}}]{Kordopatis16}
{Kordopatis}, G., {Amorisco}, N.~C., {Evans}, N.~W., {Gilmore}, G., \&
  {Koposov}, S.~E. 2016, \mnras, 457, 1299

\bibitem[{{Lemasle} {et~al.}(2012){Lemasle}, {Hill}, {Tolstoy}, {Venn},
  {Shetrone}, {Irwin}, {de Boer}, {Starkenburg}, \& {Salvadori}}]{Lemasle12}
{Lemasle}, B., {Hill}, V., {Tolstoy}, E., {et~al.} 2012, \aap, 538, A100

\bibitem[{{Li} {et~al.}(2021){Li}, {Hammer}, {Babusiaux}, {Pawlowski}, {Yang},
  {Arenou}, {Du}, \& {Wang}}]{Li2021pm}
{Li}, H., {Hammer}, F., {Babusiaux}, C., {et~al.} 2021, arXiv e-prints,
  arXiv:2104.03974

\bibitem[{Li \& {S5 Collaboration}(2021)}]{s5release}
Li, T.~S., \& {S5 Collaboration}. 2021, Southern Stellar Stream Spectroscopic
  Survey: The First Public Data Release,  Zenodo, doi:10.5281/zenodo.4695135

\bibitem[{{Li} {et~al.}(2017){Li}, {Simon}, {Drlica-Wagner}, {Bechtol}, {Wang},
  {Garc{\'\i}a-Bellido}, {Frieman}, {Marshall}, {James}, {Strigari}, {Pace},
  {Balbinot}, {Zhang}, {Abbott}, {Allam}, {Benoit-L{\'e}vy}, {Bernstein},
  {Bertin}, {Brooks}, {Burke}, {Carnero Rosell}, {Carrasco Kind}, {Carretero},
  {Cunha}, {D'Andrea}, {da Costa}, {DePoy}, {Desai}, {Diehl}, {Eifler},
  {Flaugher}, {Goldstein}, {Gruen}, {Gruendl}, {Gschwend}, {Gutierrez},
  {Krause}, {Kuehn}, {Lin}, {Maia}, {March}, {Menanteau}, {Miquel}, {Plazas},
  {Romer}, {Sanchez}, {Santiago}, {Schubnell}, {Sevilla-Noarbe}, {Smith},
  {Sobreira}, {Suchyta}, {Tarle}, {Thomas}, {Tucker}, {Walker}, {Wechsler},
  {Wester}, {Yanny}, \& {DES Collaboration}}]{Li17}
{Li}, T.~S., {Simon}, J.~D., {Drlica-Wagner}, A., {et~al.} 2017, \apj, 838, 8

\bibitem[{{Li} {et~al.}(2019){Li}, {Koposov}, {Zucker}, {Lewis}, {Kuehn},
  {Simpson}, {Ji}, {Shipp}, {Mao}, {Geha}, {Pace}, {Mackey}, {Allam}, {Tucker},
  {Da Costa}, {Erkal}, {Simon}, {Mould}, {Martell}, {Wan}, {De Silva},
  {Bechtol}, {Balbinot}, {Belokurov}, {Bland-Hawthorn}, {Casey}, {Cullinane},
  {Drlica-Wagner}, {Sharma}, {Vivas}, {Wechsler}, {Yanny}, \& {S5
  Collaboration}}]{Li19}
{Li}, T.~S., {Koposov}, S.~E., {Zucker}, D.~B., {et~al.} 2019, \mnras, 490,
  3508

\bibitem[{{Li} {et~al.}(2020){Li}, {Koposov}, {Erkal}, {Ji}, {Shipp}, {Pace},
  {Hilmi}, {Kuehn}, {Lewis}, {Mackey}, {Simpson}, {Wan}, {Zucker},
  {Bland-Hawthorn}, {Cullinane}, {Da Costa}, {Drlica-Wagner}, {Hattori},
  {Martell}, \& {Sharma}}]{Li20}
{Li}, T.~S., {Koposov}, S.~E., {Erkal}, D., {et~al.} 2020, arXiv e-prints,
  arXiv:2006.10763

\bibitem[{{Lim} {et~al.}(2020){Lim}, {C{\^o}t{\'e}}, {Peng}, {Ferrarese},
  {Roediger}, {Durrell}, {Mihos}, {Wang}, {Gwyn}, {Cuillandre}, {Liu},
  {S{\'a}nchez-Janssen}, {Toloba}, {Sales}, {Guhathakurta}, {Lan{\c{c}}on}, \&
  {Puzia}}]{Lim20}
{Lim}, S., {C{\^o}t{\'e}}, P., {Peng}, E.~W., {et~al.} 2020, \apj, 899, 69

\bibitem[{{Lindegren} {et~al.}(2018){Lindegren}, {Hern{\'a}ndez}, {Bombrun},
  {Klioner}, {Bastian}, {Ramos-Lerate}, {de Torres}, {Steidelm{\"u}ller},
  {Stephenson}, {Hobbs}, {Lammers}, {Biermann}, {Geyer}, {Hilger}, {Michalik},
  {Stampa}, {McMillan}, {Casta{\~n}eda}, {Clotet}, {Comoretto}, {Davidson},
  {Fabricius}, {Gracia}, {Hambly}, {Hutton}, {Mora}, {Portell}, {van Leeuwen},
  {Abbas}, {Abreu}, {Altmann}, {Andrei}, {Anglada}, {Balaguer-N{\'u}{\~n}ez},
  {Barache}, {Becciani}, {Bertone}, {Bianchi}, {Bouquillon}, {Bourda},
  {Br{\"u}semeister}, {Bucciarelli}, {Busonero}, {Buzzi}, {Cancelliere},
  {Carlucci}, {Charlot}, {Cheek}, {Crosta}, {Crowley}, {de Bruijne}, {de
  Felice}, {Drimmel}, {Esquej}, {Fienga}, {Fraile}, {Gai}, {Garralda},
  {Gonz{\'a}lez-Vidal}, {Guerra}, {Hauser}, {Hofmann}, {Holl}, {Jordan},
  {Lattanzi}, {Lenhardt}, {Liao}, {Licata}, {Lister}, {L{\"o}ffler},
  {Marchant}, {Martin-Fleitas}, {Messineo}, {Mignard}, {Morbidelli}, {Poggio},
  {Riva}, {Rowell}, {Salguero}, {Sarasso}, {Sciacca}, {Siddiqui}, {Smart},
  {Spagna}, {Steele}, {Taris}, {Torra}, {van Elteren}, {van Reeven}, \&
  {Vecchiato}}]{Lindegren18}
{Lindegren}, L., {Hern{\'a}ndez}, J., {Bombrun}, A., {et~al.} 2018, \aap, 616,
  A2

\bibitem[{{Lindegren} {et~al.}(2020){Lindegren}, {Klioner}, {Hern{\'a}ndez},
  {Bombrun}, {Ramos-Lerate}, {Steidelm{\"u}ller}, {Bastian}, {Biermann}, {de
  Torres}, {Gerlach}, {Geyer}, {Hilger}, {Hobbs}, {Lammers}, {McMillan},
  {Stephenson}, {Casta{\~n}eda}, {Davidson}, {Fabricius}, {Gracia-Abril},
  {Portell}, {Rowell}, {Teyssier}, {Torra}, {Bartolom{\'e}}, {Clotet},
  {Garralda}, {Gonz{\'a}lez-Vidal}, {Torra}, {Abbas}, {Altmann}, {Anglada
  Varela}, {Balaguer-N{\'u}{\~n}ez}, {Balog}, {Barache}, {Becciani}, {Bernet},
  {Bertone}, {Bianchi}, {Bouquillon}, {Brown}, {Bucciarelli}, {Busonero},
  {Butkevich}, {Buzzi}, {Cancelliere}, {Carlucci}, {Charlot}, {Cioni},
  {Crosta}, {Crowley}, {del Peloso}, {del Pozo}, {Drimmel}, {Esquej}, {Fienga},
  {Fraile}, {Gai}, {Garcia-Reinaldos}, {Guerra}, {Hambly}, {Hauser},
  {Jan{\ss}en}, {Jordan}, {Kostrzewa-Rutkowska}, {Lattanzi}, {Liao}, {Licata},
  {Lister}, {L{\"o}ffler}, {Marchant}, {Masip}, {Mignard}, {Mints}, {Molina},
  {Mora}, {Morbidelli}, {Murphy}, {Pagani}, {Panuzzo}, {Pe{\~n}alosa Esteller},
  {Poggio}, {Re Fiorentin}, {Riva}, {Sagrist{\`a} Sell{\'e}s}, {Sanchez
  Gimenez}, {Sarasso}, {Sciacca}, {Siddiqui}, {Smart}, {Souami}, {Spagna},
  {Steele}, {Taris}, {Utrilla}, {van Reeven}, \& {Vecchiato}}]{Lindegren20}
{Lindegren}, L., {Klioner}, S.~A., {Hern{\'a}ndez}, J., {et~al.} 2020, arXiv
  e-prints, arXiv:2012.03380

\bibitem[{{Lynden-Bell}(1975)}]{LyndenBell75}
{Lynden-Bell}, D. 1975, Vistas in Astronomy, 19, 299

\bibitem[{{Martin} {et~al.}(2016){Martin}, {Ibata}, {Lewis}, {McConnachie},
  {Babul}, {Bate}, {Bernard}, {Chapman}, {Collins}, {Conn}, {Crnojevi{\'c}},
  {Fardal}, {Ferguson}, {Irwin}, {Mackey}, {McMonigal}, {Navarro}, \&
  {Rich}}]{Martin16}
{Martin}, N.~F., {Ibata}, R.~A., {Lewis}, G.~F., {et~al.} 2016, \apj, 833, 167

\bibitem[{{McConnachie} \& {Venn}(2020a)}]{McConnachie20}
{McConnachie}, A.~W., \& {Venn}, K.~A. 2020a, \aj, 160, 124

\bibitem[{{McConnachie} \& {Venn}(2020b)}]{McConnachie20b}
---. 2020b, Research Notes of the American Astronomical Society, 4, 229

\bibitem[{{McGaugh}(2016)}]{McGaugh16}
{McGaugh}, S.~S. 2016, \apjl, 832, L8

\bibitem[{{McGaugh} \& {de Blok}(1998)}]{McGaugh98a}
{McGaugh}, S.~S., \& {de Blok}, W.~J.~G. 1998, \apj, 499, 41

\bibitem[{{McMillan}(2017)}]{McMillan17}
{McMillan}, P.~J. 2017, \mnras, 465, 76

\bibitem[{Miszalski {et~al.}(2006)Miszalski, Shortridge, Saunders, Parker, \&
  Croom}]{Miszalski:2006ef}
Miszalski, B., Shortridge, K., Saunders, W., Parker, Q.~A., \& Croom, S.~M.
  2006, \mnras, 371, 1537.
\newblock
  \url{https://academic.oup.com/mnras/article-lookup/doi/10.1111/j.1365-2966.2006.10777.x}

\bibitem[{{Mu{\~n}oz} {et~al.}(2018){Mu{\~n}oz}, {C{\^o}t{\'e}}, {Santana},
  {Geha}, {Simon}, {Oyarz{\'u}n}, {Stetson}, \& {Djorgovski}}]{Munoz18}
{Mu{\~n}oz}, R.~R., {C{\^o}t{\'e}}, P., {Santana}, F.~A., {et~al.} 2018, \apj,
  860, 66

\bibitem[{{M{\"u}ller} {et~al.}(2020){M{\"u}ller}, {Marleau}, {Duc}, {Habas},
  {Fensch}, {Emsellem}, {Poulain}, {Lim}, {Agnello}, {Durrell}, {Paudel},
  {S{\'a}nchez-Janssen}, \& {van der Burg}}]{Muller20}
{M{\"u}ller}, O., {Marleau}, F.~R., {Duc}, P.-A., {et~al.} 2020, \aap, 640,
  A106

\bibitem[{{Nidever} {et~al.}(2020){Nidever}, {Dey}, {Fasbender}, {Juneau},
  {Meisner}, {Wishart}, {Scott}, {Matt}, {Nikutta}, \& {Pucha}}]{Nidever20}
{Nidever}, D.~L., {Dey}, A., {Fasbender}, K., {et~al.} 2020, arXiv e-prints,
  arXiv:2011.08868

\bibitem[{{Pace} \& {Li}(2019)}]{Pace19}
{Pace}, A.~B., \& {Li}, T.~S. 2019, \apj, 875, 77

\bibitem[{{Pace} {et~al.}(2020){Pace}, {Kaplinghat}, {Kirby}, {Simon},
  {Tollerud}, {Mu{\~n}oz}, {C{\^o}t{\'e}}, {Djorgovski}, \& {Geha}}]{Pace20}
{Pace}, A.~B., {Kaplinghat}, M., {Kirby}, E., {et~al.} 2020, \mnras, 495, 3022

\bibitem[{{Pe{\~n}arrubia} {et~al.}(2008){Pe{\~n}arrubia}, {Navarro}, \&
  {McConnachie}}]{Penarrubia08}
{Pe{\~n}arrubia}, J., {Navarro}, J.~F., \& {McConnachie}, A.~W. 2008, \apj,
  673, 226

\bibitem[{{Petersen} \& {Pe{\~n}arrubia}(2021)}]{Petersen21}
{Petersen}, M.~S., \& {Pe{\~n}arrubia}, J. 2021, Nature Astronomy, 5, 251

\bibitem[{{Pietrzy{\'n}ski} {et~al.}(2013){Pietrzy{\'n}ski}, {Graczyk},
  {Gieren}, {Thompson}, {Pilecki}, {Udalski}, {Soszy{\'n}ski}, {Koz{\l}owski},
  {Konorski}, {Suchomska}, {Bono}, {Moroni}, {Villanova}, {Nardetto},
  {Bresolin}, {Kudritzki}, {Storm}, {Gallenne}, {Smolec}, {Minniti}, {Kubiak},
  {Szyma{\'n}ski}, {Poleski}, {Wyrzykowski}, {Ulaczyk}, {Pietrukowicz},
  {G{\'o}rski}, \& {Karczmarek}}]{Pietrzyski13}
{Pietrzy{\'n}ski}, G., {Graczyk}, D., {Gieren}, W., {et~al.} 2013, \nat, 495,
  76

\bibitem[{Price-Whelan {et~al.}(2020)Price-Whelan, Sipőcz, Lenz, Greco, Major,
  Koposov, Oh, \& Lim}]{gala:zenodo}
Price-Whelan, A., Sipőcz, B., Lenz, D., {et~al.} 2020, adrn/gala: v1.3, vv1.3,
   Zenodo, doi:10.5281/zenodo.4159870.
\newblock \url{https://doi.org/10.5281/zenodo.593786}

\bibitem[{Price-Whelan(2017)}]{gala}
Price-Whelan, A.~M. 2017, The Journal of Open Source Software, 2,
  doi:10.21105/joss.00388.
\newblock \url{https://doi.org/10.21105%2Fjoss.00388}

\bibitem[{{Price-Whelan} {et~al.}(2018){Price-Whelan}, {Sip{\H{o}}cz},
  {G{\"u}nther}, {Lim}, {Crawford}, {Conseil}, {Shupe}, {Craig}, {Dencheva},
  {Ginsburg}, {VanderPlas}, {Bradley}, {P{\'e}rez-Su{\'a}rez}, {de Val-Borro},
  {Paper Contributors}, {Aldcroft}, {Cruz}, {Robitaille}, {Tollerud},
  {Coordination Committee}, {Ardelean}, {Babej}, {Bach}, {Bachetti}, {Bakanov},
  {Bamford}, {Barentsen}, {Barmby}, {Baumbach}, {Berry}, {Biscani}, {Boquien},
  {Bostroem}, {Bouma}, {Brammer}, {Bray}, {Breytenbach}, {Buddelmeijer},
  {Burke}, {Calderone}, {Cano Rodr{\'\i}guez}, {Cara}, {Cardoso}, {Cheedella},
  {Copin}, {Corrales}, {Crichton}, {D{\textquoteright}Avella}, {Deil},
  {Depagne}, {Dietrich}, {Donath}, {Droettboom}, {Earl}, {Erben}, {Fabbro},
  {Ferreira}, {Finethy}, {Fox}, {Garrison}, {Gibbons}, {Goldstein}, {Gommers},
  {Greco}, {Greenfield}, {Groener}, {Grollier}, {Hagen}, {Hirst}, {Homeier},
  {Horton}, {Hosseinzadeh}, {Hu}, {Hunkeler}, {Ivezi{\'c}}, {Jain}, {Jenness},
  {Kanarek}, {Kendrew}, {Kern}, {Kerzendorf}, {Khvalko}, {King}, {Kirkby},
  {Kulkarni}, {Kumar}, {Lee}, {Lenz}, {Littlefair}, {Ma}, {Macleod},
  {Mastropietro}, {McCully}, {Montagnac}, {Morris}, {Mueller}, {Mumford},
  {Muna}, {Murphy}, {Nelson}, {Nguyen}, {Ninan}, {N{\"o}the}, {Ogaz}, {Oh},
  {Parejko}, {Parley}, {Pascual}, {Patil}, {Patil}, {Plunkett}, {Prochaska},
  {Rastogi}, {Reddy Janga}, {Sabater}, {Sakurikar}, {Seifert}, {Sherbert},
  {Sherwood-Taylor}, {Shih}, {Sick}, {Silbiger}, {Singanamalla}, {Singer},
  {Sladen}, {Sooley}, {Sornarajah}, {Streicher}, {Teuben}, {Thomas},
  {Tremblay}, {Turner}, {Terr{\'o}n}, {van Kerkwijk}, {de la Vega}, {Watkins},
  {Weaver}, {Whitmore}, {Woillez}, {Zabalza}, \& {Contributors}}]{astropy:2018}
{Price-Whelan}, A.~M., {Sip{\H{o}}cz}, B.~M., {G{\"u}nther}, H.~M., {et~al.}
  2018, \aj, 156, 123

\bibitem[{{Rey} {et~al.}(2019){Rey}, {Pontzen}, {Agertz}, {Orkney}, {Read},
  {Saintonge}, \& {Pedersen}}]{Rey19}
{Rey}, M.~P., {Pontzen}, A., {Agertz}, O., {et~al.} 2019, \apjl, 886, L3

\bibitem[{{Riello} {et~al.}(2020){Riello}, {De Angeli}, {Evans}, {Montegriffo},
  {Carrasco}, {Busso}, {Palaversa}, {Burgess}, {Diener}, {Davidson}, {Rowell},
  {Fabricius}, {Jordi}, {Bellazzini}, {Pancino}, {Harrison}, {Cacciari}, {van
  Leeuwen}, {Hambly}, {Hodgkin}, {Osborne}, {Altavilla}, {Barstow}, {Brown},
  {Castellani}, {Cowell}, {De Luise}, {Gilmore}, {Giuffrida}, {Hidalgo},
  {Holland}, {Marinoni}, {Pagani}, {Piersimoni}, {Pulone}, {Ragaini}, {Rainer},
  {Richards}, {Sanna}, {Walton}, {Weiler}, \& {Yoldas}}]{Riello20}
{Riello}, M., {De Angeli}, F., {Evans}, D.~W., {et~al.} 2020, arXiv e-prints,
  arXiv:2012.01916

\bibitem[{{Sameie} {et~al.}(2020){Sameie}, {Chakrabarti}, {Yu},
  {Boylan-Kolchin}, {Vogelsberger}, {Zavala}, \& {Hernquist}}]{Sameie20}
{Sameie}, O., {Chakrabarti}, S., {Yu}, H.-B., {et~al.} 2020, arXiv e-prints,
  arXiv:2006.06681

\bibitem[{{Sanders} {et~al.}(2018){Sanders}, {Evans}, \& {Dehnen}}]{Sanders18}
{Sanders}, J.~L., {Evans}, N.~W., \& {Dehnen}, W. 2018, \mnras, 478, 3879

\bibitem[{{Schlafly} \& {Finkbeiner}(2011)}]{Schlafly11}
{Schlafly}, E.~F., \& {Finkbeiner}, D.~P. 2011, \apj, 737, 103

\bibitem[{{Schlegel} {et~al.}(1998){Schlegel}, {Finkbeiner}, \&
  {Davis}}]{Schlegel:1998}
{Schlegel}, D.~J., {Finkbeiner}, D.~P., \& {Davis}, M. 1998, \apj, 500, 525

\bibitem[{{Sch{\"o}nrich} {et~al.}(2010){Sch{\"o}nrich}, {Binney}, \&
  {Dehnen}}]{Schonrich10}
{Sch{\"o}nrich}, R., {Binney}, J., \& {Dehnen}, W. 2010, \mnras, 403, 1829

\bibitem[{{Shipp} {et~al.}(2019){Shipp}, {Li}, {Pace}, {Erkal},
  {Drlica-Wagner}, {Yanny}, {Belokurov}, {Wester}, {Koposov}, {Kuehn}, {Lewis},
  {Simpson}, {Wan}, {Zucker}, {Martell}, {Wang}, \& {S5
  Collaboration}}]{Shipp19}
{Shipp}, N., {Li}, T.~S., {Pace}, A.~B., {et~al.} 2019, \apj, 885, 3

\bibitem[{{Simon}(2019)}]{Simon19}
{Simon}, J.~D. 2019, \araa, 57, 375

\bibitem[{{Speagle}(2020)}]{dynesty}
{Speagle}, J.~S. 2020, \mnras, 493, 3132

\bibitem[{{Spencer} {et~al.}(2018){Spencer}, {Mateo}, {Olszewski}, {Walker},
  {McConnachie}, \& {Kirby}}]{Spencer18}
{Spencer}, M.~E., {Mateo}, M., {Olszewski}, E.~W., {et~al.} 2018, \aj, 156, 257

\bibitem[{{Tanoglidis} {et~al.}(2021){Tanoglidis}, {Drlica-Wagner}, {Wei},
  {Li}, {S{\'a}nchez}, {Zhang}, {Peter}, {Feldmeier-Krause}, {Prat}, {Casey},
  {Palmese}, {S{\'a}nchez}, {DeRose}, {Conselice}, {Gagnon}, {Abbott},
  {Aguena}, {Allam}, {Avila}, {Bechtol}, {Bertin}, {Bhargava}, {Brooks},
  {Burke}, {Rosell}, {Kind}, {Carretero}, {Chang}, {Costanzi}, {da Costa}, {De
  Vicente}, {Desai}, {Diehl}, {Doel}, {Eifler}, {Everett}, {Evrard},
  {Flaugher}, {Frieman}, {Garc{\'\i}a-Bellido}, {Gerdes}, {Gruendl},
  {Gschwend}, {Gutierrez}, {Hartley}, {Hollowood}, {Huterer}, {James},
  {Krause}, {Kuehn}, {Kuropatkin}, {Maia}, {March}, {Marshall}, {Menanteau},
  {Miquel}, {Ogando}, {Paz-Chinch{\'o}n}, {Romer}, {Roodman}, {Sanchez},
  {Scarpine}, {Serrano}, {Sevilla-Noarbe}, {Smith}, {Suchyta}, {Tarle},
  {Thomas}, {Tucker}, {Walker}, \& {DES Collaboration}}]{Tanoglidis21}
{Tanoglidis}, D., {Drlica-Wagner}, A., {Wei}, K., {et~al.} 2021, \apjs, 252, 18

\bibitem[{{Torrealba} {et~al.}(2016){Torrealba}, {Koposov}, {Belokurov}, \&
  {Irwin}}]{Torrealba16}
{Torrealba}, G., {Koposov}, S.~E., {Belokurov}, V., \& {Irwin}, M. 2016,
  \mnras, 459, 2370

\bibitem[{{Torrealba} {et~al.}(2019){Torrealba}, {Belokurov}, {Koposov}, {Li},
  {Walker}, {Sanders}, {Geringer-Sameth}, {Zucker}, {Kuehn}, {Evans}, \&
  {Dehnen}}]{Torrealba19}
{Torrealba}, G., {Belokurov}, V., {Koposov}, S.~E., {et~al.} 2019, \mnras, 488,
  2743

\bibitem[{{Tremmel} {et~al.}(2020){Tremmel}, {Wright}, {Brooks}, {Munshi},
  {Nagai}, \& {Quinn}}]{Tremmel20}
{Tremmel}, M., {Wright}, A.~C., {Brooks}, A.~M., {et~al.} 2020, \mnras, 497,
  2786

\bibitem[{{van der Marel} {et~al.}(2002){van der Marel}, {Alves}, {Hardy}, \&
  {Suntzeff}}]{vanderMarel02}
{van der Marel}, R.~P., {Alves}, D.~R., {Hardy}, E., \& {Suntzeff}, N.~B. 2002,
  \aj, 124, 2639

\bibitem[{{van~der~Walt} {et~al.}(2011){van~der~Walt}, Colbert, \&
  Varoquaux}]{numpy}
{van~der~Walt}, S., Colbert, S.~C., \& Varoquaux, G. 2011, Computing in Science
  \& Engineering, 13, 22.
\newblock
  \url{http://scitation.aip.org/content/aip/journal/cise/13/2/10.1109/MCSE.2011.37}

\bibitem[{{van Dokkum} {et~al.}(2015){van Dokkum}, {Abraham}, {Merritt},
  {Zhang}, {Geha}, \& {Conroy}}]{vanDokkum15}
{van Dokkum}, P.~G., {Abraham}, R., {Merritt}, A., {et~al.} 2015, \apjl, 798,
  L45

\bibitem[{{Vasiliev} {et~al.}(2021){Vasiliev}, {Belokurov}, \&
  {Erkal}}]{Vasiliev21}
{Vasiliev}, E., {Belokurov}, V., \& {Erkal}, D. 2021, \mnras, 501, 2279

\bibitem[{{Vivas} {et~al.}(2020){Vivas}, {Walker}, {Mart{\'\i}nez-V{\'a}zquez},
  {Monelli}, {Bono}, {Dorta}, {Nidever}, {Fiorentino}, {Gallart}, {Andreuzzi},
  {Braga}, {Dall'Ora}, {Olsen}, \& {Stetson}}]{Vivas20}
{Vivas}, A.~K., {Walker}, A.~R., {Mart{\'\i}nez-V{\'a}zquez}, C.~E., {et~al.}
  2020, \mnras, 492, 1061

\bibitem[{{Walker} {et~al.}(2019){Walker}, {Mart{\'\i}nez-V{\'a}zquez},
  {Monelli}, {Vivas}, {Bono}, {Gallart}, {Cassisi}, {Andreuzzi}, {Bernard},
  {Dall'Ora}, {Fiorentino}, {Nidever}, {Olsen}, {Pietrinferni}, \&
  {Stetson}}]{Walker19}
{Walker}, A.~R., {Mart{\'\i}nez-V{\'a}zquez}, C.~E., {Monelli}, M., {et~al.}
  2019, \mnras, 490, 4121

\bibitem[{{Walker} {et~al.}(2008){Walker}, {Mateo}, \& {Olszewski}}]{Walker08}
{Walker}, M.~G., {Mateo}, M., \& {Olszewski}, E.~W. 2008, \apjl, 688, L75

\bibitem[{{Walker} {et~al.}(2016){Walker}, {Mateo}, {Olszewski}, {Koposov},
  {Belokurov}, {Jethwa}, {Nidever}, {Bonnivard}, {Bailey}, {Bell}, \&
  {Loebman}}]{Walker16}
{Walker}, M.~G., {Mateo}, M., {Olszewski}, E.~W., {et~al.} 2016, \apj, 819, 53

\bibitem[{{Wan} {et~al.}(2020){Wan}, {Lewis}, {Li}, {Simpson}, {Martell},
  {Zucker}, {Mould}, {Erkal}, {Pace}, {Mackey}, {Ji}, {Koposov}, {Kuehn},
  {Shipp}, {Balbinot}, {Bland-Hawthorn}, {Casey}, {Da Costa}, {Kafle},
  {Sharma}, \& {De Silva}}]{Wan20}
{Wan}, Z., {Lewis}, G.~F., {Li}, T.~S., {et~al.} 2020, \nat, 583, 768

\bibitem[{Waskom {et~al.}(2016)Waskom, Botvinnik, drewokane, Hobson, Halchenko,
  Lukauskas, Warmenhoven, Cole, Hoyer, Vanderplas, gkunter, Villalba, Quintero,
  Martin, Miles, Meyer, Augspurger, Yarkoni, Bachant, Evans, Fitzgerald, Nagy,
  Ziegler, Megies, Wehner, St-Jean, Coelho, Hitz, Lee, \& Rocher}]{seaborn}
Waskom, M., Botvinnik, O., drewokane, {et~al.} 2016, seaborn: v0.7.0 (January
  2016), , , doi:10.5281/zenodo.45133.
\newblock \url{http://dx.doi.org/10.5281/zenodo.45133}

\bibitem[{{Wenger} {et~al.}(2000){Wenger}, {Ochsenbein}, {Egret}, {Dubois},
  {Bonnarel}, {Borde}, {Genova}, {Jasniewicz}, {Lalo{\"e}}, {Lesteven}, \&
  {Monier}}]{Simbad}
{Wenger}, M., {Ochsenbein}, F., {Egret}, D., {et~al.} 2000, \aaps, 143, 9

\bibitem[{{Wheeler} {et~al.}(2017){Wheeler}, {Pace}, {Bullock},
  {Boylan-Kolchin}, {O{\~n}orbe}, {Elbert}, {Fitts}, {Hopkins}, \&
  {Kere{\v{s}}}}]{Wheeler17}
{Wheeler}, C., {Pace}, A.~B., {Bullock}, J.~S., {et~al.} 2017, \mnras, 465,
  2420

\bibitem[{{Wilson}(1955)}]{Wilson55}
{Wilson}, A.~G. 1955, \pasp, 67, 27

\bibitem[{{Wolf} {et~al.}(2010){Wolf}, {Martinez}, {Bullock}, {Kaplinghat},
  {Geha}, {Mu{\~n}oz}, {Simon}, \& {Avedo}}]{Wolf10}
{Wolf}, J., {Martinez}, G.~D., {Bullock}, J.~S., {et~al.} 2010, \mnras, 406,
  1220

\bibitem[{{Wright} {et~al.}(2021){Wright}, {Tremmel}, {Brooks}, {Munshi},
  {Nagai}, {Sharma}, \& {Quinn}}]{Wright21}
{Wright}, A.~C., {Tremmel}, M., {Brooks}, A.~M., {et~al.} 2021, \mnras,
  arXiv:2005.07634

\bibitem[{{Zivick} {et~al.}(2020){Zivick}, {Kallivayalil}, \& {van der
  Marel}}]{Zivick20}
{Zivick}, P., {Kallivayalil}, N., \& {van der Marel}, R.~P. 2020, arXiv
  e-prints, arXiv:2011.02525

\bibitem[{Zonca {et~al.}(2019)Zonca, Singer, Lenz, Reinecke, Rosset, Hivon, \&
  Gorski}]{Zonca2019}
Zonca, A., Singer, L., Lenz, D., {et~al.} 2019, Journal of Open Source
  Software, 4, 1298.
\newblock \url{https://doi.org/10.21105/joss.01298}

\end{thebibliography}

\appendix

\section{Data Tables}\label{sec:tables}
Tables \ref{tab:ant2} and \ref{tab:cra2} contain machine-readable tables for properties of stars in Antlia 2 and Crater 2, respectively.

\begin{rotatetable*}
\begin{deluxetable*}{lrrrrrrrrrrrcrr}
\centerwidetable
\tablecolumns{14}
\tabletypesize{\footnotesize}
\tablecaption{\label{tab:ant2}Antlia2 Data}
\tablehead{Gaia Source ID & RA & Dec & Gaia G & $\mu_{\alpha^*}$ & $\mu_\delta$ & $v_{\rm{hel}}$ & $e_v$ & [Fe/H] & $e_{\text{Fe}}$ & [Fe/H] & $e_{\text{Fe}}$ & binary & $p_{\text{mem}}$ & S/N \\
 & (deg) & (deg) & (mag) & (\masyr) & (\masyr) & (\kms) & (\kms) & (DR2.2) & (DR2.2) & (CaT) & (CaT) & & }
\startdata
 5430123686295270528 & 142.53357 & -37.91439 & 19.45 & 0.880 & 1.683 & 79.01 & 2.49 & 0.06 & 0.16 & -0.93 & 0.43 & N & 0.00 & 7.4 \\
 5430127019191414144 & 142.53799 & -37.79130 & 19.00 & -0.352 & -0.771 & 38.98 & 1.96 & -1.12 & 0.22 & -1.52 & 0.36 & N & 0.00 & 7.1 \\
 5430144851898821120 & 142.37813 & -37.56172 & 17.76 & -0.239 & 0.021 & 180.92 & 1.16 & 0.10 & 0.09 & -1.99 & 0.56 & N & 0.00 & 10.0 \\
 5432707744779139200 & 145.32899 & -38.46708 & 18.81 & -0.212 & -0.019 & 305.01 & 2.25 & -1.14 & 0.12 & -2.42 & 0.20 & N & 0.99 & 9.9 \\
 5432712035451068672 & 145.21509 & -38.39652 & 18.72 & -0.087 & 0.256 & 291.02 & 2.03 & -0.00 & 99.00 & -0.81 & 0.30 & N & 1.00 & 9.0 \\
 5432719255291437952 & 145.42085 & -38.47635 & 18.15 & -1.494 & -1.091 & 11.57 & 0.91 & -0.34 & 0.01 & -2.03 & 0.14 & N & 0.00 & 32.2 \\
 5432719697672405248 & 145.43874 & -38.44401 & 18.83 & -0.156 & 0.361 & 59.50 & 1.65 & -0.17 & 0.14 & -0.98 & 0.56 & N & 0.00 & 6.7 \\
 5432721660472955008 & 145.53543 & -38.36941 & 19.53 & -0.528 & 1.769 & 267.07 & 4.21 & -2.69 & 0.49 & -2.30 & 0.40 & N & 0.00 & 4.1 \\
 5432724542395731200 & 145.81868 & -38.34123 & 19.30 & 0.194 & -1.032 & 42.57 & 3.45 & -0.55 & 0.20 & -0.44 & 0.73 & N & 0.00 & 5.2 \\
 5432730246112749056 & 145.68201 & -38.19851 & 19.06 & -0.457 & -0.228 & 20.93 & 2.42 & -0.28 & 0.13 & -0.67 & 0.49 & N & 0.00 & 4.3 \\
 5432735159555276800 & 145.44373 & -38.23879 & 19.58 & -0.067 & -0.293 & 290.96 & 4.71 & -0.69 & 0.40 & -1.44 & 0.59 & N & 0.96 & 4.1 \\
 5432736057203390208 & 145.36721 & -38.30410 & 19.12 & 0.526 & 0.889 & 5.46 & 2.58 & -0.69 & 0.43 & -0.82 & 0.59 & N & 0.00 & 7.4 \\
 5432800619153094784 & 144.62609 & -38.57772 & 18.89 & -0.096 & -0.001 & 295.91 & 3.33 & -2.29 & 0.38 & -1.95 & 0.59 & N & 1.00 & 5.0 \\
 5432815943594883456 & 144.32467 & -38.48502 & 18.81 & 0.094 & 0.269 & 296.32 & 2.10 & 0.10 & 99.00 & -1.72 & 0.45 & N & 1.00 & 7.9 \\
 5432828618045034496 & 143.98730 & -38.75532 & 18.19 & -0.576 & -0.685 & 1.43 & 2.20 & -1.02 & 0.12 & -1.52 & 0.43 & N & 0.00 & 10.8 \\
 5432829824929895424 & 143.95847 & -38.70626 & 19.69 & -0.513 & 0.825 & 55.46 & 5.62 & 0.10 & 0.20 & -1.46 & 0.97 & N & 0.00 & 4.7 \\
 5432847584621060352 & 143.88772 & -38.55068 & 19.16 & -0.845 & -0.183 & -29.79 & 3.78 & -1.23 & 0.19 & -1.12 & 0.46 & N & 0.00 & 5.5 \\
 5432851982667756160 & 143.79150 & -38.46329 & 19.48 & -0.735 & 0.301 & 19.97 & 3.70 & -0.53 & 0.31 & -2.29 & 1.43 & N & 0.00 & 4.0 \\
 5432856896110223872 & 144.25280 & -38.51619 & 18.79 & 0.108 & -0.749 & 40.59 & 2.08 & 0.06 & 0.31 & -0.42 & 0.33 & N & 0.00 & 6.8 \\
 5432859026413982720 & 144.10070 & -38.52739 & 19.13 & -0.230 & 0.207 & -10.08 & 1.88 & -0.08 & 0.15 & -0.73 & 0.53 & N & 0.00 & 6.2 \\
\enddata
\tablecomments{The first 20 rows are shown here. The full table is available online as a machine-readable table.}
\end{deluxetable*}
\end{rotatetable*}

\begin{rotatetable*}
\begin{deluxetable*}{lrrrrrrrrrrrcrr}
\centerwidetable
\tablecolumns{14}
\tabletypesize{\footnotesize}
\tablecaption{\label{tab:cra2}Crater2 Data}
\tablehead{Gaia Source ID & RA & Dec & Gaia G & $\mu_{\alpha^*}$ & $\mu_\delta$ & $v_{\rm{hel}}$ & $e_v$ & [Fe/H] & $e_{\text{Fe}}$ & [Fe/H] & $e_{\text{Fe}}$ & binary & $p_{\text{mem}}$ & S/N \\
 & (deg) & (deg) & (mag) & (\masyr) & (\masyr) & (\kms) & (\kms) & (DR2.2) & (DR2.2) & (CaT) & (CaT) & & }
\startdata
 3543024835796029056 & 177.59227 & -19.35520 & 20.15 & -0.292 & -0.469 & 18.76 & 3.77 & -1.02 & 0.41 & -0.93 & 0.32 & N & 0.00 & 4.1 \\
 3543053487522870912 & 177.62997 & -19.19029 & 19.39 & -0.818 & -1.319 & 44.80 & 3.70 & -1.35 & 0.27 & -1.74 & 0.55 & N & 0.00 & 3.2 \\
 3543056309318642304 & 177.75071 & -19.05828 & 20.08 & 1.705 & -0.197 & -17.76 & 2.28 & -1.12 & 0.13 & -1.49 & 1.38 & N & 0.00 & 4.5 \\
 3543057164015770880 & 177.73569 & -19.02361 & 18.46 & -0.189 & -0.082 & 94.03 & 1.29 & -2.09 & 0.20 & -1.86 & 0.20 & N & 1.00 & 13.6 \\
 3543058435326076800 & 177.93394 & -19.06145 & 17.85 & -0.307 & 0.314 & 18.36 & 1.49 & -0.92 & 0.07 & -2.25 & 0.21 & N & 0.00 & 10.8 \\
 3543064516999805056 & 177.90168 & -18.94614 & 19.94 & -5.778 & 1.410 & 159.86 & 4.89 & -1.28 & 0.21 & -1.39 & 0.61 & N & 0.00 & 3.2 \\
 3543142822843457280 & 178.17990 & -18.96932 & 19.04 & -1.569 & 0.279 & 12.82 & 4.55 & -1.02 & 0.19 & -1.35 & 0.42 & N & 0.00 & 4.1 \\
 3543156193077156608 & 178.05288 & -18.88107 & 19.00 & -0.608 & 0.213 & 17.78 & 1.58 & -1.11 & 0.14 & -1.74 & 0.34 & N & 0.00 & 7.2 \\
 3543158357740679168 & 178.12481 & -18.81081 & 16.51 & 0.209 & -0.409 & -6.91 & 1.19 & -0.69 & 0.03 & -2.04 & 0.20 & N & 0.00 & 15.7 \\
 3543195152725167744 & 178.31004 & -18.51465 & 19.69 & -0.298 & -0.231 & 228.00 & 6.57 & -1.99 & 0.39 & -0.95 & 0.76 & N & 0.00 & 3.6 \\
 3543195771200464128 & 178.32564 & -18.48732 & 19.63 & 0.140 & 0.883 & 22.13 & 3.94 & -1.63 & 0.24 & -1.87 & 0.42 & N & 0.00 & 3.7 \\
 3543399455728835200 & 176.97970 & -19.40391 & 20.21 & -3.184 & -0.093 & 50.80 & 3.43 & -1.17 & 0.20 & -0.45 & 0.51 & N & 0.00 & 4.3 \\
 3543412271911301504 & 176.98540 & -19.31376 & 19.10 & -1.015 & -1.042 & 281.00 & 2.89 & -1.15 & 0.19 & -0.79 & 0.46 & N & 0.00 & 5.1 \\
 3543415364287322368 & 176.74756 & -19.30076 & 19.70 & -0.296 & -0.407 & 220.88 & 6.35 & -1.56 & 0.36 & -2.27 & 0.49 & N & 0.00 & 3.2 \\
 3543480445928587904 & 176.54002 & -19.06992 & 19.17 & -1.122 & -0.912 & 258.82 & 5.14 & -0.91 & 0.29 & -1.56 & 0.52 & N & 0.00 & 3.7 \\
 3543774423553961600 & 177.08996 & -19.36856 & 19.43 & -2.134 & -0.708 & 216.75 & 8.53 & -1.59 & 0.32 & -2.18 & 0.45 & N & 0.00 & 4.1 \\
 3543796211924443520 & 177.00119 & -19.02632 & 17.87 & 0.041 & 0.084 & 21.92 & 0.92 & -0.52 & 0.05 & -1.33 & 0.15 & N & 0.00 & 17.2 \\
 3543807065305006464 & 177.51300 & -19.04849 & 19.92 & -3.646 & -1.318 & 366.55 & 7.40 & -2.02 & 0.44 & -2.19 & 0.68 & N & 0.00 & 3.1 \\
 3543810741797032960 & 177.41296 & -19.00933 & 18.99 & 0.022 & 0.159 & 91.90 & 2.13 & -2.31 & 0.22 & -2.21 & 0.26 & N & 1.00 & 8.5 \\
 3543811600790524288 & 177.45042 & -18.95351 & 19.09 & -1.197 & 0.025 & -12.52 & 1.99 & -0.63 & 0.12 & -1.56 & 0.48 & N & 0.00 & 5.7 \\
\enddata
\tablecomments{The first 20 rows are shown here. The full table is available online as a machine-readable table.}
\end{deluxetable*}
\end{rotatetable*}

\section{Marginalizing Over Known Parallax}\label{sec:pmcorrection}

The \Gaia astrometric solution provides a full 5-dimensional covariance matrix for sky position, proper motions, and parallax.
If a true value is known for one of these parameters, we can marginalize over the covariance matrix given the true value, which can introduce small corrections to the mean values and uncertainties.
We now show the procedure to use the known distance to our galaxies as a known parallax to correct the proper motions.
First we define the vector $V$ and matrix $M$ containing the measured astrometric information from \Gaia:
\begin{align}
    V &= \begin{pmatrix}
      \mu_{\alpha^*} \\
      \mu_{\delta} \\
      \omega
    \end{pmatrix} \\
    M &= \begin{pmatrix}
      \sigma_{\mu\alpha}^2 & \sigma_{\mu\alpha}\sigma_{\mu\delta}\rho_{\mu\alpha,\mu\delta} & \sigma_{\mu\alpha}\sigma_{\omega}\rho_{\mu\alpha,\omega} \\
      \sigma_{\mu\delta}\sigma_{\mu\alpha}\rho_{\mu\alpha,\mu\delta} & \sigma_{\mu\delta}^2 & \sigma_{\mu\delta}\sigma_{\omega}\rho_{\mu\delta,\omega} \\
      \sigma_{\omega}\sigma_{\mu\alpha}\rho_{\mu\alpha,\omega} & \sigma_{\omega}\sigma_{\mu\delta}\rho_{\mu\delta,\omega} & \sigma_{\omega}^2
    \end{pmatrix}
\end{align}
Also define $V_0$ and $M_0$ as vectors and matrices containing the true parallax $\omega_{\text{gal}}$ and uncertainty $\sigma_{\omega,\text{gal}}$ (based on the distance modulus and uncertainty):
\begin{align}
    V_0 &= \begin{pmatrix}
      0 \\
      0 \\
      \omega_{\text{gal}}
    \end{pmatrix} \\
    M_0 &= \begin{pmatrix}
      0 & 0 & 0 \\
      0 & 0 & 0 \\
      0 & 0 & \sigma_{\omega,\text{gal}}^2
    \end{pmatrix}
\end{align}
Then if all distributions are multivariate Gaussians, marginalizing over the known values is given by:
\begin{align}
    M_{\text{new}} &= (M^{-1} + M_0^{-1})^{-1} \\
    V_{\text{new}} &= M_{\text{new}} \left(M^{-1}V + M_0^{-1}V_0 \right)
\end{align}
We use the first two components of $V_{\text{new}}$ and $M_{\text{new}}$ as the proper motions for our kinematic modeling.
Including the known distance affects the proper motions of Ant2 by less than 0.01 \masyr, but it decreases $\mu_\delta$ for Cra2 by 0.01 \masyr.

\end{document}